\documentclass[twocolumn,showpac,aps,prb,superscriptaddress]{revtex4-2}

\usepackage{graphicx}
\usepackage{dcolumn}
\usepackage{bm}
\usepackage{amssymb}
\usepackage{amsmath}
\usepackage{xcolor}
\usepackage[T1]{fontenc}
\usepackage[utf8]{inputenc}
\usepackage{xcolor}

\usepackage{tagging}

\DeclareUnicodeCharacter{2212}{-}

\begin{document}
\title{Role of intercalated cobalt in the electronic structure of Co$_{1/3}$NbS$_2$}

  \author{Petar Pop\v{c}evi\'{c}}
  \email{ppopcevic@ifs.hr}
  \affiliation{Institute of Physics, Bijeni\v{c}ka c. 46, 10000 Zagreb, Croatia}
  \author{Yuki Utsumi}
  \email{yutsumi@ifs.hr}
  \affiliation{Institute of Physics, Bijeni\v{c}ka c. 46, 10000 Zagreb, Croatia} 
  \author{Izabela Bia\l o}
  \affiliation{Institute of Solid State Physics, TU Wien, 1040 Vienna, Austria}
  \affiliation{AGH University of Science and Technology, Faculty of Physics and Applied Computer Science, 30-059 Krakow, Poland}
   \author{Wojciech Tabis}
  \affiliation{Institute of Solid State Physics, TU Wien, 1040 Vienna, Austria}
  \affiliation{AGH University of Science and Technology, Faculty of Physics and Applied Computer Science, 30-059 Krakow, Poland}
  \author{Mateusz A. Gala}
  \affiliation{AGH University of Science and Technology, Faculty of Physics and Applied Computer Science, 30-059 Krakow, Poland}
  \author{Marcin Rosmus}
  \affiliation{Solaris National synchrotron Radiation Centre, Jagiellonian University, Czerwone Maki 98, 30-392, Krakow, Poland}
  \affiliation{Faculty of Physics, Astronomy, and Applied Computer Science, Jagiellonian University, Lojasiewicza 11, 30-348 Krakow, Poland}
  \author{Jacek J. Kolodziej}
  \affiliation{Solaris National synchrotron Radiation Centre, Jagiellonian University, Czerwone Maki 98, 30-392, Krakow, Poland}
  \affiliation{Faculty of Physics, Astronomy, and Applied Computer Science, Jagiellonian University, Lojasiewicza 11, 30-348 Krakow, Poland}
  \author{Natalia Tomaszewska}
  \affiliation{Solaris National synchrotron Radiation Centre, Jagiellonian University, Czerwone Maki 98, 30-392, Krakow, Poland}
   \author{Mariusz Garb}
  \affiliation{Faculty of Physics, Astronomy, and Applied Computer Science, Jagiellonian University, Lojasiewicza 11, 30-348 Krakow, Poland}
  \author{Helmuth Berger}
\affiliation{Laboratory of Physics of Complex Matter, \'Ecole polytechnique f\'{e}d\'{e}rale de Lausanne, 1015 Lausanne, Switzerland}
  \author{Ivo Batisti\'{c}}
  \affiliation{Department of Physics, Faculty of Science, University of Zagreb, Bijeni\v{c}ka c. 32, 10000 Zagreb, Croatia}
  \author{Neven Bari\v{s}i\'{c}}
  \affiliation{Institute of Solid State Physics, TU Wien, 1040 Vienna, Austria}
  \affiliation{Department of Physics, Faculty of Science, University of Zagreb, Bijeni\v{c}ka c. 32, 10000 Zagreb, Croatia}
  \author{L\'{a}szl\'{o} Forr\'{o}}
  \affiliation{Laboratory of Physics of Complex Matter, \'Ecole polytechnique f\'{e}d\'{e}rale de Lausanne, 1015 Lausanne, Switzerland}
  \affiliation{Stavropoulos Center for Complex Quantum Matter, University of Notre Dame, Notre Dame, Indiana 46556, USA}
  \author{Eduard Tuti\v{s}}
  \email{etutis@ifs.hr}
  \affiliation{Institute of Physics, Bijeni\v{c}ka c. 46, 10000 Zagreb, Croatia}

\date{\today}

\begin{abstract}
  
  Co$_{1/3}$NbS$_2$ is the magnetic intercalate of 2H-NbS$_2$ where electronic itinerant and magnetic properties strongly influence each other throughout the phase diagram.  
  Here we report the angle-resolved photoelectron spectroscopy (ARPES) study in Co$_{1/3}$NbS$_2$. 
  In agreement with previous reports, the observed electronic structure seemingly resembles the one of the parent material 2H-NbS$_2$, with the shift in Fermi energy of 0.5 eV accounting for the charge transfer of approximately two electrons from each Co ion into the NbS$_2$ layers. 
  However, in addition, and in contrast to previous reports, we observe significant departures that cannot be explained by the rigid band shift accompanied by minor deformation of bands: 
  First, entirely unrelated to the 2H-NbS$_2$ electronic structure, a shallow electronic band is found crossing the Fermi level near the boundary of the first Brillouin zone of Co$_{1/3}$NbS$_2$.
  The evolution of the experimental spectra upon varying the incident photon energy suggests the Co origin of this band. 
  Second, the Nb bonding band, found deeply submerged below the Fermi level at the $\Gamma$ point, indicates that the interlayer-hybridization is significantly amplified by intercalation, with Co magnetic ions probably acting as strong covalent bridges between NbS$_2$ layers.   
  The strong hybridization between orbitals that support the itinerant states and the orbitals hosting the local magnetic moments indicates the importance of strong electronic correlations, with the interlayer coupling playing an exquisite role.

\end{abstract}
\maketitle

\section{Introduction}  

   For decades, the transition metal dichalcogenides (TMDs) have been fascinating the scientific community by the diversity of electronic phases in their phase diagrams.
   The phases that have been discussed include various charge-density-wave states, superconducting phases, Mott, band Jahn-Teller, excitonic and magnetic insulator states \cite{Wilson1969, Wilson1975, Tosatti1976, Naito1982, Morosan2006, Cercellier2007, Sipos2008, Guo2017, Liu2021}.
     More recently, this fascination by bulk properties  has been accompanied by the surge of interest in atomically thin TMD systems, which offer variety of promising technologies and fundamental particularities \cite{Manzeli2017, Choi2017, Wang2018, Hsu2017}. 

  As TMDs consist of layers separated by the van der Waals gaps, they also offer the possibility of being intercalated by various types of atoms and molecules \cite{Friend1977}, further contributing to the abundance of phases.
  The intercalation by magnetic ions has proven particularly attractive, as it leads to rather complex magnetic structures in $M_x$TiS$_2$, $M_x$TaS$_2$, $M_x$NbS$_2$, and $M_x$NbSe$_2$ ($M$: $3d$ transition-metal ions) \cite{Anzenhofer1970, Friend1977, Parkin1980a, Parkin1980b, Parkin1983, Marseglia1983, Negishi1987, VanderBerg1986}. 
  An example is found in 2H-NbS$_2$, which becomes superconducting below 6 K \cite{Maaren1966}.
  If intercalated by Fe, Co and Ni, the superconductivity is suppressed in favor of the antiferromagnetic ground state.  
  The intercalation with V, Mn, or Cr results in the ferromagnetic order \cite{Friend1977, Parkin1980a, Parkin1980b, Parkin1983}. 
  Moreover, the chiral magnetic soliton lattice has been proposed in the case of Cr and Mn intercalation \cite{Togawa2012, Dai2019}, as well as some other exotic behaviors \cite{Morosan2007, Ghimire2018}.
   A topological semimetal phase has been suggested for certain compounds recently \cite{Inoshita2019}.
  

  \begin{figure*}[t!] 
  \includegraphics[width=\textwidth]{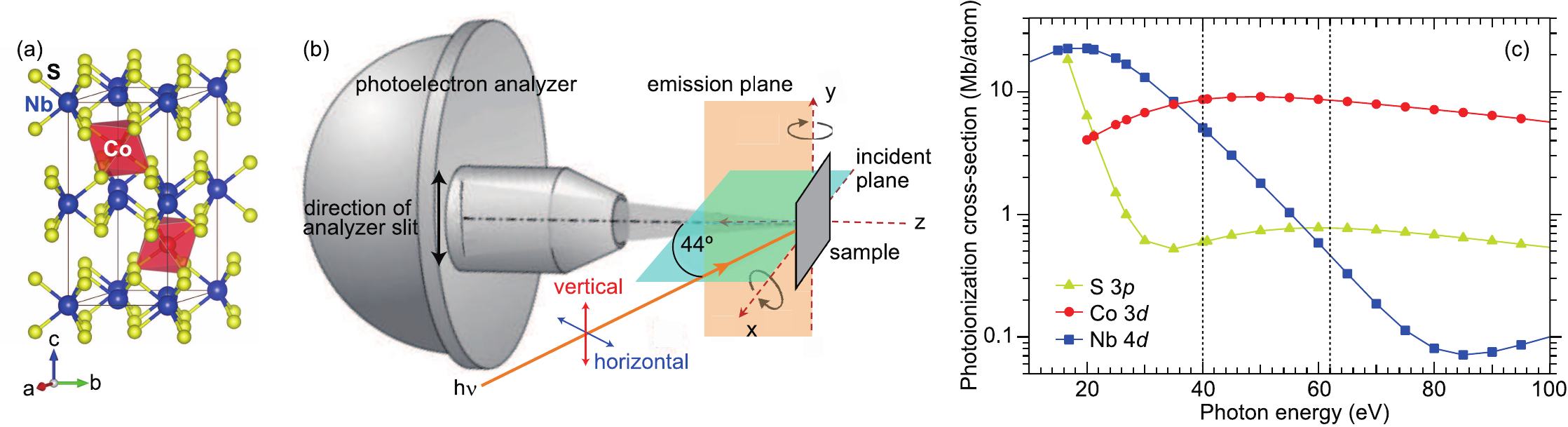}

  \caption{(Color online) (a) Crystal structure of Co$_{1/3}$NbS$_2$,   with two NbS$_2$ layers and two Co-layers entering the crystal unit cell, drawn by VESTA \cite{Momma2011}.    
  (b) The orientations of the photoelectron analyzer and the incident photon beam at the UARPES beamline of the SOLARIS synchrotron. The {\it horizontal} and {\it vertical} light polarizations are defined by blue and red arrows. 
   The horizontal and vertical polarizations correspond to the $p$- and $s$-polarizations of Ref.~\onlinecite{Himpsel1983}, respectively. 
   (c) Atomic subshell photoionization cross-sections for S $3p$, Co $3d$, and Nb $4d$, from Ref.~\onlinecite{Yeh1985}}.  
  \label{Lfig01} 
\end{figure*}

   The simplest view of the electronic and magnetic properties of TMDs intercalated by $3d$ transition-metal ions comes in the form of the rigid-band model \cite{Parkin1980a, Parkin1980b}. 
  According to this model, a portion of the $3d$ electrons from the intercalated ions are transferred to transition metal 4$d$ bands, which remain unaffected in shape. 
  The remaining $3d$ electrons form the magnetic moments at the intercalated ions. 
  They are considered as interacting only weakly with the conduction band and with each other through the Ruderman-Kittel-Kasuya-Yosida (RKKY), and super-exchange mechanisms \cite{Clark1976, Friend1977, Parkin1980a, Parkin1980b}. 
  
   The rigid-band-model view was questioned already early on when the optical conductivity data suggested conducting band broadening as a result of increased interlayer bonding caused by intercalation \cite{Parkin1980c}.
  More recently, the deformation of conduction bands and {\it non-uniform doping} across the Fermi surface has been reported \cite{Battaglia2007, Sirica2016}.
   But indeed, some deformation of the lattice of the host material is inevitably expected upon intercalation. 
   Depending on ionic radii and the ionicity of the intercalated ions, they can act in direction of increasing or decreasing the average distance between layers and deform the host lattice in their close vicinity \cite{Anzenhofer1970, Parkin1980a}.
    This static deformation is likely to affect the band structure of the host material, along with the periodic electrostatic potential coming from the extra ions.   
    However, given the robustness of TMD atomic layers \cite{Manzeli2017}, the effect is expected to be minor leading to a   {\it"quasi-rigid"} 
  response of the electronic bands \footnote{In fact, the mentioned effects  are  routinely taken into account within {\it ab initio} calculations, and were rather thoroughly addressed by exploring the results of multiple DFT calculations that we made in relation this paper.}.
    A more profound effect originates from the {\it overlap} between electronic wavefunctions of the intercalated ions and those of the host material. 
   These overlaps are expected to further deform the conduction bands of the host material and can change the character of the Fermi surface of TMDs from quasi-2D to 3D.
  Recently however, the signs of additional states, unrelated to the band structure of the host material, were reported in the vicinity and at the Fermi level in Cr$_{1/3}$NbS$_2$ \cite{Sirica2016, Sirica2020}.
  The evidence also started to show that the usual assumption of the weak coupling between intercalated magnetic moments and itinerant electrons of the host material needs to be abandoned \cite{Sirica2020}.   
  The appearance of a new type of state at the Fermi level breaks the {\it quasi-rigid-band} notion in magnetically intercalated TMDs.
  The strong mixing between spin and charge degrees of freedom also represents an essential paradigm shift in these materials.

  Here we delve into this subject through the experimental observation of the electronic structure and the Fermi surface of Co$_{1/3}$NbS$_2$ by angle-resolved photoelectron spectroscopy (ARPES).
   This compound is unique from several points of view. 
  For one, it has the lowest magnetic ordering temperature, $T_{\rm N}\sim26$ K \cite{Friend1977} among the materials of the $M_{1/3}$NbS$_2$ series. 
   This ordering temperature can be additionally decreased by applying hydrostatic pressure\cite{Barisic2011}, down to complete suppression of the magnetic order above 2 GPa, the first such suppression observed in intercalated TMDs, and possibly leading to the Co quantum spin liquid embedded between metallic layers \cite{Popcevic2019}. 
     For two, the large anomalous Hall effect was recently reported in a magnetically ordered state and suggested originating from complex noncollinear magnetic textures. 
  The finding sparked a renewed interest in this particular compound\cite{Ghimire2018, Mangelsen2021}.    
  Presumably connected to the anomalous Hall effect, a small ferromagnetic (FM) canting of Co magnetic moments along the $c$-axis was suggested to occur on the top of the antiferromagnetic ordering below $T_{\rm N}$ \cite{Ghimire2018, Popcevic2019}.   
  Furthermore, the electrical resistivity along the $c$-axis is significantly lower in Co$_{1/3}$NbS$_2$ than in the parent 2H-NbS$_2$, suggesting that Co atoms act as the conduction links between layers\cite{Popcevic2019}. 
   Thus, the compound seems like an excellent candidate for testing for the radical departures from the quasi-rigid-band picture. 
  Here we compare the measured spectra with what is known about the electronic structure of 2H-NbS$_2$ and Co$_{1/3}$NbS$_2$. 
    The qualitative differences are explored through theoretical modeling.
  
  Our central claim is that magnetically intercalated TMDs should be regarded as a particular class of strongly correlated electron systems, with strong hybridization between magnetic and metallic layers being mainly responsible for many material properties. 
  
\section{Material and methods}

\begin{figure}[t]
  \includegraphics[width=\columnwidth]{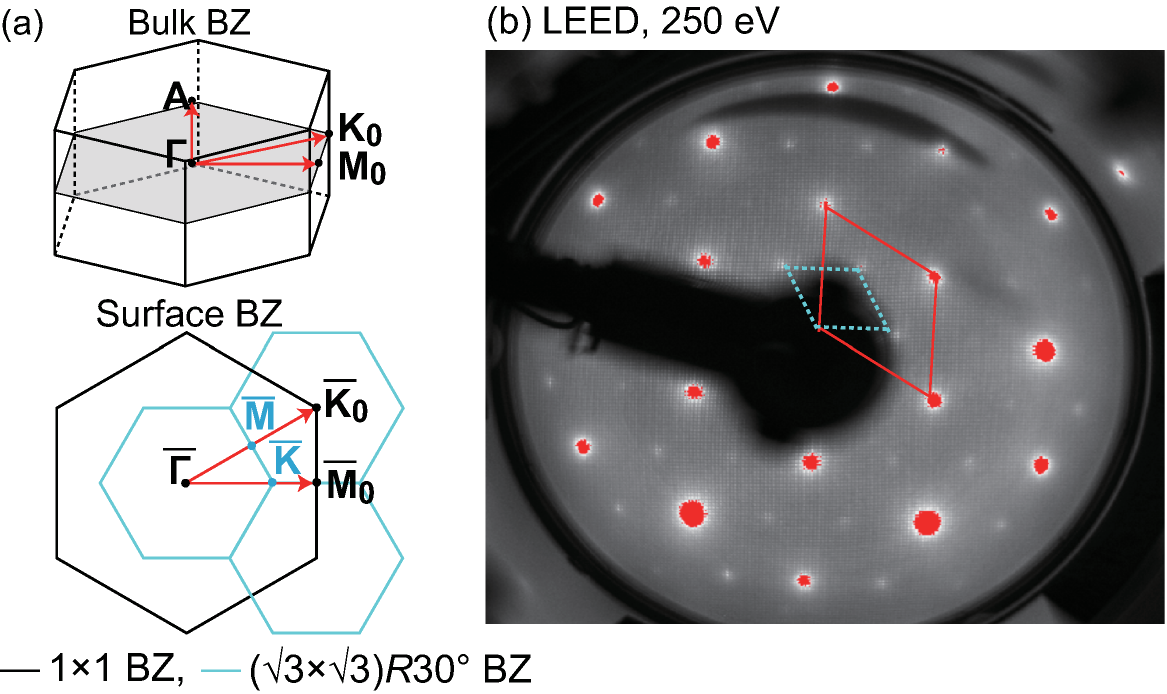}
  \caption{(Color online). (a) Bulk Brillouin zone of 2H-NbS$_2$ (top) and the surface Brillouin zones of 2H-NbS$_2$ and Co$_{1/3}$NbS$_2$ (bottom), shown in black and blue lines, respectively.   (b) The low-energy electron diffraction (LEED) pattern of Co$_{1/3}$NbS$_2$ obtained by using the 250 eV electrons. The stronger red dots correspond to the crystal structure of 2H-NbS$_2$. The weaker intensity white peaks correspond to the ($\sqrt{3}\times\sqrt{3}$)$R30^{\circ}$ superlattice of 2H-NbS$_2$. The unit cells in the reciprocal space for two lattices are shown as blue and red rhombuses, respectively.  }
  \label{Lfig02}
  \end{figure}

\begin{figure*}[t!]
  \includegraphics[width=\textwidth]{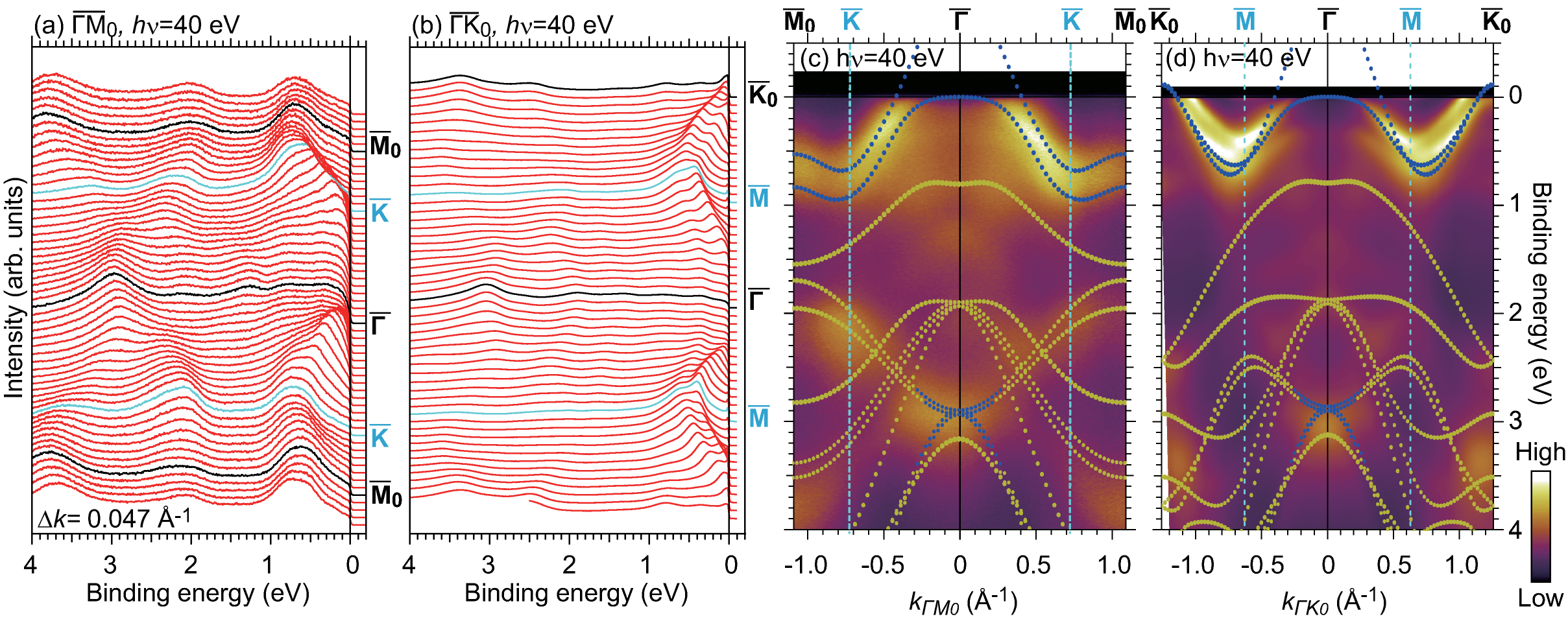}
  \caption{(Color online).  Energy distribution curves (EDCs) along (a) $\Gamma {\rm M_0}$ and (b) $\Gamma {\rm K_0}$ lines in the $k$-space. The ARPES intensity plots for Co$_{1/3}$NbS$_2$ measured along (c) $\Gamma {\rm M_0}$ and (d) $\Gamma {\rm K_0}$ directions using the $h\nu=40$ eV photons, and horizontal polarization. 
In order to cover an extended momentum range, three ARPES images collected $\pm15^{\circ}$-frame are stitched.
  The vertical dashed lines represent the boundaries of the Co$_{1/3}$NbS$_2$ first Brillouin zone.  
  The lines composed of blue and yellow dots represent the energy bands calculated for 2H-NbS$_2$, with the unit cell adjusted to Co$_{1/3}$NbS$_2$, and additional $4/3$ electrons per unit cell (see text). 
  The blue and yellow dots stand for states predominantly composed of Nb- and S-derived orbitals, respectively. 
  The calculated bands are subsequently shifted by 0.1 eV with respect to the Fermi level in the direction that increases the conduction band filling.}
  \label{Lfig03} 
  \end{figure*}

   Single crystals of Co$_{1/3}$NbS$_2$ were grown by chemical vapor transport method using iodine as a transport agent \cite{Friend1977}. 
   The Co atoms occupy the octahedral voids between NbS$_2$ layers (see Fig.~\ref{Lfig01} (a)). 
   When compared to the 1$\times$1 unit cell of 2H-NbS$_2$, the intercalated compound Co$_{1/3}$NbS$_2$ (space group $P6_322$) forms the $\sqrt{3}\times\sqrt{3}$ supercell rotated by 30$^{\circ}$ \cite{Anzenhofer1970, Parkin1983}.   The corresponding first Brillouin zones, the bigger one for 2H-NbS$_2$ and the threefold smaller for Co$_{1/3}$NbS$_2$,  are shown in Fig.~\ref{Lfig03}(a).   For brevity, these will be also referred to as {\it large} and {\it small} Brillouin zones. 
   The grown crystals were characterized by magnetic susceptibility and electrical resistivity measurements. 
   Using the Laue diffraction, the crystals were oriented along the high symmetry $\Gamma{\rm M_0}$ and $\Gamma {\rm K_0}$ directions (see Fig.~\ref{Lfig03} (a) for notation) and glued to Ti flat plates. 
  The clean surfaces of the samples were obtained by cleaving $\it{in~situ}$ within the preparation chamber under ultra-high vacuum (better than 3$\times10^{-10}$ mbar), then immediately transferred to the measurement chamber with the base pressure of 8$\times10^{-11}$ mbar. 
  The cleaved sample surface was checked before and after performing the ARPES measurement by low energy electron diffraction (LEED). 
   The LEED pattern, shown in Fig. \ref{Lfig02}, obtained by using the electrons at the incident energy of 250 eV, clearly shows the periodicity of ($\sqrt{3}\times\sqrt{3}$)$R30^{\circ}$ reciprocal lattice. 
   As usual, the {\it super-spots} in  Fig. \ref{Lfig02}, appearing in white, are considerably weaker in intensity than the red spots corresponding to the parent 2H-NbS$_2$ structure. Along with the smaller concentration of Co atoms than Nb atoms in the material, this is also the consequence of somewhat lower cross-section for elastic electron scattering for Co than for Nb (2:3 at 250 eV)  \cite{NIST}. 
  The experiment was performed at the UARPES beamline of the SOLARIS synchrotron, using the photons in the 34-79 eV energy range. 
  The photoionization cross-section of niobium $4d$ orbitals within this energy range diminishes by two orders of magnitude. 
  In contrast, the photoionization cross-sections of sulfur $3p$ and cobalt $3d$ orbitals stay unchanged, as shown in Fig.~\ref{Lfig01}(c). 
  This property was used to partially separate the features originating from S and Co orbitals from those arising from Nb orbitals.  
   The photoelectrons were collected using a hemispherical analyzer DA30 (VG Scienta) and a multi-channel plate coupled to a CCD detector. 
   The overall energy resolution was estimated to be 20 meV from the Au Fermi edge fitting.


  The linearly polarized light was used in the experiment.
  The experimental setup and two polarization directions used in our measurements are schematically shown in Fig.~\ref{Lfig01}(b). 
   The polarization is called {\it horizontal} when the electric field vector of the incident light is parallel to the {\it incident plane} (the light blue plane in Fig.~\ref{Lfig01} (b)) defined by the light propagation vector and the normal to the sample surface. 
   The polarization is called {\it vertical} when the electric field vector of the incident light is perpendicular to the incident plane and parallel to the {\it emission plane} (the light orange plane in Fig.~\ref{Lfig01} (b) defined by the direction of the analyzer slit). 
   The total intensity of each ARPES spectrum, extending over a measured span of electronic momenta and the binding energy between 0 and 4 eV, is subsequently normalized to the same value. 
  
  The band structure for 2H-NbS$_2$ was calculated using the Quantum ESPRESSO Density Functional Theory (DFT) package \cite{Giannozzi2009, Giannozzi2017}. 
  We use the kinetic energy cutoff of 70-80 Ry for wavefunctions and 500-600 Ry for charge density and potentials.
  The ultrasoft pseudopotentials are from pslibrary \cite{DalCorso2014} based on Perdew-Burke-Ernzerhof exchange-correlation functional \cite{Perdew1996}. 
   The Fermi-energy discontinuity is smeared as proposed by Marzari-Vanderbilt \cite{Marzari1999} with broadening of 0.005-0.01 Ry.
   We made several different DFT calculations, where the DFT-optimized crystal structure was used, unless otherwise stated.
   For the pristine NbS$_2$, the DFT optimized crystal lattice parameters are found to be a$_0$ = 3.349 \AA \phantom{ }and c$_0$ = 13.495 \AA, with the vertical position of sulphur atom at z = 0.133. 
   The DFT optimized crystal lattice parameters for antiferromagnetically ordered Co$_{1/3}$NbS$_2$, with orthorhombic unit cell, are a$_1$ = 5.792 \AA, b$_1$ = 10.032 \AA (= a$_1\sqrt{3})$ and  c$_1$ = 12.199 \AA \cite{Popcevic2019}.
   The crystal parameters for calculation of NbS$_2$ with 2/3 extra electrons per Nb atom are inferred from DFT optimized antiferromagnetically ordered Co$_{1/3}$NbS$_2$ crystal parameters, a$_2$ =  3.344 \AA (= a$_1/\sqrt{3})$ and c$_2$ = 12.199  \AA (= c$_1$) with vertical sulphur position z = 0.123 \cite{Popcevic2019}.
     We use the k-point mesh of $20\times20\times8$ for the Brillouin zone, without shift.


  \begin{figure*}[t!]
  \includegraphics[width=\textwidth]{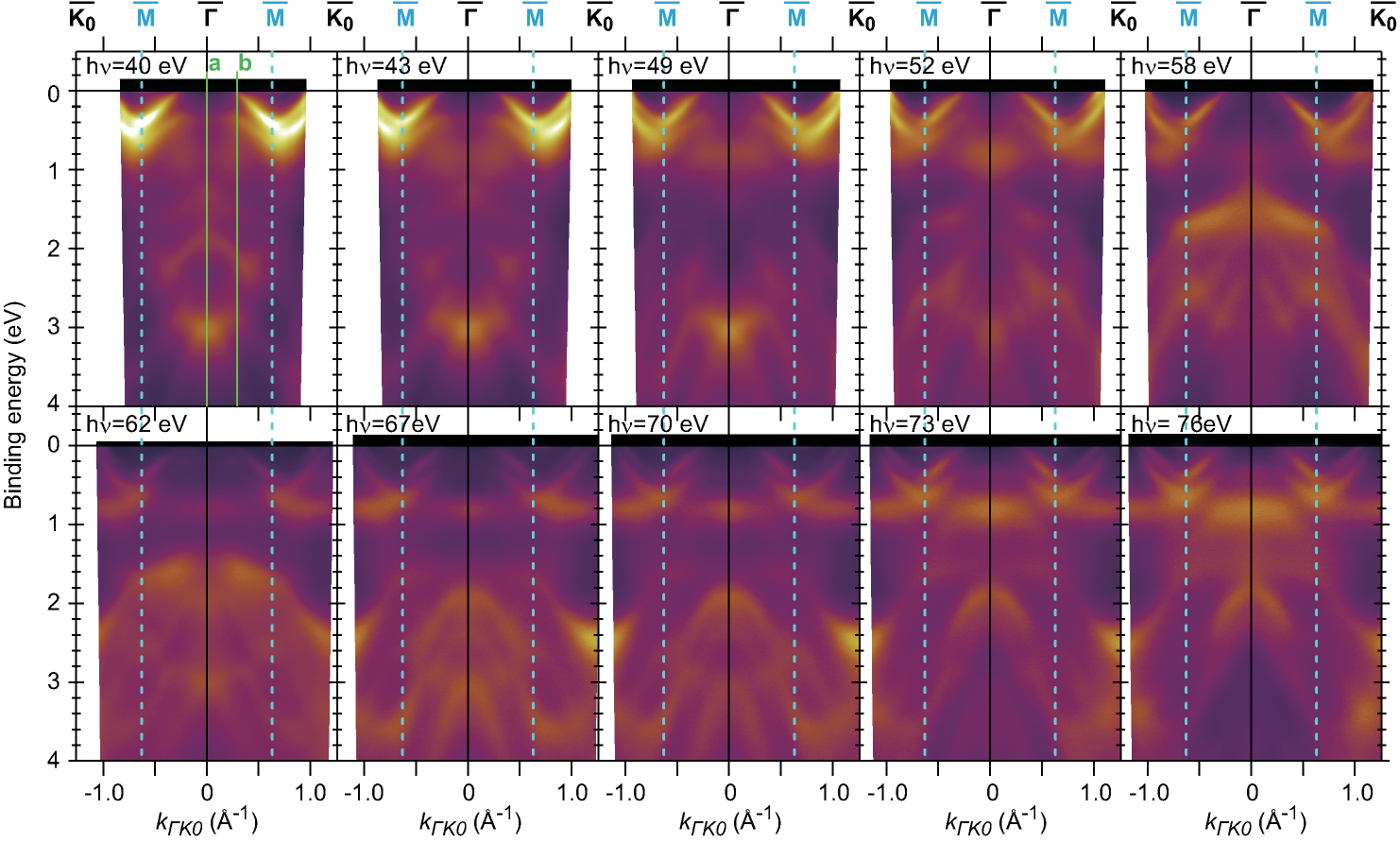}
  \caption{(Color online). The ARPES intensity plots along $\Gamma{\rm K_0}$ direction measured with different energy of the photons used in the experiment. 
  All ARPES images are measured using horizontal polarization and plotted in the same intensity scale. 
  Blue dashed lines correspond to the boundaries of the first Brillouin zone of Co$_{1/3}$NbS$_2$. The green vertical lines in the first panel denote the cuts taken in the spectrum leading to 40 eV curves in Figs.~\ref{Lfig08}(a) and (b).}
  \label{Lfig04}
  \end{figure*}

\section{Experimental results}

\subsection{Scans along cuts in k-space}

  Fig.~\ref{Lfig03} shows the ARPES spectra of Co$_{1/3}$NbS$_2$ collected along $\Gamma {\rm M_0}$ and $\Gamma {\rm K_0}$ directions of the Brillouin zone  of 2H-NbS$_2$. 
  They were measured using photons with energy $h\nu=$ 40 eV and horizontal photon polarization.
   Energy distribution curves (EDCs) along these directions are also shown. 
  The zero in binding energy ($E_B$) corresponds to the Fermi level ($E_{\rm F}$). 
   The blue dashed lines in panels (c) and (d), and full blue lines in panels (b) and (c) correspond to the Brillouin zone boundaries of Co$_{1/3}$NbS$_2$. 
   The spectra on panels (a) and (c) (along $\Gamma$K$_0$ direction) are recorded at 78 K, while the spectra at panels (b) and (d) along $\Gamma$M$_0$ direction are measured at 10 K.
  We point out that $k$-axes essentially specify the $(k_x, k_y)$ coordinates of the initial electron state, and we use the surface Brillouin zone labels to emphasize that. The corresponding labels without bars designate the points at $k_z=0$ in the bulk Brillouin zone, whereas spectra represent contribution coming from electronic states at finite $k_z$. 
    We also compared the ARPES spectra measured below and above the magnetic ordering temperature $T_{\rm N}$, taken along the same direction.    
  We have not observed any significant differences between these spectra within the experimental resolution, except for the thermal broadening around the Fermi level. 
   The details are provided in Supplemental Material \cite{SIref}.
   Some difference in data quality along $\Gamma M_0$ and  $\Gamma K_0$ directions in Fig.~\ref{Lfig03}  should be attributed primarily to a difference in crystal quality of samples used for respective scans  \cite{SIref}.

The dotted curves in panels (c) and (d) in Fig.~\ref{Lfig03}  represent the electronic band dispersions obtained from the DFT calculation for the parent compound, as detailed in Appendix B.
   The EDCs show pronounced dispersive features in the energy windows of $E_B$ = 0-1 eV and 2-4 eV.
   Two bands can be discerned from the spectra in the energy range between zero and 1 eV binding energy (more so along the $\Gamma$K$_0$ direction where the data quality is somewhat higher).
   In the region between 2 and 4 eV binding energy, the signal is more smeared, and individual bands are hard to discern without using the calculated band structure as a reference.

  In Fig.~\ref{Lfig04} we show the ARPES intensity plots of spectra measured at ten photon energies, ranging from $h\nu$ = 40 to 76 eV.     
  These spectra were normalized separately for each panel and plotted using the same intensity scale.
The most notable feature of Fig.~\ref{Lfig04} is the shift in intensity from lower to higher binding energy upon rising the photon energy. 
  The exception may be spotted with the signal at 3 eV binding energy at $\Gamma$ point, which loses the intensity upon increasing the photon energy. 
  The non-monotonous intensity dependence on photon energy is also seen for the signal around 1.5 eV binding energy.
  These features will be discussed in more detail in the next section.
  
\begin{figure}[t]
  \includegraphics[width=\columnwidth]{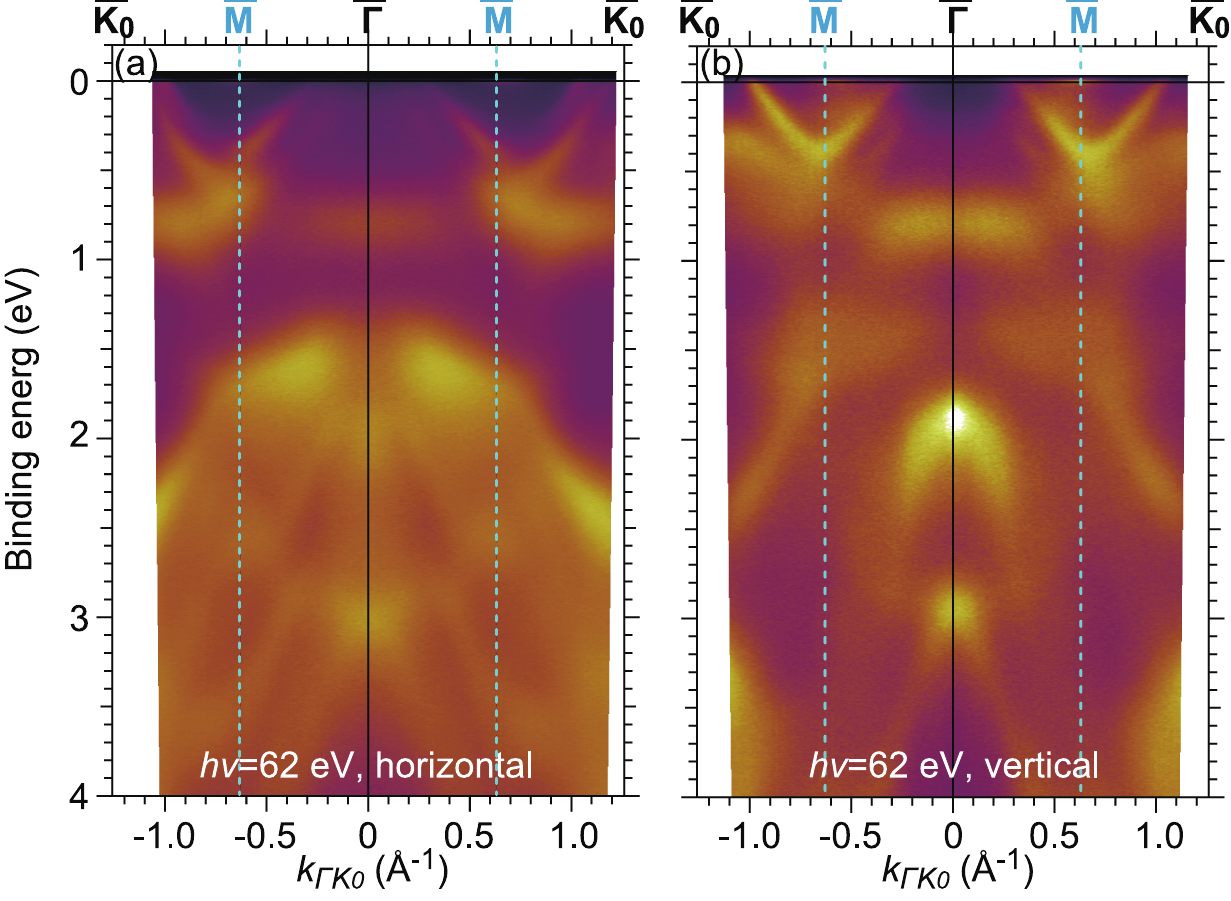} 
  \caption{(Color online). ARPES intensity plots of Co$_{1/3}$NbS$_2$ along the $\Gamma {\rm K_0}$ direction for
   (a) horizontal and (b) vertical photon polarization. The spectra are measured with $h\nu$ = 62 eV at 10 K. 
    The dashed lines in light-blue mark the M high symmetry point at the boundary of the first Brillouin zone of Co$_{1/3}$NbS$_2$. } 
  \label{Lfig05}
  \end{figure}

  Fig.~\ref{Lfig05} shows the ARPES intensity plots measured at $h\nu$ = 62 eV, using horizontal and vertical polarization setups. 
  The intensity distribution of S $3p$-derived bands in the $E_B$ = 2-4 eV region is more homogenous for the horizontal than vertical polarization. 
  A similar difference regarding polarizations can be seen in spectra taken using 40 eV photons \cite{SIref}.
The comparison between Figs.~\ref{Lfig05}(a) and \ref{Lfig03}(d), as well as the comparison between panels in Fig.~\ref{Lfig04},  shows that more bands can be observed at higher photon energy, especially in the $E_B$ = 2-4 eV region. 
	

\subsection {The constant energy scans}


  \begin{figure}[t!]
  \includegraphics[width=\columnwidth]{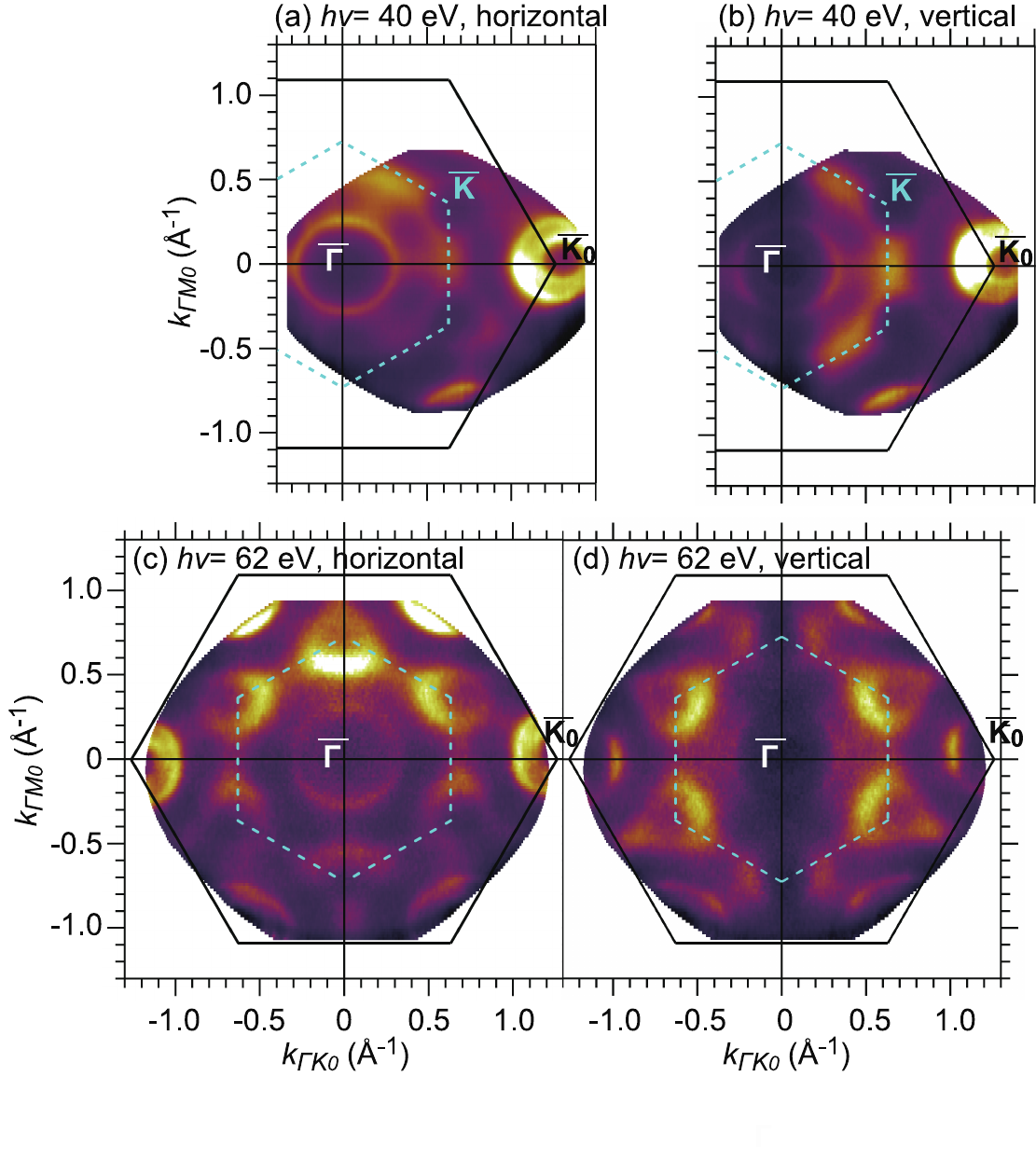} 
  \caption{(Color online) Constant energy cuts at the Fermi level, measured at 10 K using  $h\nu$ = 40 eV photons, in (a) horizontal (a) and  (b) vertical polarization setups. The panels (c) and (d) show the data obtained using $h\nu$ = 62 eV photons, respectively in horizontal and vertical polarization. 
  The large black hexagon and the small hexagons shown in light-blue dashed line represent the first Brillouin zones of NbS$_2$  and Co$_{1/3}$NbS$_2$, respectively. }
  \label{Lfig06}
  \end{figure}

  Fig.~\ref{Lfig06} shows the signal at the Fermi level of Co$_{1/3}$NbS$_2$ measured at photon energies of 40 and 62 eV using horizontal and vertical polarizations. 
   Two concentric Fermi surface sections are observed around K$_0$ point, while only one circular Fermi surface section is present around the $\Gamma$ point.
  The additional set of signals is found around the boundary of first Brillouin zone of Co$_{1/3}$NbS$_2$, or  the ($\sqrt{3}\times\sqrt{3}$)$R30^{\circ}$ superlattice of 2H-NbS$_2$.       
  The signals are visible in both photon polarizations and best seen in scans made using 62 eV energy photons.
   We call this set of signals the $\beta$-{\it band}, following Ref. \onlinecite{Sirica2016} where a somewhat similar feature was first observed in the sister compound Cr$_{1/3}$NbS$_2$ and termed the $\beta$-{\it feature}. 
    Contrary to Cr$_{1/3}$NbS$_2$, in the constant-energy scans in Co$_{1/3}$NbS$_2$ we find the $\beta$-band spreading throughout the boundary of the Brillouin zone.
  In the 40 eV photon energy scans (Fig.~\ref{Lfig06} (a) and (b)), the $\beta$-band can probably be described as another large ring around $\Gamma$-point, concentric to the antibonding Nb $d$ band Fermi surface.  
  This is how the $\beta$-feature was primarily described in Cr$_{1/3}$NbS$_2$ \cite{Sirica2020}. 
  The scans at 62 eV photon energy (Figs.~\ref{Lfig06} (c) and (d)) provide better images of the $\beta$-band, as a necklace composed of of six triangularly shaped or dewdrop-like pockets surrounding the K-points at the boundary of the small Brillouin zone.


  \begin{figure}[t!]
  \includegraphics[width=0.5\textwidth]{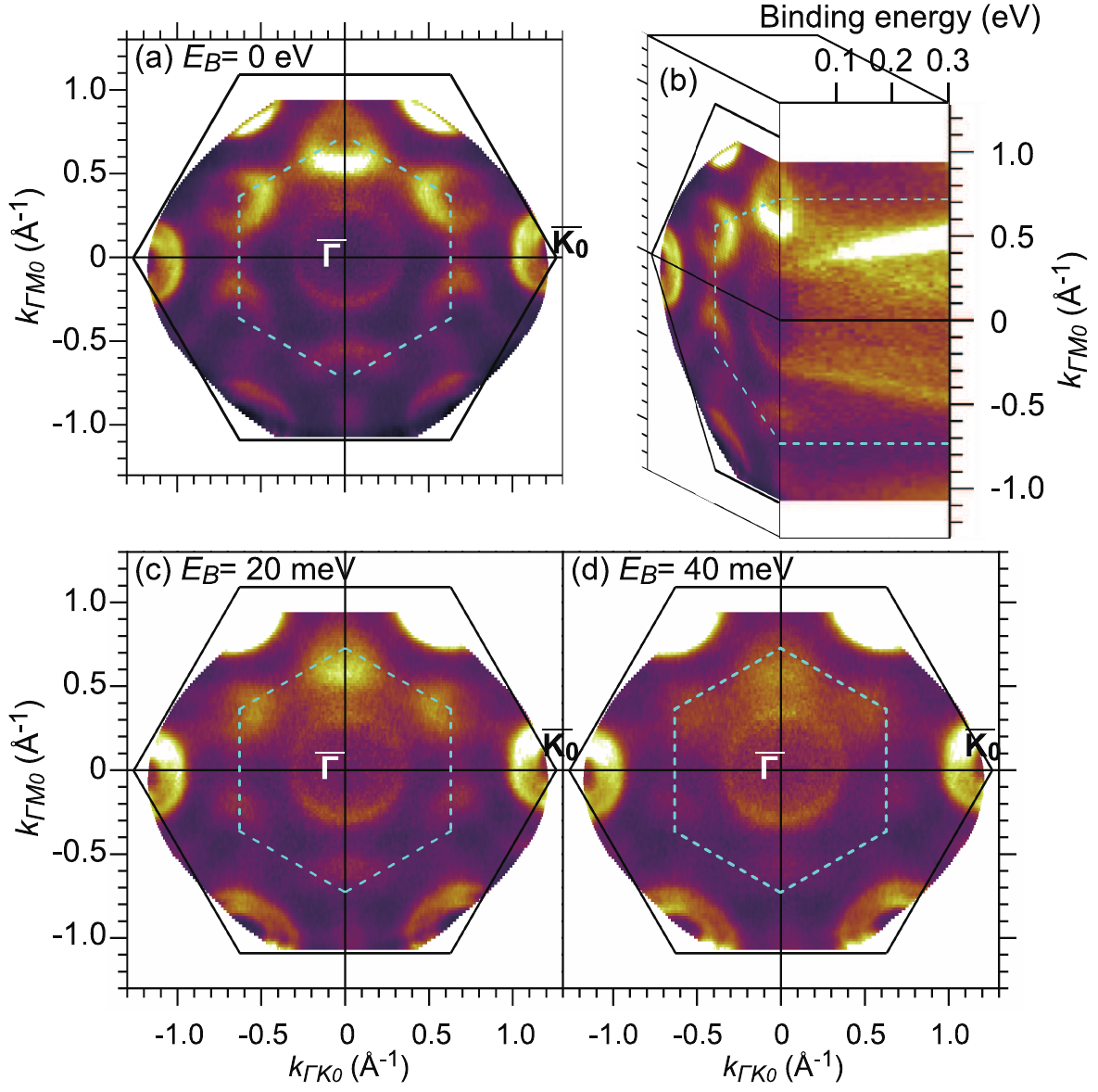}
  \caption{(Color online). Constant energy cuts at (a) $E_{\rm F}$, (b) 20 meV below $E_{\rm F}$ and (c) 40 meV below the Fermi energy of Co$_{1/3}$NbS$_2$, measured at 10 K with $h\nu$ = 62 eV in horizontal polarization. 
  The large black hexagons and the small hexagon shown in light-blue dashed line represent the first Brillouin zones of NbS$_2$  and Co$_{1/3}$NbS$_2$, respectively.
     (d) Left half of the Fermi surface shown on panel (a) is presented with ARPES spectra along $\Gamma$ - M$_0$ direction down to 0.3 eV binding energy.}
  \label{Lfig07}
  \end{figure}
  One can follow these pockets to energies below the Fermi level, assisted by the constant energy scans shown in Fig.~\ref{Lfig07}.  
  The panels (a), (c), and (d) in Fig.~\ref{Lfig07} indicate the collapse of the pockets towards the $\bar{\rm K}$-points at higher binding energies, identifying the $\beta$-band as a collection of shallow electron pockets, of few tens of meV in depth. 
   On a more quantitative side, the size of each pocket can be estimated to be around 2 \% of the large Brillouin zone, and their depth to $(40 \pm 10)$ meV. 
  Within simple parabolic approximation  $\mathcal{E}_\beta=\hbar^2 ( k_\beta)^2/2m_\beta^*$, with  $\mathcal{E}_\beta$ and $k_\beta$ denoting the depth and the radius of one pocket, this yields the effective mass $m_\beta^*=(2.4 \pm 0.6) m_e$ for these pockets, roughly two times bigger than the effective mass $m_0^*$ appropriate to conducting bands of 2H-NbS$_2$. The latter is determined from the Fermi velocity in 2H-NbS$_2$ in the usual way,  $m_0^*=\hbar k_{\rm F}/v_ {\rm F}$, with $v_ {\rm F}$ being read from the slope of the calculated electronic dispersion, and $k_ {\rm F}$ being the characteristic wave vector (radius) of the Fermi surface section. While two approaches to read the effective mass from the electronic dispersion are equivalent for quadratic band dispersions, the former is easier to apply for shallow pockets.   
  
  It should be mentioned that in the sister compound Cr$_{1/3}$NbS$_2$ the $\beta$-feature undergoes splitting into two signals below the helical/ferromagnetic ordering temperature.
   This was argued in favor of the Cr nature of the $\beta$-feature. 
   The resonant photoemission (ResPES) study suggests the same \cite{ Sirica2016,Sirica2020}.
   In Co$_{1/3}$NbS$_2$, we do not observe any significant temperature dependence of the $\beta$-band \cite{SIref}.

\section{Discussion of experimental results} \label{sec:discussion}

\subsection{Scans along cuts in k-space}
  
   The surface sensitivity of ARPES often casts doubt on the relevance of the observed band structure to the bulk electronic structure. 
   For intercalated quasi-two-dimensional TMDs, a cleavage is expected to happen at a layer defined by the intercalated ions, where the binding is weaker than within host material TMD layers. 
   By symmetry, an ideal cleavage is expected to divide the intercalate (Co) atoms equally between the two cleaved surfaces, thus leaving disordered Co at the very surface. 
   The relevance of this surface effect to the interpretation of the measured spectra as the genuine electronic structure of bulk material was assessed for the sister compound Cr$_{1/3}$NbS$_2$ \cite{Sirica2016}. 
   Two types of termination layers were observed by scanning tunneling microscopy on $in~situ$ cleaved Cr$_{1/3}$NbS$_2$.
   The first was the termination layer with disordered Cr atoms, which was connected to the observation of the non-dispersive bands in ARPES spectra \cite{Sirica2016}. 
   The second was the termination layer with ordered S atoms. 
  In that case, the band dispersions and Fermi momentum obtained by low energy ARPES were not significantly modified with respect to those of bulk Cr$_{1/3}$NbS$_2$ \cite{Sirica2016}. 
    In Co$_{1/3}$NbS$_2$, we do not observe non-dispersive bands in the measured energy range, suggesting no significant influence of Co disorder on our spectra.
 
     In spectra presented in Figs \ref{Lfig03}, \ref{Lfig04} and \ref{Lfig05} no clear evidence of the band folding or discontinuations (gaps) are observed at the boundary of the small Brillouin zone.  
 This is in accordance with previous reports in (Mn, Cr)$_{1/3}$NbS$_2$ \cite{Battaglia2007, Sirica2016}.  
  
  
  The Co-induced superstructure showing clearly in LEED and being missing in ARPES seems contradictory. 
  It should be kept in mind, however, that Co-induced superstructure does show in ARPES through the $\beta$-band, which is particular to the intercalated compound and appears long the border of its first Brillouin zone.
  It should be also remarked that the DFT calculation for Co$_{1/3}$NbS$_2$ puts the electronic bands dominated by Co orbitals, well outside the binding energy window that we address in ARPES. 
  These results are graphically summarized in Fig.~\ref{Lfig12} in the Appendix.
  In contrast, all electronic bands contribute to LEED irrespective of their binding energy, as much as they add to the charge density variation in space. 
  Finally, how the electrons perceive the Co-induced charge superstructure depends very much on (Co vs. Nb) elastic-scattering cross-section dependence on the electron energy.
  The additional LEED images  that  illustrate this point are provided in Supplemental Material \cite{SIref}.
  It is also worth mentioning that the band folding/gapping is not always observed by ARPES even when the superstructure spots are well detected by LEED. 
  An example may be found in the Cs ions intercalated between graphene and Ir(111) substrate and forming the  ($\sqrt{3}\times\sqrt{3}$)$R30^{\circ}$ superstructure. Although the formation of the superlattice was clearly observed by LEED, the ARPES results did not show any band folding to reflect the corresponding periodicity of Brillouin zone \cite{Ehlen2020}. 
  On the other hand, the both superlattice LEED spots and the band folding effects in ARPES were observed in the Cs intercalated graphene with ($2\times 2$) superstructure \cite{Ehlen2020, Hell2020}.  
  
  The lack of significant marks of the periodicity of Co$_{1/3}$NbS$_2$ in observed spectra motivates us to start our discussion by comparing measured ARPES spectra to the electronic bands calculated for the parent material 2H-NbS$_2$.
  
  The Nb $4d$ states play a major role in the metallic properties of 2H-NbS$_2$. 
  In 2H-NbS$_2$ the Nb atom is surrounded by six S atoms, forming the NbS$_6$ prism with the $D_{3h}$ symmetry (see  the Fig.~\ref{Lfig01} (a)), and leading to Nb $4d$ orbitals split into $a_1'$ ($d_{z^2}$), $e'$ ($d_{xz, yz}$), and $e''$ ($d_{x^2-y^2, xy}$) states.
     According to electronic structure calculation, the bands crossing the Fermi level are primarily composed of Nb $4d_{z^2}$ orbitals, and, to a smaller extent, of S $p_z$ orbitals \cite{Popcevic2019}.  
    With two Nb atoms per  2H-NbS$_2$ unit cell, two conduction bands emerge from the calculation, as shown in Fig.~\ref{Lfig03} (d) and in more details in Fig. ~\ref{Lfig12} in Appendix.  
      These two bands can also be discerned from the experimental spectra in Fig.~\ref{Lfig03} (d), covering most of the energy range between zero and 1 eV binding energy. 
    The lower band corresponds to the bonding, whereas the upper band corresponds to the antibonding combination of Nb $4d_{z^2}$ orbitals of the two Nb layers in the unit cell.
   One noticeable difference between the experiment and the calculated bands shows in the central part of the figure.  
  As the $\Gamma$ point is being approached, the observed splitting between bonding and antibonding bands is much bigger than for calculated bands. For example, the bonding band in ARPES spectra reaches the $\Gamma$ point around 0.3 eV below the Fermi level. On the other hand, the calculations foresee the bonding band approaching and crossing the Fermi level before reaching the $\Gamma$ point.
  Further on, the dispersions of the bands measured in the energy window of $E_B$ = 2-4 eV, presented in Figs. \ref{Lfig03} (c) and (d), agree rather well with the bands calculated for 2H-NbS$_2$.    
  According to our calculations, these bands are mostly built from sulfur orbitals, with a small contribution from Nb orbitals appearing around $\Gamma$-point at $E_B\sim$ 3 eV, as indicated in the Figs. \ref{Lfig03} (c) and (d).

\subsection {Effect of photon polarization and photon energy on ARPES spectra}

  Several factors influence the intensity variation within a single ARPES spectrum and the differences between spectra taken at different setups. 
  Those differences are usually attributed to the dependence of the transition matrix element on photon energy, light polarization, symmetry of the underlying electronic wavefunctions, as well as the geometry of the experimental setup \cite{Moser2017}.
   The influence of light polarization on ARPES intensity  in Co$_{1/3}$NbS$_2$ is exemplified in Fig.~\ref{Lfig05}.
   The spectra recorded using different photon polarizations are usually used to analyze the orbital content of the electronic bands \cite{Cao2013, Sirica2016}. 
   For this, it is advantageous to align the emission plane of the experimental setup to the mirror plane of the crystal.  
  Unfortunately, this condition cannot be fulfilled in our case, as the intercalated compound has no mirror symmetry planes.	

   The influence of variation in photon energy on ARPES spectra is shown in Fig.~\ref{Lfig04}. 
     It primarily manifests through the shift in intensity from lower to higher binding energy upon rising the photon energy.
   This shift is in accordance with Fig.~\ref{Lfig01} (c) and our expectations regarding the atomic character of different bands, encoded through colors of the calculated curves in Figs.~\ref{Lfig03}(c) and (d), with Nb orbitals contributing mainly to bands up to 1 eV binding energy, and sulfur orbitals dominating the bands in the remaining part of the spectrum.

  In principle, the variation of photon energy can also offer information on band dispersion in the direction perpendicular to the surface, i.e., the $k_z$ dispersion \cite{Damascelli2004}.
  In TMDs, which are usually described as quasi-two-dimensional systems, this dispersion is expected to be weak compared to band dispersion in the direction parallel to layers. 
  On the other hand, the $k_z$ dispersion may be particularly interesting upon changing the coupling between layers through intercalation. 
  Indeed, the Co-intercalation in 2H-NbS$_2$ was found to reduce the electrical resistivity in the direction perpendicular to layers by approximately two times, as well as to reduce the anisotropy of the electrical resistivity \cite{Popcevic2019}. 
  The DFT calculations in Co$_{1/3}$NbS$_2$ predict the bonding Nb $d_{z^2}$ band crossing the Fermi level along the $k_x=0$, $k_y=0$ axis with considerable $k_z$ dispersion \cite{Popcevic2019}.
  The same calculations also predict the $k_z$ dispersion for the topmost sulfur band as the most pronounced \cite{Popcevic2019, SIref}.
  One can expect these dispersion effects to be observable in Fig.~\ref{Lfig04}, as different photon energies should reflect initial electronic states at different $k_z$'s, for any fixed values of $k_x$ and $k_y$.
  The momentum along the $z$ direction may be calculated using the expression \cite{Damascelli2004}: 
  \begin{equation} \label{eq:kz}
  k_z=\sqrt{\frac{2m}{\hbar^2}(E_k{\rm cos}^2\theta+V_0)}.
  \end{equation}
  Here $E_k$ stands for the kinetic energy of photoelectrons at the detector, $\theta$ represents the angle between the surface normal and photoelectron wave vector, $m$ denotes the free electron mass, and V$_0$ is the inner potential.
  The value of V$_0$ is generally determined experimentally, $V_0$ = 14 eV being recently proposed for the sister compound Cr$_{1/3}$NbS$_2$ \cite{Sirica2016}. 
  With $V_0$ of this magnitude, and with the photon energy range from 34 to 79 eV, one expects to scan over the $k_z$ range exceeding the span of $2\pi/c$ of the first Brillouin zone.
  

  \begin{figure}[!ht]
  \includegraphics[width=\columnwidth]{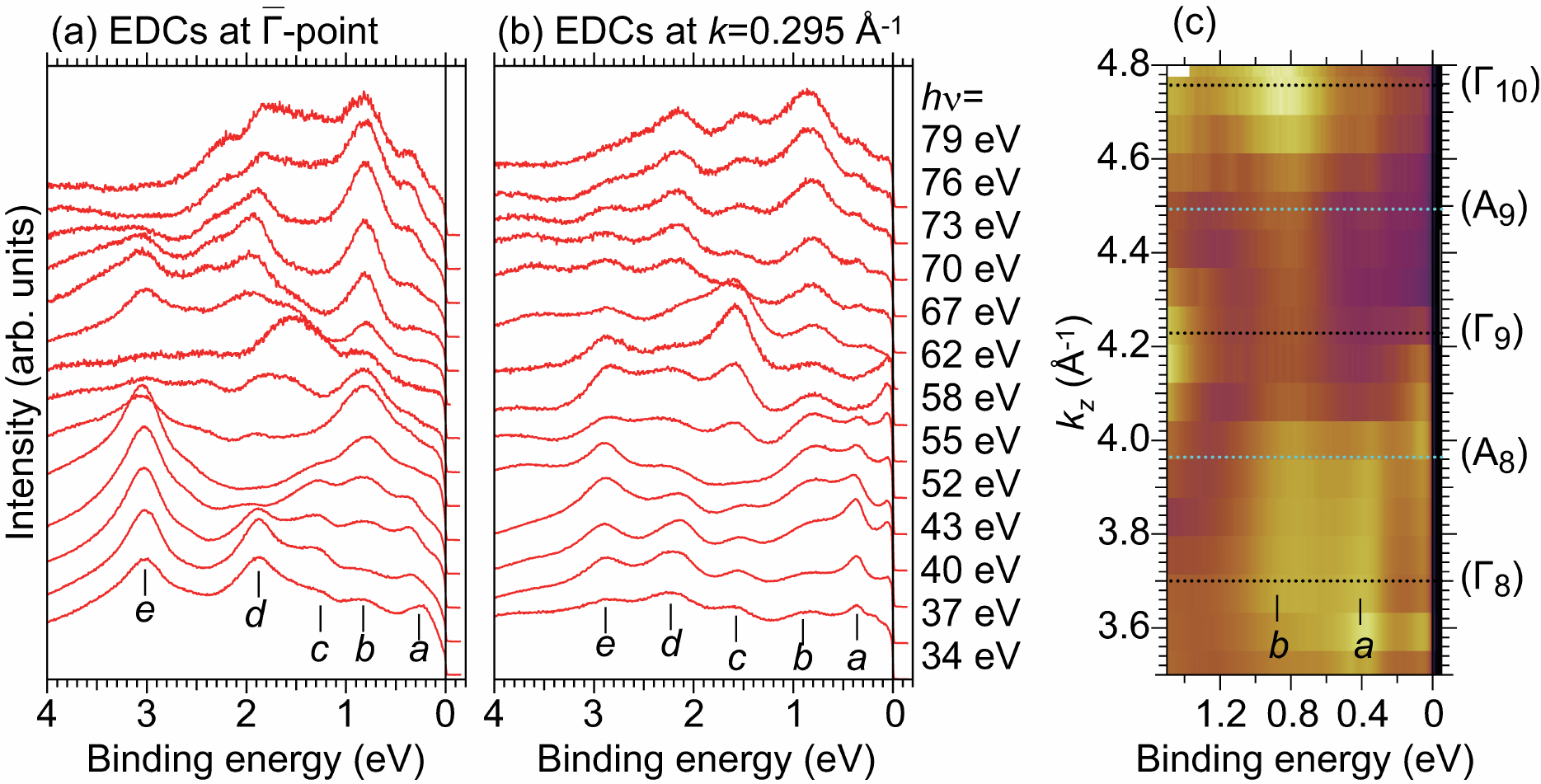} 
  \caption{(Color online) The energy distribution curves (EDCs) at two position in $k$-space are shown in panels (a) and (b). These positions are marked by green full vertical lines in the first panel of Fig.~\ref{Lfig04}, labeled by $a$ and $b$, respectively. 
  (a) The EDCs at $ (k_x,k_y)$ = $(0, 0)$ $\equiv \bar{\Gamma}$ point, measured for various energies of incident photons. 
   The main spectral features are indicated by vertical bars and labels below the bottommost curve. 
   (b) The EDCs at  $k$ = 0.295 \AA $^{-1}$ along the $\Gamma K_0$ direction. 
  (c) The data at the same $k$-point, $k_{\Gamma K_0}$ = 0.295 \AA $^{-1}$ shown as the function the binding energy and $k_z$, given by Eq.~(\ref{eq:kz}). 
   The labels ($\Gamma_n$) next to black dashed lines mark the $k_z$ levels of  $\Gamma_n\equiv (0,0, 2\pi/c \times n)$, whereas the label $A_n$ stands for the midpoint between $\Gamma_n$ and $\Gamma_{n+1}$, as usual. 
 }
  \label{Lfig08}
  \end{figure}

  Another view at data underlying Fig.~\ref{Lfig04} can be obtained from Fig.~\ref{Lfig08}.  
  It emphasizes the non-monotonous variation of the intensity of several bands upon changing the photon energy. 
  An important example of this behavior is the Nb-bonding band, clearly showing at 42 eV in Fig.~\ref{Lfig04}, disappearing in 58 and 62 eV spectra, and reappearing in the 76 eV panel.
  In Fig.~\ref{Lfig08} (c), we follow the intensity variation of the bonding-band signal at a particular ($k_x,k_y$) point as the function of $k_z$, calculated through Eq.~(\ref{eq:kz}),  extending over the range of several Brillouin zone replicas.
  The signal, labeled by $a$, shows well for $k_z$ around $\Gamma_8$, then almost vanishes around $\Gamma_9$, and reappears around $\Gamma_{10}$.   
  The same signal can also be followed in energy distribution curves (EDC) in Fig.~\ref{Lfig08} (b).
  The observed sequence reflects the internal structure of the bonding band wavefunction in the unit cell \cite{Moreschini2014, Moser2017}.
  The internal structure of the wavefunction enters into photoemission through the photon-absorption matrix element. 
  It is approximately given by the projection of the initial electronic state wavefunction, at the crystal wave-vector $k$, onto the plane-wave at the wave-vector $k'\approx k+G$ \cite{Moser2017}, with $G$ denoting the wave-vector of the reciprocal lattice. 
  This projection depends on the variation of the electronic wavefunction within the unit cell.  
  It reflects the positions of the atoms involved, as well as on the variation of the electronic wavefunction within particular orbitals, duly taken into account through the orbital photoemission cross-section shown in Fig.\ref{Lfig01}. 
  Without going into formulae, it can be understood that in TMDs this internal structure of the wavefunction comes primarily from the layers participating in the crystal structure unit cell. 
  In particular, for Co$_{1/3}$NbS$_2$ and Nb-dominated bands, it comes from the Nb layers, uniformly separated by $d=c/2$ in the crystal, with two (2) of them participating in the unit cell (see Fig.~\ref{Lfig01}).  
  In the absence of the Nb-interlayer hybridization, the ($k_x$, $k_y$ ) dispersions are expected to be the same for all separate Nb layers.  
  A hybridization between neighboring layers leads to $-\cos(k_z d)=-\cos(2k_z c)$  dispersion in the $k_z$-direction, with the band bottom, for a fixed $(k_x, k_y)$  showing at $k_z=n\times 2\pi/d= 2n\times 2\pi/c$ in the extended zone scheme, and the band tops appearing at $k_z= (2n+1)\times 2\pi/c$, with $n$ standing for any whole number.  
  The particularity of having two Nb-layers per unit cell in 2H-NbS$_2$ and similar TMDs leads to folding the conduction band spectra into the $(-\pi/c,\pi/c)$ $k_z$-range of the first Brillouin zone. 
  The folding produces the bonding and the antibonding bands, initially corresponding to the lower and the upper parts of the single extended Nb-band. 
  Still, taking the photoemission matrix elements into account that is expected to {\it rediscover} the Nb bonding band minima and the Nb antibonding band maxima around $\Gamma_{2n}$  and  $\Gamma_{2n+1}$ points, respectively. 
  The simple argument given above is not a real substitute for explicate calculations of matrix elements, which also take into account the inequivalence of two Nb-layers in the unit cell produced by their surrounding atoms.  
  Upon looking at Fig.~\ref{Lfig08} (a) and (b), one may keep in mind that a similar qualitative argument applies for sulfur-dominated bands, where two S-double-layers appear stacked in the unit cell.     
  The range and the density of the photon energy points used in the current experiment limit our exploration in that direction.      
  However, the effect was observed for the Se signal in 2H-NbSe$_2$ \cite{weber2018}.  
  
  At this point, we may conclude that change in ARPES spectra upon varying the photon energy originates at least from two contributions.  
  The change in the photoionization intensity of involved elements produces the slowly varying background.
  On a more detailed scale, the variation comes from the internal structure of wavefunctions within the unit cell.
  
  The third contribution is generally expected to produce the changes at a still finer scale. 
  It is expected to come from $k_z$ varying within the single $2\pi/c$ period as the photon energy changes, with gradual repositioning of signals reflecting the $k_z$ dispersion of particular bands. 
  These changes in band position, foreseen by bands structure calculation, cannot be discerned in Fig.~\ref{Lfig04}. 
  There are several possible explanations.  
  On the side of the physics of the material, it is possible that the DFT calculation very much overestimates the interlayer-coherence upon assuming the frozen magnetic order in the material where the interlayer coupling is primarily coming from magnetic ions. 
  The Co spin fluctuations, presumably present down to very low temperatures in Co$_{1/3}$NbS$_2$, provide an excellent mechanism for destroying the interlayer coherence. 
  However, the temperature dependence of resistivity in the direction perpendicular to layers does not point in the same direction \cite{Popcevic2019}.   
  Instead, the drop in resistivity upon Co intercalation speaks in favor of the Co intercalation increasing the interlayer coupling coherently, enough to enhance the interlayer charge transport.
   More probably, the explanation is to be sought  
on the side of the experimental technique. 
  The low $k_z$ sensitivity in ARPES performed at photon energies below one hundred eVs \cite{Strocov2003,Moser2017} has been argued to be the consequence  to rather short electronic {\it escape length} $\lambda$. 
  This parameter measures the depth from which the electron can be extracted without being scattered. 
  In our case,  $\lambda$  
  is estimated by using the Tanuma-Powell-Penn (TPP-2M) formula \cite{Tanuma1994} to $ 4.5$ \AA, less than half a value of the c-axis lattice constant.
   According to the scenario, the measured signal essentially averages the contributions of $k_z$ states within the  $1/\lambda$ range around the particular $k_z$ value given by Eq.~(\ref{eq:kz}). 
  For obvious reasons, the result tends to pin at minima or maxima of a band's $k_z$ dispersion at fixed $(k_x,k_y)$.   
  
  Finally, the sample quality has been seen to influence the quality of the signal \cite{SIref}, where weaker signals originating from more dispersive bands are likely to suffer the most. 

\subsection {Rigid-band picture and beyond} 

  The energy distribution curves presented in Fig.~\ref{Lfig08} provide a real {\it features inventory}, which we are going to inspect now to determine to which degree the ARPES spectra conform the picture of rigidly shifted NbS${_2}$ bands.
  In panel  Fig.~\ref{Lfig08} (a), we focus on signals at the $\Gamma$ point, whereas panel (b)  helps to follow these features away from the $\Gamma$ point.  Both figures include 14 spectra taken at different photon energies. 
  
   Five features can be discerned in most curves in Fig.~\ref{Lfig08} (a), labeled by $a$ through $e$, starting from the one closest to the Fermi energy. 
  
   We start our inspection from feature $e$, found at $E_B$ = 3 eV in panel (a).  
  The intensity of this signal has a general trend decreasing upon increasing the photon energy, although some non-monotonous behavior can be observed as well.
  The general behavior follows the niobium character of the signal, foreseen by the DFT calculation shown in Figs.~\ref{Lfig03}. 
   The DFT calculations predict several bands crossing the $\Gamma$ point at this energy, with one group of bands curved upward and another group of bands curved downward around the $\Gamma$ point.
   In Fig.~\ref{Lfig08}(b), we can identify the upward curved bands while downward curved ones are not readily observable further away from the $\Gamma$ point.

     The feature $d$, located at $E_B$ = 2 eV, also shows non-monotonous photon energy dependence of intensity, but no overall decrease, indicating its dominantly sulfur character. 
  Again, this is consistent with the DFT calculations predicting four dominantly sulfur bands merging at $\Gamma$ at $E_B$ = 2 eV.   
   These bands, all showing maxima at or around $\Gamma$  in Fig.~\ref{Lfig03}, cannot be individually resolved from measured spectra.
  
  Although the signal is less abundant in the $E_B\approx$ 1-2 eV energy range, at least two bands can be discerned in this range from most of the panels in Fig.~\ref{Lfig04}.
   The first one, labeled $c$ in Fig.~\ref{Lfig08}, is visible mostly around $\Gamma$ point at $E_B$ = 1.3 eV.
   The second one, labeled $b$ in Fig.~\ref{Lfig08}, can be observed at $E_B$ = 0.8 eV. 
  Only one band in this energy range appears in 2H-NbS$_2$ band structure calculations.
  This is the highest lying sulfur $p$-band, whose position and the energy span appear to be particularly sensitive to changes in the c-axis lattice constant, as shown in Appendix ~B.   
  The calculations also predict the pronounced $k_z$ dispersion for this band, which we cannot confirm from the experiment. 
  
  The deviations from simple rigid-band picture become more pronounced upon addressing the feature $a$, appearing at $\Gamma$ point at 0.3 eV below the Fermi energy, and persisting in most of the EDCs in Fig.~\ref{Lfig08} (a). 
  This feature does not appear in the DFT spectra and, as announced earlier, can be identified as the Nb $4d$ bonding band. 
  We have also announced that the DFT calculated bands for pristine 2H-NbS$_2$, (Appendix, Fig.~\ref{Lfig12}(a)), rigidity shifted to account for the charge transfer from Co ions, do not predict the bonding Nb $4d$ band getting below Fermi level at $\Gamma$ point. 
  The same applies for DFT calculations for 2H-NbS$_2$ with two extra electrons per three NbS$_2$ units, simulating the charge transfer from Co atoms, added to NbS$_2$ layers from the outset (see Appendix, Fig.~\ref{Lfig12}(b)). 
   This indicates that Co intercalation goes well beyond the charge transfer, strongly increasing the interlayer hybridization and the splitting between Nb $4d$ bands around $\Gamma$ point, to the level of pushing one conduction band maxima much below the Fermi level.  
  The same was observed in ARPES spectra in Cr$_{1/3}$NbS$_2$ \cite{Sirica2016}. 
  This is also the robust feature of DFT results in Co$_{1/3}$NbS$_2$ (Appendix, Fig.~\ref{Lfig12}(c)),  as discussed in Ref.~\onlinecite{Popcevic2019},  and expanded upon in the next section. 
  The effects of the extra band splitting diminish upon moving away from the zone center, leading to the {\it non-uniform doping} across the Fermi surface, as shown in Ref.~\onlinecite{Battaglia2007} for Mn$_{1/3}$NbS$_2$ and Ni$_{1/3}$NbS$_2$.   
  Therefore, the pronounced non-uniform deformation of conduction bands appears to be the general consequence of intercalation of 2H-NbS$_2$ by transition metal ions.

\subsection{Signal at the Fermi level}  

  For the Fermi level, the rigid-band picture foretells two hole-like pockets appearing around $\Gamma$ point, as well as around K$_0$ point.
   The only expected difference from the Fermi surfaces measured in pristine 2H-NbS$_2$ \cite{Youbi2020} would be the reduction in the size of those pockets caused by the electron transfer from Co into NbS$_2$ layers. 
   In fact, this kind of reduction in size was already reported for sister materials Mn$_{1/3}$NbS$_2$ and Ni $_{1/3}$NbS$_2$   \cite{Battaglia2007}.
   Given the submergence of bonding Nb $4d$ band maximum below the Fermi level, one 
  expects to lose one pocket in the central part of the Brillouin zone, leaving only one circular hole pocket around  $\Gamma$, related to the antibonding band. 
    This circular Fermi surface around $\Gamma$ is what we observe for all photon energies, along with the pockets around $K_0$, as pictured in   Fig.~\ref{Lfig06}.

  However, Figs.~\ref{Lfig06} and \ref{Lfig07} also bring in the most significant difference between electronic structures of Co$_{1/3}$NbS$_2$ and the parent compound 2H-NbS$_2$. 
  Entirely unforeseen by the latter, the $\beta$-band appears in the vicinity of the Fermi level, in the region too narrow in energy to be easily spotted in Fig.~\ref{Lfig04} and in other spectra that present data in the entire binding energy range covered by the experiment.  
  In the $k$-space, the $\beta$-band shows as the necklace of shallow and wide electron pockets positioned around K-points of the small Brillouin zone imposed by the intercalation, weakly overlapping.
  As mentioned in the previous section, a somewhat similar $\beta$-feature was first observed recently in sister compound Cr$_{1/3}$NbS$_2$ \cite{ Sirica2016, Sirica2020}.
   Despite this feature receiving much attention in Ref. \onlinecite{Sirica2020}, and being convincingly attributed to Cr orbitals, its origin remains elusive. 
  In Co$_{1/3}$NbS$_2$, we observe the feature much more clearly, fully developed.
  The fact that the $\beta$-band gets increasingly pronounced at higher photon energies corroborates its cobalt character. 
  This goes well in hand with the Cr-origin of the $\beta$-feature in Cr$_{1/3}$NbS$_2$ concluded upon by other means. 
  The possibility of the $\beta$-band/feature coming from magnetic atoms remaining on the surface upon cleavage seems very unlikely.  
  The structured ARPES signal, well away from the $\Gamma$ point, would demand the surface reconstruction neatly ordering these atoms, which was not seen in Cr$_{1/3}$NbS$_2$.  
\section{Modeling the electronic structure of ${\rm Co}_{1/3}{\rm NbS}_2$}

\subsection{Unfolded DFT spectra} \label{sec:udft}

  As stated towards the end of the previous section, no explanation of the $\beta$-band can be found in published {\it ab initio} electronic structure calculations in Co$_{1/3}$NbS$_2$ \cite{Nakayama2006, Mankovsky2016, Inoshita2019, Popcevic2019}.
   Still, these calculations contain several features worth noticing, and the comparison between such a DFT calculation and our ARPES spectra seems worth doing. 
  Subsequently, we discuss the ways to go beyond such DFT calculations in order to account for the observed $\beta$-band.

  To start with, the spin 3/2 magnetic state of Co convincingly shapes several experimental properties of Co$_{1/3}$NbS$_2$, and in particular the Curie-Weiss behavior of the magnetic susceptibility that extends over a wide temperature range \cite{Anzenhofer1970}. 
  The DFT can address this property only by considering the magnetically ordered state. 
  Alternatively, the electronic states emerging from the calculation are characterized by the same density of spin-up and spin-down electrons at all Co sites.   
  The DFT+U approach \cite{Anisimov1993, Liechtenstein1995, Cococcioni2005, Martin2004, Tolba2018}, which includes the Coulomb Hubbard correction $U$ to local interactions, turns helpful in better addressing the electronic correlations on Co atom and in reproducing the magnetic moment measured experimentally.  
  Our self-consistent determination of $U$ within the DFT+U approach  \cite{Cococcioni2005,Timrov2018}  
yields $U\approx 5$ eV at Co atom.
  This value of $U$ was also used in several previous studies in compounds with Co atoms \cite{Juhin2010,Chen2011,Mann2016}, as well in our electronic {\it ab initio} calculations for Co$_{1/3}$NbS$_2$ in  Ref. ~\onlinecite{Popcevic2019}. 
   The results of that calculation, also briefly presented in Fig.~\ref{Lfig12}(c) in the Appendix, will be used in this section for the comparison  with our experimental data.   
  
   The calculation was performed for the particular antiferromagnetically ordered state termed the {\it hexagonal order of the first kind} (HOFK), determined for Co$_{1/3}$NbS$_2$ by neutron scattering a long time ago \cite{Parkin1983}.
  The first Brillouin zone appropriate for this state is six-fold smaller than the first Brillouin zone of 2H-NbS$_2$ and significantly smaller than the $k$-vector range covered by experimental scans.   
  For this reason, as well as to account for the signal variation in $k$-space introduced by  {\it super-structuring} induced by intercalation, the DFT-calculated spectra are  {\it unfolded} into the first Brillouin zone of 2H-NbS2 before comparing them to our ARPES results.   
    As usual, the unfolding determines the spectral weights in the extended $k$-space range based on the calculated wavefunctions. 

    We confine our comparison to the energy range of approximately 1 eV below and above the Fermi level, dominantly covered by bands originating from Nb orbitals,  but also include the lowest Co-dominated bands, starting at approximately 1 eV above the Fermi level (see Appendix, Fig.~\ref{Lfig12}) .  
  We choose to make the unfolding starting from the tight-binding (TB) parametrization of the DFT-calculated bands, obtained by using the Wannier90 code \cite{Wannier902014, SIref}.
  The unfolding procedure essentially follows the one outlined in Ref. \onlinecite{Wu2017}, with the in-house code developed for the occasion. 
  The choice of using the effective tight-binding approach has some advantages. 
  First, the TB parameters have direct physical significance, especially when only a few are required to shape the bands.
  Second, the properties of the bands calculated for 2H-NbS$_2$ and Co$_{1/3}$NbS$_2$  can be easily compared through these parameters. 
  Finally, the differences between calculated and measured spectra can be explored by adjusting these parameters, possibly leading to improved electronic modeling of the material.


  \begin{figure}[t!] 
  \includegraphics[width=\columnwidth]{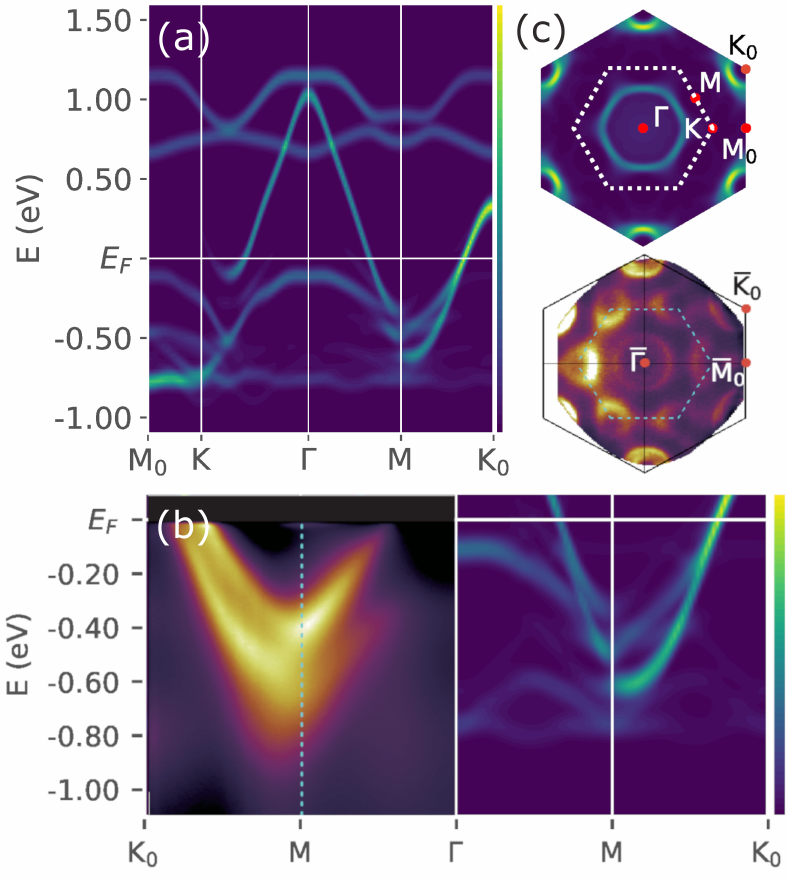}
  \caption{(Color online). (a) The electronic spectra calculated for Co$_{1/3}$NbS$_2$ (see Ref. \onlinecite{Popcevic2019}) in the particular antiferromagnetically ordered state, termed  the {\it hexagonal order of the first kind} (HOFK).  The electronic bands are shown unfolded along M$_0$$\Gamma$K$_0$ line of Brillouin zone of 2H-NbS$_2$, averaged over three equivalent directions by the hexagonal symmetry.  
   (b) The comparison between measured and calculated spectra along the $\Gamma-K_0$ line. 
  (c) The comparison between the calculated and the experimental spectra at the Fermi level.
  The most obvious qualitative difference is the lack of the $\beta$-band signal in the DFT-calculated spectra along the boundary of the first Brillouin zone of the of Co$_{1/3}$NbS$_2$, shown in a dotted line.
  The color bar indicates the intensity scale.  }
  \label{Lfig09}
  \end{figure}

  The unfolded spectrum is shown in Fig.~\ref{Lfig09}(a).
  The overall spectral distribution below the Fermi level broadly follows the spectral shape of the 2H-NbS$_2$ conduction bands.  
  Along with commonalities, some differences are readily spotted. 
  The most apparent difference emerges through the bonding Nb  $d$-band found submerged below the Fermi level near the $\Gamma$ point. 
  At this point, the {\it ab initio} electronic structure calculations in Co$_{1/3}$NbS$_2$ qualitatively agree with experimental spectra. 
   The comparison of spectra along the $\Gamma-K_0$ line is shown in Fig.~\ref{Lfig09}(b). 
  The agreement is not perfect regarding the band's position, being off by some 0.15 eV. 
  Shifting the calculated bands to higher binding energies by this amount would improve the agreement.       
  The physical origin of the boding band appearing below the Fermi level at the $\Gamma$ point can be readily traced to the big Co-Nb orbital overlap, $t_{Co}\approx 0.23$ eV. 
  It emerges as the dominant hybridization integral in the whole tight-binding parametrization of Co$_{1/3}$NbS$_2$.
  At the same time, the position of the antibonding band is saved from the shift for symmetry reasons. 

  On the other hand, in Fig.~\ref{Lfig09}  (a)  one observes a significant reconstruction of spectra calculated for Co$_{1/3}$NbS$_2$ relative to 2H-NbS$_2$ conduction bands, with the band reorganization being particularly pronounced around the K and M points, i.e., at the boundary of the first Brillouin zone of Co$_{1/3}$NbS$_2$.
  These effects are difficult to observe in experimental spectra.
  On the level of TB modeling, the dominant source of this reconstruction of bands can be traced back to the variation of the in-plane Nb-Nb hybridization integrals, imposed by the antiferromagnetic superstructure, as well as to the strong spin polarization at Co orbitals.
  Both mechanisms can be verified by exploring other magnetic structures, e.g., the ferromagnetically ordered state, or by calculating the unfolded spectra upon reducing the spatial variation of Nb-Nb hybridization integrals within the TB parametrization.
  These changes lead to the Nb $4d$ conduction bands smoothly varying throughout the large Brillouin zone.  
  
  Interestingly, and more importantly, no extra features appear at the Fermi level in the same region of the $k$-space. 
   This comparison with the experimental spectrum is shown in Fig.~\ref{Lfig09} (c).   
  With no signal around K and M points, the calculated spectrum at the Fermi level bears no marks of the $\beta$-band.   
  At the Fermi level the only qualitative departure from the picture of rigidly-filled 2H-NbS$_2$ bands appears though only one ring encircling the $\Gamma$ point, instead of two.


  \begin{figure}[t!]
  \includegraphics[width=\columnwidth]{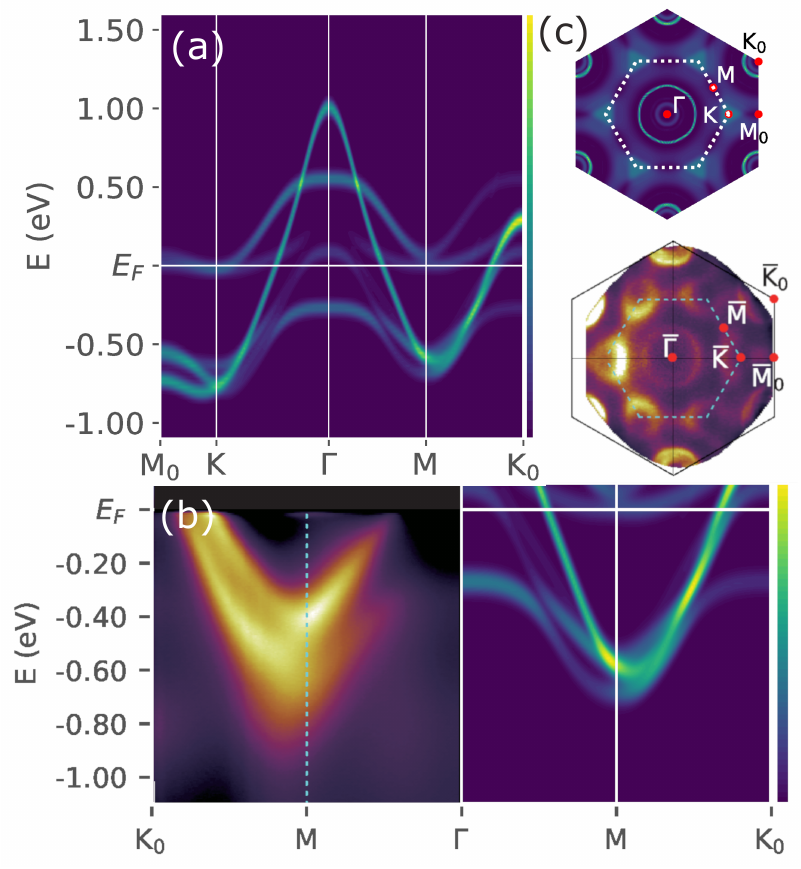}

  \caption{(Color online). (a) The unfolded spectra of the system modeled through the tight-binding reparameterization, with the parameters initially taken from our DFT/Wannier90 calculation. Very few modifications, detailed in the text, are inspired by studies of the effects of strong electron correlations in other systems. The image shows the spectral density along the M$_0$$\Gamma$K$_0$ line in the $k$-space. 
    (b) The comparison between measured and calculated spectra along the $\Gamma-K_0$ line. The faint $\beta$-band signal appears on both sides, near the top of the M vertical line.  
  The bonding band signals approach the $\Gamma$ point at approximately the same energy from both sides.
(c) The comparison between the calculated and the experimental at the Fermi level.
  The dotted lines mark the boundary of the first Brillouin zone of the of Co$_{1/3}$NbS$_2$. 
  Both subpanels show the $\beta$-band signal that spreads along this boundary, 
  particularly enhanced around the K point.  }
  \label{Lfig10}
  \end{figure}

\subsection{Beyond DFT}   

    For a particular magnetic state offered as a seed, the spin-polarized DFT calculation tends to self-consistently determine the magnetic state on Co atoms in the specific magnetic configuration and produce the appropriate electronic bands.  
  The calculation results with Co atoms with sizable magnetic moments, with some lower and upper {\it Hubbard states} populated by spin up or spin down electrons. 
  The resulting bands conform to the TB description where, apart from the spin character at particular Co orbitals, the other parameters stay essentially unaltered among different magnetic states \cite{SIref}.
  The TB parametrization of electronic bands in various magnetic states leads to results where
  the hybridization integral between Co and Nb orbitals ($t_{Co}$) is approximately two times bigger than any other hybridization integral in the model, including those responsible for the Nb $4d$ bands dispersions in directions parallel and perpendicular to layers. 
  Further on, the separation between spin-down and spin-up $3d$ states at Co atoms below and above the Fermi level is one order of magnitude bigger than any hybridization integral in the TB model and approximately six times bigger than the Co-Nb orbital energy level separation.
  The Nb orbital energy level and the calculated Fermi energy lie in-between energy levels of the full Co orbital of one spin projection, and the empty Co orbital of the opposite spin projection.   
  This description conforms with the picture of a strongly correlated electron system.
  It may be safely said that the DFT producing an array of local magnetic moments, with their hosting orbitals strongly hybridized to relatively narrow metallic bands, is a good indication of a strongly correlated electron system.
  Conversely, strongly correlated electron systems are poorly treated by common DFT,  or by any approach where the determinant of single-particle states is used to describe the many-body wavefunction. 
  Whereas some properties may be caught better than others, the description of low energy dynamics of a mixture of magnetic moments and itinerant electrons is obliged to fail in such approaches.
  
  The methods to deal with strongly correlated electron systems have been developing intensively in recent decades.  A proper theoretical framework for the present problem may be possibly sought within the combination of the density functional theory and the dynamical mean-field theory (DFT+DMFT) \cite{Georges1996, Kotliar2006},  with the Hund's rule coupling properly included to account for the spin-3/2 Co state \cite{Medici2011}.
  With the implementation particular to Co$_{1/3}$NbS$_2$ still being awaited for, the poor man's alternative would be to apply the qualitative recipes based on previous works.  
  The hint may be sought in the previous treatments of centers with big local Coulomb repulsion in contact with conducting electrons, appearing within the Anderson impurity model, in heavy fermion systems, and in electronic models for copper oxide superconductors.
  Several approaches to strongly correlated electron systems result in effectively reducing the hopping integral between conduction band states and the large-$U$ orbital, accompanied by the big shift in the orbital energy at the strong-U center level towards the Fermi level \cite{Kotliar1988, Coleman1984, Coleman2007, Niksic1995, Tutis1994, Tutis1997}.
  These two changes, appearing within the {\it slave-boson} approach to strongly correlated electron systems, possibly represent the shortest path to what is called the Kondo-Anderson-Suhl resonance in systems with magnetic impurities, or to resonance bands appearing at the Fermi level in crystalline systems.  
  Physically they are related to the electron of one spin projection blocking the electrons of other spin projection from hopping into the large-$U$ orbital, resulting in the formation of the low-energy resonance related to the local spin-flip process. 
  The width of the resonance (band) depends very much on the initial parameters like bare overlap integral.  
  It is expected to be much smaller in the systems involving $f$ orbitals than in the systems based on $d$ orbitals.  
  In our case, these changes amount to reducing $t_{Co}$ and shifting the Co orbital energy level towards the Fermi level.
  Also, all Co are to be treated equivalently, to account for a system without magnetic order.
  Performed together, the changes in $t_{Co}$ and Co orbital energy can keep the position of the Nb bonding-band at the $\Gamma$ point below the Fermi level, in the place indicated by experimental spectra.
  At the same time, the reparameterization shifts the bottom of the narrow Co-band to the Fermi level. 
  The resulting spectra are shown in Fig.~\ref{Lfig10}. 
    The Nb bands in Fig.~\ref{Lfig10} (a) are found to run continuously throughout the first Brillouin zone of2H-NbS$_2$, similarly as in measured spectra. 
   In contrast to bare DFT results shown in Fig.~\ref{Lfig09} (c), Fig.~\ref{Lfig10}  (c) shows the calculated signal appearing  along  the boundary of the first Brillouin zone of  Co$_{1/3}$NbS$_2$, getting particularly strong near the K points, similarly to observed $\beta$-bands.
  The comparison between the experimental and calculated spectra along the $\Gamma-K_0$ line is shown in  Fig.~\ref{Lfig10}  (b).

  Not unseen in the previous instances in strongly correlated electron systems, the required renormalizations of two tight-binding parameters are relatively big, 
  with the Co-Nb level separation requiring reduction from 0.83 eV to approximately 0.25 eV,
  whereas the Co-Nb hybridization demands the resizing by approximately one-half.
  Although the shift in energy level, much studied within the slave-boson approach to strongly correlated systems, produces the resonance band at the Fermi level, it should be kept in mind that Fig.~\ref{Lfig10} represents only the partial result. 
  The figure represents the spectrum related to auxiliary fermion fields, whose features remain present upon returning to real fermions, but accompanied by the significant transfer of spectral weight into the energy range of the unrenormalized energy levels \cite{Tutis1994, Niksic1995, Tutis1997}. 
  As announced, a more detailed study of this particular topic is yet to follow. 
  
  It remains to comment upon another possible mechanism that can influence the TB parameters in the same direction. 
  This mechanism is the relaxation of crystal layers in the close vicinity of the sample surface.
  An increase of the interlayer separation near the sample surface is expected to locally decrease the Co-Nb hybridization integral, as well as the separation between Co and Nb energy levels. 
  The latter would come from the change in electrostatic potential of the electron in the Co orbital, as the Co atom moves further away from negatively charged Nb atoms. 
  We have not studied numerically the surface relaxation in Co$_{1/3}$NbS$_2$, and it remains to be seen if the required significant changes in parameters may be accounted for in this way. 
  However, one must keep in mind that even within the scenario based on surface relaxation, the problem of strong correlations remains present and unattended, both in bulk and near the surface.

\section{Conclusions}

 We have presented the first ARPES study on Co$_{1/3}$NbS$_2$ accompanied by a detailed theoretical analysis of the observed electronic structure.
   At first, the measured electronic structure shows remarkable similarity to the parent material 2H-NbS$_2$, whose electronic structure is known from previous measurements and is well described by DFT electronic structure calculations. 
  In the observed spectra, we find no signs of the 2H-NbS$_2$ bands getting folded or gapped at the boundary of the Brillouin zone of Co$_{1/3}$NbS$_2$, i.e.  the superlattice of 2H-NbS$_2$ imposed by the intercalation.
  Motivated by that and by comparisons between electronic structures of intercalates and the host materials in the literature, we systematically compare the features observed in Co$_{1/3}$NbS$_2$ with those appearing in 2H-NbS$_2$ spectra. 
  The DFT calculation for 2H-NbS$_2$ with additional 4/3 electrons per unit cell, simulating the charge transfer of two electrons per Co into NbS$_2$ layers, produces the shift in Fermi level of approximately 0.5 eV. 
  The shift accounts rather well for the observed position Fermi level relative to what we identify as the Nb $4d$ bands in Co$_{1/3}$NbS$_2$. 
  All this goes in favor of the 2H-NbS$_2$  quasi-rigid-band 
   picture. 
  
  However, several deviations from the rigid-band picture will entirely change our view of magnetically intercalated TMDs. 
   One of these deviations corresponds to the local maximum of the bonding Nb $d$ band being found submerged deep below the Fermi level at $\Gamma$ point, whereas it is expected to stand above the Fermi level in the rigid-band picture. 
  The observed change in the bonding band corresponds to very strong Co-Nb hybridization, bigger than any other parameter shaping the 2H-NbS$_2$ conduction bands. 
  This overwhelmingly strong hybridization questions the applicability of the low-order perturbative concepts to the calculation of magnetic interactions in intercalates \cite{Sirica2020}. 
  The second, even more important, deviation from the 2H-NbS$_2$ electronic structure is appearance of a new band unforeseen by DFT calculations, referred in this paper as the $\beta$-band.
   In our spectra, it appears as the necklace of shallow electron pockets centered at six K-point corners of the first Brillouin zone of Co$_{1/3}$NbS$_2$. 
  The strong electron correlations impose as a possible explanation for the appearance of this new band at the Fermi level, which we also explore to a certain degree.
  The strong hybridization and unexpected shallow band at the Fermi level speak together in favor of Co$_{1/3}$NbS$_2$ being considered a strongly correlated electron system,  with the coupling between metallic and magnetic layer playing the essential role.
  These substantial shifts in view on magnetically intercalated TMDs need to be further explored experimentally and theoretically. 
\section*{Acknowledgements}

Y. U. thanks Z. Chen, J. Dong, E. Papalazarou, and L. Perfetti for their experimental support to perform preliminary ARPES measurement using He I$\alpha$ resonance line, and O. Bari\v{s}i\'{c} for helpful discussions. This work has been supported in part by the Croatian Science Foundation under the project numbers IP-2016-06-7258 and IP 2018-01-7828. These researches have been performed at the National synchrotron radiation center SOLARIS at the UARPES beamline (proposals No. 181PH009 and 201046). The experiment has been performed thanks to the collaboration of the SOLARIS staff. The research leading to this result has been supported by the project CALIPSOplus under the Grand Agreement 730872 from the EU framework programme for research and innovation HORIZON 2020. The work in Lausanne was supported by the Swiss National Foundation. The work at the TU Wien was supported by the European Research Council (ERC Consolidator Grant No 725521), while the work at the University of Zagreb was supported by project CeNIKS co-financed by the Croatian Government and the European Union through the European Regional Development Fund - Competitiveness and Cohesion Operational Programme (Grant No. KK.01.1.1.02.0013). I.B., W.T., and M.A.G. acknowledge financing from the Polish National Agency for Academic Exchange under the "Polish Returns 2019" Program, PPN/PPO/2019/1/00014/U/0001, and subsidy of the Ministry of Science and Higher Education of Poland.  The authors thank 
M. Petrovi{\'c} for the helpful discussion of the LEED data. 

While in the review process, we became aware of two ARPES studies of Co$_{1/3}$NbS$_2$ \cite{Tanaka2022, Yang2022}. These studies do not conflict with our discoveries.

P.P. and  Y.U.  have equally contributed to the paper.

\section{Appendix}

\subsection{The $\beta$-band and the $k_z$ dispersion}


  \begin{figure}[b!]
  \includegraphics[width=0.5\textwidth]{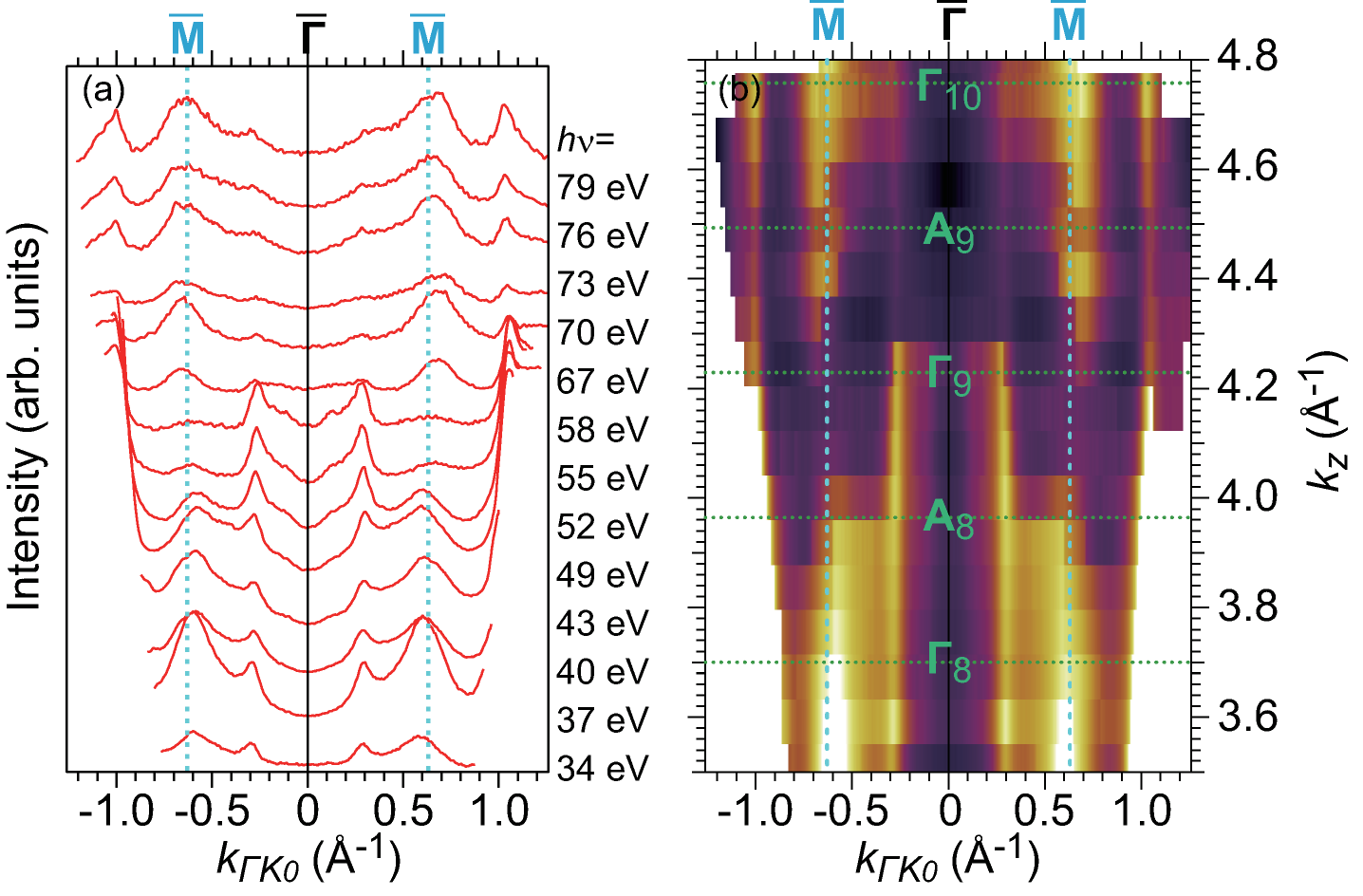}
  \caption{(Color online). (a) The photon energy dependence of the momentum distribution curves (MDCs) along the $\Gamma$K$_0$ direction at the Fermi level. The MDCs are taken at the same photon energies as in Fig.~\ref{Lfig08}. The same normalization applies as in Fig.~\ref{Lfig04}. (b) ARPES intensity plot of the $k_z$-$k_{\Gamma K_0}$ band dispersion at Fermi level. The labels $\Gamma_n$ and $A_n$ have been defined in relation to Fig~\ref{Lfig07}. Two vertical dashed lines shown in cyan in both panels represent the boundary of the Brillouin zone of Co$_{1/3}$NbS$_2$ and correspond to the M point at $k_z$ = 0.}
  \label{Lfig11}
\end{figure}
     Fig.~\ref{Lfig11}(a) shows the momentum distribution curves (MDCs) at the Fermi level, measured by using horizontally polarized photons at various photon energies. 
   Fig.~\ref{Lfig11}(b)  shows the ARPES intensity plot at the Fermi level within the rectangular section in $k-$-space, extending along  $k_z$ and $k_{\Gamma K_0}$ directions. 
  This plot is also produced from the ARPES spectra taken at various photon energies.
    In Fig.~\ref{Lfig11}(a) the antibonding Nb $4d$ band can be observed at $\pm$0.3 \AA$^{-1}$, and as well defined vertical lines in Fig.~\ref{Lfig11}(b). 
     The signal related to the $\beta$-band is observed in both panels along the vertical cyan lines marking the boundary of the first BZ of Co$_{1/3}$NbS$_2$. 
  The latter signal dominates Fig.~\ref{Lfig11}(b).


  \begin{figure*}[t!]
  \includegraphics[width=\textwidth]{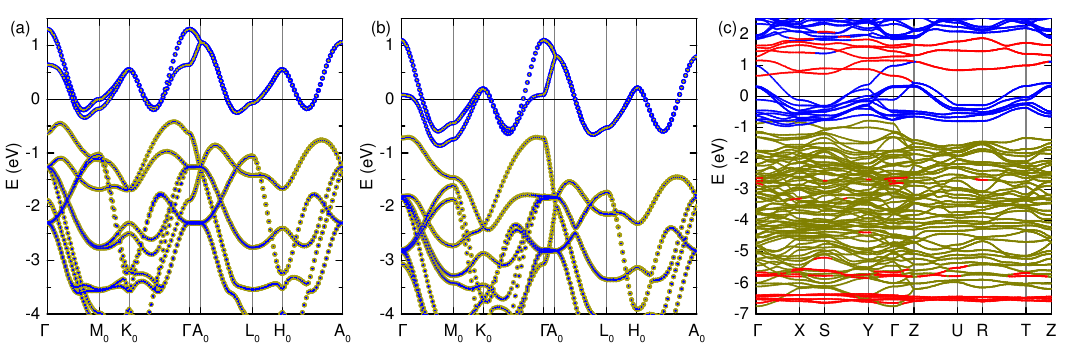}
  \caption{(Color online). (a) DFT-calculated electronic bands of 2H-NbS$_2$ with DFT-relaxed lattice parameters. (b) DFT-calculated electronic bands of 2H-NbS$_2$ with additional 2/3 electrons per Nb atom and the lattice parameters borrowed form DFT-relaxed Co$_{1/3}$NbS$_2$ structure. Blue and yellow colors reflect the relative weights of niobium and sulfur orbitals in particular states. 
 (c) DFT-calculated electronic bands of Co$_{1/3}$NbS$_2$ in antiferromagnetically ordered state. 
  The colors mark the dominant character of electronic states:  blue indicates that the electronic state is predominantly composed of niobium orbitals, yellow represents the dominance of sulfur orbitals, and red specifies the dominance of cobalt orbitals in the electronic state. }
  \label{Lfig12}
  \end{figure*}
  
  The momentum along the $z$ direction for Fig.~\ref{Lfig11}(b) is calculated using the expression in Eq.~(\ref{eq:kz}). 
   For the parameter V$_0$ in Eq.~(\ref{eq:kz}) we adopt the value proposed previously for the sister compound Cr$_{1/3}$NbS$_2$, i.e. $V_0$ = 14 eV \cite{Sirica2016}.
   It may be remarked that the periodicity of $2\pi/c$ along $k_z$ is not observed in Fig.~\ref{Lfig11}(b). 
  Some possible reasons have been already discussed in Sec. ~\ref{sec:discussion}, in relation to Figs~\ref{Lfig04} and \ref{Lfig08}. 
  In that respect, a wider photon energy range is needed to observe the periodicity in $k_z$.

\subsection{DFT calculations in 2H-NbS$_2$}\label{sec:AppDFT1}

  We made a long way to compare the measured electronic structure of Co$_{1/3}$NbS$_2$  with the calculated electronic structures in 2H-NbS$_2$  and  Co$_{1/3}$NbS$_2$. 
  Here we enclose three of several DFT calculations that we made for this comparison. Some more can be found in Supplemental Material \cite{SIref}.
  
  In Figs.~\ref{Lfig12} (a) and (b) we show the DFT results for the electronic band structure of 2H-NbS$_2$. The spectra are colored to reflect the dominant and sub-dominant contributions of Nb and S atomic orbitals. 
   The results depicted in panel (a) stand for the unit cell obtained upon relaxing the atomic forces within the DFT calculation. 
  The calculation did not include the electronic charge transfer from Co atoms into NbS$_2$ layers. 
  After the rigid band shift, introduced to account for the transfer, this electronic structure differs from the one observed in Co$_{1/3}$NbS$_2$, with all points of disagreement detailed in the main text. 
  
    Fig.~\ref{Lfig12} (b) shows the electronic structure of 2H-NbS$_2$ with the lattice parameters borrowed from DFT-relaxed AF-ordered Co$_{1/3}$NbS$_2$. 
  Also, 2/3 electrons per Nb were added into the calculation to account for the charge transfer from Co ions, compensated, as usual,  by an appropriate positive homogenous background charge.
  Entirely expected, the added electrons shift the bands towards higher binding energies.
   Apart from the shift, the main difference relative to the panel (a)  is the increase in separation between bonding and antibonding Nb $4d$ bands and the change in position of the highest-laying sulfur band. 
  The comparison between (a) and (b) also points to the very high sensitivity of the $k_z$ dispersive bands (one niobium and one sulfur dominated) to the interlayer distance, particularly affecting the spectra around the $\Gamma$ point. 
  Despite these differences, the bonding Nb $4d$ band at the $\Gamma$ point remains above the Fermi level for both cases, revealing that the explicit Co-Nb hybridization is essential for the qualitative and quantitative understanding of the bonding band in Co$_{1/3}$NbS$_2$. 
  Still, the results shown in panel (b) have been chosen for our first comparison with experimental spectra in Fig.~\ref{Lfig03}.  
  For a better match, in Fig.~\ref{Lfig03} the spectra in Fig.~\ref{Lfig12} (b) were shifted to higher binding energy by 0.1 eV.
  On that occasion and for the better match, the spectrum in Fig.~\ref{Lfig12} (b) was further shifted by 0.1 eV to the higher binding energy.

   Fig. ~\ref{Lfig12} (c) shows the result of the DFT calculation for Co$_{1/3}$NbS$_2$ in the antiferromagnetically ordered state  (HOFK). 
  The calculation uses the Hubbard $U=5$ eV and the DFT-relaxed atomic structure.
  These results have been used in Sec.~\ref{sec:udft} to compare against the experimental spectra upon unfolding into six times bigger first Brillouin zone of 2H-NbS$_2$. 
  The spectra are colored to reflect the dominant contributions of Nb, S, and Co orbitals. 
  For better visibility, the red points (the Co-dominated states) were painted last to avoid being covered by other colors.
  According to the calculation, the main contribution of Co orbitals to occupied states is seen around 6 eV below the Fermi level, well outside the binding energy range covered by our ARPES data. 
  The Co-dominated states closest to the Fermi level are the empty states above the Fermi level. 
  Co orbitals also contribute to the Nb conduction bands, particularly the bonding band, without becoming dominant.  
  More details can be found in Ref.~\onlinecite{Popcevic2019}.

\vfill
\eject 

\bibliographystyle{apsrev4-2}
\bibliography{CoNbS2-ARPES}

\begin{thebibliography}{80}%
\makeatletter
\providecommand \@ifxundefined [1]{%
 \@ifx{#1\undefined}
}%
\providecommand \@ifnum [1]{%
 \ifnum #1\expandafter \@firstoftwo
 \else \expandafter \@secondoftwo
 \fi
}%
\providecommand \@ifx [1]{%
 \ifx #1\expandafter \@firstoftwo
 \else \expandafter \@secondoftwo
 \fi
}%
\providecommand \natexlab [1]{#1}%
\providecommand \enquote  [1]{``#1''}%
\providecommand \bibnamefont  [1]{#1}%
\providecommand \bibfnamefont [1]{#1}%
\providecommand \citenamefont [1]{#1}%
\providecommand \href@noop [0]{\@secondoftwo}%
\providecommand \href [0]{\begingroup \@sanitize@url \@href}%
\providecommand \@href[1]{\@@startlink{#1}\@@href}%
\providecommand \@@href[1]{\endgroup#1\@@endlink}%
\providecommand \@sanitize@url [0]{\catcode `\\12\catcode `\$12\catcode
  `\&12\catcode `\#12\catcode `\^12\catcode `\_12\catcode `\%12\relax}%
\providecommand \@@startlink[1]{}%
\providecommand \@@endlink[0]{}%
\providecommand \url  [0]{\begingroup\@sanitize@url \@url }%
\providecommand \@url [1]{\endgroup\@href {#1}{\urlprefix }}%
\providecommand \urlprefix  [0]{URL }%
\providecommand \Eprint [0]{\href }%
\providecommand \doibase [0]{https://doi.org/}%
\providecommand \selectlanguage [0]{\@gobble}%
\providecommand \bibinfo  [0]{\@secondoftwo}%
\providecommand \bibfield  [0]{\@secondoftwo}%
\providecommand \translation [1]{[#1]}%
\providecommand \BibitemOpen [0]{}%
\providecommand \bibitemStop [0]{}%
\providecommand \bibitemNoStop [0]{.\EOS\space}%
\providecommand \EOS [0]{\spacefactor3000\relax}%
\providecommand \BibitemShut  [1]{\csname bibitem#1\endcsname}%
\let\auto@bib@innerbib\@empty
\bibitem [{\citenamefont {Wilson}\ and\ \citenamefont
  {Yoffe}(1969)}]{Wilson1969}%
  \BibitemOpen
  \bibfield  {author} {\bibinfo {author} {\bibfnamefont {J.}~\bibnamefont
  {Wilson}}\ and\ \bibinfo {author} {\bibfnamefont {A.}~\bibnamefont {Yoffe}},\
  }\href {https://doi.org/10.1080/00018736900101307} {\bibfield  {journal}
  {\bibinfo  {journal} {Advances in Physics}\ }\textbf {\bibinfo {volume}
  {18}},\ \bibinfo {pages} {193} (\bibinfo {year} {1969})}\BibitemShut
  {NoStop}%
\bibitem [{\citenamefont {Wilson}\ \emph {et~al.}(1975)\citenamefont {Wilson},
  \citenamefont {Salvo},\ and\ \citenamefont {Mahajan}}]{Wilson1975}%
  \BibitemOpen
  \bibfield  {author} {\bibinfo {author} {\bibfnamefont {J.}~\bibnamefont
  {Wilson}}, \bibinfo {author} {\bibfnamefont {F.~D.}\ \bibnamefont {Salvo}},\
  and\ \bibinfo {author} {\bibfnamefont {S.}~\bibnamefont {Mahajan}},\ }\href
  {https://doi.org/10.1080/00018737500101391} {\bibfield  {journal} {\bibinfo
  {journal} {Advances in Physics}\ }\textbf {\bibinfo {volume} {24}},\ \bibinfo
  {pages} {117} (\bibinfo {year} {1975})}\BibitemShut {NoStop}%
\bibitem [{\citenamefont {Tosatti}\ and\ \citenamefont
  {Fazekas}(1976)}]{Tosatti1976}%
  \BibitemOpen
  \bibfield  {author} {\bibinfo {author} {\bibfnamefont {E.}~\bibnamefont
  {Tosatti}}\ and\ \bibinfo {author} {\bibfnamefont {P.}~\bibnamefont
  {Fazekas}},\ }\href {https://doi.org/10.1051/jphyscol:1976426} {\bibfield
  {journal} {\bibinfo  {journal} {Le Journal de Physique Colloques}\ }\textbf
  {\bibinfo {volume} {37}},\ \bibinfo {pages} {C4} (\bibinfo {year}
  {1976})}\BibitemShut {NoStop}%
\bibitem [{\citenamefont {Naito}\ and\ \citenamefont
  {Tanaka}(1982)}]{Naito1982}%
  \BibitemOpen
  \bibfield  {author} {\bibinfo {author} {\bibfnamefont {M.}~\bibnamefont
  {Naito}}\ and\ \bibinfo {author} {\bibfnamefont {S.}~\bibnamefont {Tanaka}},\
  }\href {https://doi.org/10.1143/JPSJ.51.219} {\bibfield  {journal} {\bibinfo
  {journal} {Journal of the Physical Society of Japan}\ }\textbf {\bibinfo
  {volume} {51}},\ \bibinfo {pages} {219} (\bibinfo {year} {1982})}\BibitemShut
  {NoStop}%
\bibitem [{\citenamefont {Morosan}\ \emph {et~al.}(2006)\citenamefont
  {Morosan}, \citenamefont {Zandbergen}, \citenamefont {Dennis}, \citenamefont
  {Bos}, \citenamefont {Onose}, \citenamefont {Klimczuk}, \citenamefont
  {Ramirez}, \citenamefont {Ong},\ and\ \citenamefont {Cava}}]{Morosan2006}%
  \BibitemOpen
  \bibfield  {author} {\bibinfo {author} {\bibfnamefont {E.}~\bibnamefont
  {Morosan}}, \bibinfo {author} {\bibfnamefont {H.~W.}\ \bibnamefont
  {Zandbergen}}, \bibinfo {author} {\bibfnamefont {B.~S.}\ \bibnamefont
  {Dennis}}, \bibinfo {author} {\bibfnamefont {J.~W.~G.}\ \bibnamefont {Bos}},
  \bibinfo {author} {\bibfnamefont {Y.}~\bibnamefont {Onose}}, \bibinfo
  {author} {\bibfnamefont {T.}~\bibnamefont {Klimczuk}}, \bibinfo {author}
  {\bibfnamefont {A.~P.}\ \bibnamefont {Ramirez}}, \bibinfo {author}
  {\bibfnamefont {N.~P.}\ \bibnamefont {Ong}},\ and\ \bibinfo {author}
  {\bibfnamefont {R.~J.}\ \bibnamefont {Cava}},\ }\href
  {https://doi.org/10.1038/nphys360} {\bibfield  {journal} {\bibinfo  {journal}
  {Nature Physics}\ }\textbf {\bibinfo {volume} {2}},\ \bibinfo {pages} {544}
  (\bibinfo {year} {2006})}\BibitemShut {NoStop}%
\bibitem [{\citenamefont {Cercellier}\ \emph {et~al.}(2007)\citenamefont
  {Cercellier}, \citenamefont {Monney}, \citenamefont {Clerc}, \citenamefont
  {Battaglia}, \citenamefont {Despont}, \citenamefont {Garnier}, \citenamefont
  {Beck}, \citenamefont {Aebi}, \citenamefont {Patthey}, \citenamefont
  {Berger},\ and\ \citenamefont {Forró}}]{Cercellier2007}%
  \BibitemOpen
  \bibfield  {author} {\bibinfo {author} {\bibfnamefont {H.}~\bibnamefont
  {Cercellier}}, \bibinfo {author} {\bibfnamefont {C.}~\bibnamefont {Monney}},
  \bibinfo {author} {\bibfnamefont {F.}~\bibnamefont {Clerc}}, \bibinfo
  {author} {\bibfnamefont {C.}~\bibnamefont {Battaglia}}, \bibinfo {author}
  {\bibfnamefont {L.}~\bibnamefont {Despont}}, \bibinfo {author} {\bibfnamefont
  {M.~G.}\ \bibnamefont {Garnier}}, \bibinfo {author} {\bibfnamefont
  {H.}~\bibnamefont {Beck}}, \bibinfo {author} {\bibfnamefont {P.}~\bibnamefont
  {Aebi}}, \bibinfo {author} {\bibfnamefont {L.}~\bibnamefont {Patthey}},
  \bibinfo {author} {\bibfnamefont {H.}~\bibnamefont {Berger}},\ and\ \bibinfo
  {author} {\bibfnamefont {L.}~\bibnamefont {Forró}},\ }\href
  {https://doi.org/10.1103/PhysRevLett.99.146403} {\bibfield  {journal}
  {\bibinfo  {journal} {Physical Review Letters}\ }\textbf {\bibinfo {volume}
  {99}},\ \bibinfo {pages} {146403} (\bibinfo {year} {2007})}\BibitemShut
  {NoStop}%
\bibitem [{\citenamefont {Sipos}\ \emph {et~al.}(2008)\citenamefont {Sipos},
  \citenamefont {Kusmartseva}, \citenamefont {Akrap}, \citenamefont {Berger},
  \citenamefont {Forr{\'{o}}},\ and\ \citenamefont {Tuti{\v{s}}}}]{Sipos2008}%
  \BibitemOpen
  \bibfield  {author} {\bibinfo {author} {\bibfnamefont {B.}~\bibnamefont
  {Sipos}}, \bibinfo {author} {\bibfnamefont {A.~F.}\ \bibnamefont
  {Kusmartseva}}, \bibinfo {author} {\bibfnamefont {A.}~\bibnamefont {Akrap}},
  \bibinfo {author} {\bibfnamefont {H.}~\bibnamefont {Berger}}, \bibinfo
  {author} {\bibfnamefont {L.}~\bibnamefont {Forr{\'{o}}}},\ and\ \bibinfo
  {author} {\bibfnamefont {E.}~\bibnamefont {Tuti{\v{s}}}},\ }\href
  {https://doi.org/10.1038/nmat2318} {\bibfield  {journal} {\bibinfo  {journal}
  {Nature Materials}\ }\textbf {\bibinfo {volume} {7}},\ \bibinfo {pages} {960}
  (\bibinfo {year} {2008})}\BibitemShut {NoStop}%
\bibitem [{\citenamefont {Guo}\ \emph {et~al.}(2017)\citenamefont {Guo},
  \citenamefont {Deng}, \citenamefont {Sun}, \citenamefont {Li}, \citenamefont
  {Zhao}, \citenamefont {Wu}, \citenamefont {Chu}, \citenamefont {Zhang},
  \citenamefont {Pan}, \citenamefont {Zheng}, \citenamefont {Wu}, \citenamefont
  {Jin}, \citenamefont {Wu},\ and\ \citenamefont {Xie}}]{Guo2017}%
  \BibitemOpen
  \bibfield  {author} {\bibinfo {author} {\bibfnamefont {Y.}~\bibnamefont
  {Guo}}, \bibinfo {author} {\bibfnamefont {H.}~\bibnamefont {Deng}}, \bibinfo
  {author} {\bibfnamefont {X.}~\bibnamefont {Sun}}, \bibinfo {author}
  {\bibfnamefont {X.}~\bibnamefont {Li}}, \bibinfo {author} {\bibfnamefont
  {J.}~\bibnamefont {Zhao}}, \bibinfo {author} {\bibfnamefont {J.}~\bibnamefont
  {Wu}}, \bibinfo {author} {\bibfnamefont {W.}~\bibnamefont {Chu}}, \bibinfo
  {author} {\bibfnamefont {S.}~\bibnamefont {Zhang}}, \bibinfo {author}
  {\bibfnamefont {H.}~\bibnamefont {Pan}}, \bibinfo {author} {\bibfnamefont
  {X.}~\bibnamefont {Zheng}}, \bibinfo {author} {\bibfnamefont
  {X.}~\bibnamefont {Wu}}, \bibinfo {author} {\bibfnamefont {C.}~\bibnamefont
  {Jin}}, \bibinfo {author} {\bibfnamefont {C.}~\bibnamefont {Wu}},\ and\
  \bibinfo {author} {\bibfnamefont {Y.}~\bibnamefont {Xie}},\ }\href
  {https://doi.org/10.1002/adma.201700715} {\bibfield  {journal} {\bibinfo
  {journal} {Advanced Materials}\ }\textbf {\bibinfo {volume} {29}},\ \bibinfo
  {pages} {1700715} (\bibinfo {year} {2017})}\BibitemShut {NoStop}%
\bibitem [{\citenamefont {Liu}\ \emph {et~al.}(2021)\citenamefont {Liu},
  \citenamefont {Leveillee}, \citenamefont {Lu}, \citenamefont {Yu},
  \citenamefont {Kim}, \citenamefont {Tian}, \citenamefont {Shi}, \citenamefont
  {Lai}, \citenamefont {Zhang}, \citenamefont {Giustino},\ and\ \citenamefont
  {Shih}}]{Liu2021}%
  \BibitemOpen
  \bibfield  {author} {\bibinfo {author} {\bibfnamefont {M.}~\bibnamefont
  {Liu}}, \bibinfo {author} {\bibfnamefont {J.}~\bibnamefont {Leveillee}},
  \bibinfo {author} {\bibfnamefont {S.}~\bibnamefont {Lu}}, \bibinfo {author}
  {\bibfnamefont {J.}~\bibnamefont {Yu}}, \bibinfo {author} {\bibfnamefont
  {H.}~\bibnamefont {Kim}}, \bibinfo {author} {\bibfnamefont {C.}~\bibnamefont
  {Tian}}, \bibinfo {author} {\bibfnamefont {Y.}~\bibnamefont {Shi}}, \bibinfo
  {author} {\bibfnamefont {K.}~\bibnamefont {Lai}}, \bibinfo {author}
  {\bibfnamefont {C.}~\bibnamefont {Zhang}}, \bibinfo {author} {\bibfnamefont
  {F.}~\bibnamefont {Giustino}},\ and\ \bibinfo {author} {\bibfnamefont
  {C.-K.}\ \bibnamefont {Shih}},\ }\href
  {https://doi.org/10.1126/sciadv.abi6339} {\bibfield  {journal} {\bibinfo
  {journal} {Science Advances}\ }\textbf {\bibinfo {volume} {7}},\ \bibinfo
  {pages} {eabi6339} (\bibinfo {year} {2021})}\BibitemShut {NoStop}%
\bibitem [{\citenamefont {Manzeli}\ \emph {et~al.}(2017)\citenamefont
  {Manzeli}, \citenamefont {Ovchinnikov}, \citenamefont {Pasquier},
  \citenamefont {Yazyev},\ and\ \citenamefont {Kis}}]{Manzeli2017}%
  \BibitemOpen
  \bibfield  {author} {\bibinfo {author} {\bibfnamefont {S.}~\bibnamefont
  {Manzeli}}, \bibinfo {author} {\bibfnamefont {D.}~\bibnamefont
  {Ovchinnikov}}, \bibinfo {author} {\bibfnamefont {D.}~\bibnamefont
  {Pasquier}}, \bibinfo {author} {\bibfnamefont {O.~V.}\ \bibnamefont
  {Yazyev}},\ and\ \bibinfo {author} {\bibfnamefont {A.}~\bibnamefont {Kis}},\
  }\href {https://doi.org/10.1038/natrevmats.2017.33} {\bibfield  {journal}
  {\bibinfo  {journal} {Nature Reviews Materials}\ }\textbf {\bibinfo {volume}
  {2}},\ \bibinfo {pages} {17033} (\bibinfo {year} {2017})}\BibitemShut
  {NoStop}%
\bibitem [{\citenamefont {Choi}\ \emph {et~al.}(2017)\citenamefont {Choi},
  \citenamefont {Choudhary}, \citenamefont {Han}, \citenamefont {Park},
  \citenamefont {Akinwande},\ and\ \citenamefont {Lee}}]{Choi2017}%
  \BibitemOpen
  \bibfield  {author} {\bibinfo {author} {\bibfnamefont {W.}~\bibnamefont
  {Choi}}, \bibinfo {author} {\bibfnamefont {N.}~\bibnamefont {Choudhary}},
  \bibinfo {author} {\bibfnamefont {G.~H.}\ \bibnamefont {Han}}, \bibinfo
  {author} {\bibfnamefont {J.}~\bibnamefont {Park}}, \bibinfo {author}
  {\bibfnamefont {D.}~\bibnamefont {Akinwande}},\ and\ \bibinfo {author}
  {\bibfnamefont {Y.~H.}\ \bibnamefont {Lee}},\ }\href
  {https://doi.org/10.1016/J.MATTOD.2016.10.002} {\bibfield  {journal}
  {\bibinfo  {journal} {Materials Today}\ }\textbf {\bibinfo {volume} {20}},\
  \bibinfo {pages} {116} (\bibinfo {year} {2017})}\BibitemShut {NoStop}%
\bibitem [{\citenamefont {Wang}\ \emph {et~al.}(2018)\citenamefont {Wang},
  \citenamefont {Chernikov}, \citenamefont {Glazov}, \citenamefont {Heinz},
  \citenamefont {Marie}, \citenamefont {Amand},\ and\ \citenamefont
  {Urbaszek}}]{Wang2018}%
  \BibitemOpen
  \bibfield  {author} {\bibinfo {author} {\bibfnamefont {G.}~\bibnamefont
  {Wang}}, \bibinfo {author} {\bibfnamefont {A.}~\bibnamefont {Chernikov}},
  \bibinfo {author} {\bibfnamefont {M.~M.}\ \bibnamefont {Glazov}}, \bibinfo
  {author} {\bibfnamefont {T.~F.}\ \bibnamefont {Heinz}}, \bibinfo {author}
  {\bibfnamefont {X.}~\bibnamefont {Marie}}, \bibinfo {author} {\bibfnamefont
  {T.}~\bibnamefont {Amand}},\ and\ \bibinfo {author} {\bibfnamefont
  {B.}~\bibnamefont {Urbaszek}},\ }\href
  {https://doi.org/10.1103/RevModPhys.90.021001} {\bibfield  {journal}
  {\bibinfo  {journal} {Reviews of Modern Physics}\ }\textbf {\bibinfo {volume}
  {90}},\ \bibinfo {pages} {021001} (\bibinfo {year} {2018})}\BibitemShut
  {NoStop}%
\bibitem [{\citenamefont {Hsu}\ \emph {et~al.}(2017)\citenamefont {Hsu},
  \citenamefont {Vaezi}, \citenamefont {Fischer},\ and\ \citenamefont
  {Kim}}]{Hsu2017}%
  \BibitemOpen
  \bibfield  {author} {\bibinfo {author} {\bibfnamefont {Y.-T.}\ \bibnamefont
  {Hsu}}, \bibinfo {author} {\bibfnamefont {A.}~\bibnamefont {Vaezi}}, \bibinfo
  {author} {\bibfnamefont {M.~H.}\ \bibnamefont {Fischer}},\ and\ \bibinfo
  {author} {\bibfnamefont {E.-A.}\ \bibnamefont {Kim}},\ }\href
  {https://doi.org/10.1038/ncomms14985} {\bibfield  {journal} {\bibinfo
  {journal} {Nature Communications}\ }\textbf {\bibinfo {volume} {8}},\
  \bibinfo {pages} {14985} (\bibinfo {year} {2017})}\BibitemShut {NoStop}%
\bibitem [{\citenamefont {Friend}\ \emph {et~al.}(1977)\citenamefont {Friend},
  \citenamefont {Beal},\ and\ \citenamefont {Yoffe}}]{Friend1977}%
  \BibitemOpen
  \bibfield  {author} {\bibinfo {author} {\bibfnamefont {R.~H.}\ \bibnamefont
  {Friend}}, \bibinfo {author} {\bibfnamefont {A.~R.}\ \bibnamefont {Beal}},\
  and\ \bibinfo {author} {\bibfnamefont {A.~D.}\ \bibnamefont {Yoffe}},\ }\href
  {https://doi.org/10.1080/14786437708232952} {\bibfield  {journal} {\bibinfo
  {journal} {The Philosophical Magazine: A Journal of Theoretical Experimental
  and Applied Physics}\ }\textbf {\bibinfo {volume} {35}},\ \bibinfo {pages}
  {1269} (\bibinfo {year} {1977})}\BibitemShut {NoStop}%
\bibitem [{\citenamefont {Anzenhofer}\ \emph {et~al.}(1970)\citenamefont
  {Anzenhofer}, \citenamefont {{van den Berg}}, \citenamefont {Cossee},\ and\
  \citenamefont {Helle}}]{Anzenhofer1970}%
  \BibitemOpen
  \bibfield  {author} {\bibinfo {author} {\bibfnamefont {K.}~\bibnamefont
  {Anzenhofer}}, \bibinfo {author} {\bibfnamefont {J.~M.}\ \bibnamefont {{van
  den Berg}}}, \bibinfo {author} {\bibfnamefont {P.}~\bibnamefont {Cossee}},\
  and\ \bibinfo {author} {\bibfnamefont {J.~N.}\ \bibnamefont {Helle}},\ }\href
  {https://doi.org/https://doi.org/10.1016/0022-3697(70)90315-X} {\bibfield
  {journal} {\bibinfo  {journal} {Journal of Physics and Chemistry of Solids}\
  }\textbf {\bibinfo {volume} {31}},\ \bibinfo {pages} {1057 } (\bibinfo {year}
  {1970})}\BibitemShut {NoStop}%
\bibitem [{\citenamefont {Parkin}\ and\ \citenamefont
  {Friend}(1980{\natexlab{a}})}]{Parkin1980a}%
  \BibitemOpen
  \bibfield  {author} {\bibinfo {author} {\bibfnamefont {S.~S.~P.}\
  \bibnamefont {Parkin}}\ and\ \bibinfo {author} {\bibfnamefont {R.~H.}\
  \bibnamefont {Friend}},\ }\href {https://doi.org/10.1080/13642818008245370}
  {\bibfield  {journal} {\bibinfo  {journal} {Philosophical Magazine B}\
  }\textbf {\bibinfo {volume} {41}},\ \bibinfo {pages} {65} (\bibinfo {year}
  {1980}{\natexlab{a}})}\BibitemShut {NoStop}%
\bibitem [{\citenamefont {Parkin}\ and\ \citenamefont
  {Friend}(1980{\natexlab{b}})}]{Parkin1980b}%
  \BibitemOpen
  \bibfield  {author} {\bibinfo {author} {\bibfnamefont {S.~S.~P.}\
  \bibnamefont {Parkin}}\ and\ \bibinfo {author} {\bibfnamefont {R.~H.}\
  \bibnamefont {Friend}},\ }\href {https://doi.org/10.1080/13642818008245371}
  {\bibfield  {journal} {\bibinfo  {journal} {Philosophical Magazine B}\
  }\textbf {\bibinfo {volume} {41}},\ \bibinfo {pages} {95} (\bibinfo {year}
  {1980}{\natexlab{b}})}\BibitemShut {NoStop}%
\bibitem [{\citenamefont {Parkin}\ \emph {et~al.}(1983)\citenamefont {Parkin},
  \citenamefont {Marseglia},\ and\ \citenamefont {Brown}}]{Parkin1983}%
  \BibitemOpen
  \bibfield  {author} {\bibinfo {author} {\bibfnamefont {S.~S.~P.}\
  \bibnamefont {Parkin}}, \bibinfo {author} {\bibfnamefont {E.~A.}\
  \bibnamefont {Marseglia}},\ and\ \bibinfo {author} {\bibfnamefont {P.~J.}\
  \bibnamefont {Brown}},\ }\href {https://doi.org/10.1088/0022-3719/16/14/016}
  {\bibfield  {journal} {\bibinfo  {journal} {Journal of Physics C: Solid State
  Physics}\ }\textbf {\bibinfo {volume} {16}},\ \bibinfo {pages} {2765}
  (\bibinfo {year} {1983})}\BibitemShut {NoStop}%
\bibitem [{\citenamefont {Marseglia}(1983)}]{Marseglia1983}%
  \BibitemOpen
  \bibfield  {author} {\bibinfo {author} {\bibfnamefont {E.~A.}\ \bibnamefont
  {Marseglia}},\ }\href {https://doi.org/10.1080/01442358309353343} {\bibfield
  {journal} {\bibinfo  {journal} {International Reviews in Physical Chemistry}\
  }\textbf {\bibinfo {volume} {3}},\ \bibinfo {pages} {177} (\bibinfo {year}
  {1983})}\BibitemShut {NoStop}%
\bibitem [{\citenamefont {Negishi}\ \emph {et~al.}(1987)\citenamefont
  {Negishi}, \citenamefont {Shoube}, \citenamefont {Takahashi}, \citenamefont
  {Ueda}, \citenamefont {Sasaki},\ and\ \citenamefont {Inoue}}]{Negishi1987}%
  \BibitemOpen
  \bibfield  {author} {\bibinfo {author} {\bibfnamefont {H.}~\bibnamefont
  {Negishi}}, \bibinfo {author} {\bibfnamefont {A.}~\bibnamefont {Shoube}},
  \bibinfo {author} {\bibfnamefont {H.}~\bibnamefont {Takahashi}}, \bibinfo
  {author} {\bibfnamefont {Y.}~\bibnamefont {Ueda}}, \bibinfo {author}
  {\bibfnamefont {M.}~\bibnamefont {Sasaki}},\ and\ \bibinfo {author}
  {\bibfnamefont {M.}~\bibnamefont {Inoue}},\ }\href
  {https://doi.org/10.1016/0304-8853(87)90227-7} {\bibfield  {journal}
  {\bibinfo  {journal} {Journal of Magnetism and Magnetic Materials}\ }\textbf
  {\bibinfo {volume} {67}},\ \bibinfo {pages} {179 } (\bibinfo {year}
  {1987})}\BibitemShut {NoStop}%
\bibitem [{\citenamefont {{van den Berg}}\ and\ \citenamefont
  {Cossee}(1968)}]{VanderBerg1986}%
  \BibitemOpen
  \bibfield  {author} {\bibinfo {author} {\bibfnamefont {J.}~\bibnamefont {{van
  den Berg}}}\ and\ \bibinfo {author} {\bibfnamefont {P.}~\bibnamefont
  {Cossee}},\ }\href
  {https://doi.org/https://doi.org/10.1016/S0020-1693(00)87012-7} {\bibfield
  {journal} {\bibinfo  {journal} {Inorganica Chimica Acta}\ }\textbf {\bibinfo
  {volume} {2}},\ \bibinfo {pages} {143 } (\bibinfo {year} {1968})}\BibitemShut
  {NoStop}%
\bibitem [{\citenamefont {{Van Maaren}}\ and\ \citenamefont
  {Schaeffer}(1966)}]{Maaren1966}%
  \BibitemOpen
  \bibfield  {author} {\bibinfo {author} {\bibfnamefont {M.}~\bibnamefont {{Van
  Maaren}}}\ and\ \bibinfo {author} {\bibfnamefont {G.}~\bibnamefont
  {Schaeffer}},\ }\href
  {https://doi.org/https://doi.org/10.1016/0031-9163(66)90902-4} {\bibfield
  {journal} {\bibinfo  {journal} {Physics Letters}\ }\textbf {\bibinfo {volume}
  {20}},\ \bibinfo {pages} {131} (\bibinfo {year} {1966})}\BibitemShut
  {NoStop}%
\bibitem [{\citenamefont {Togawa}\ \emph {et~al.}(2012)\citenamefont {Togawa},
  \citenamefont {Koyama}, \citenamefont {Takayanagi}, \citenamefont {Mori},
  \citenamefont {Kousaka}, \citenamefont {Akimitsu}, \citenamefont {Nishihara},
  \citenamefont {Inoue}, \citenamefont {Ovchinnikov},\ and\ \citenamefont
  {Kishine}}]{Togawa2012}%
  \BibitemOpen
  \bibfield  {author} {\bibinfo {author} {\bibfnamefont {Y.}~\bibnamefont
  {Togawa}}, \bibinfo {author} {\bibfnamefont {T.}~\bibnamefont {Koyama}},
  \bibinfo {author} {\bibfnamefont {K.}~\bibnamefont {Takayanagi}}, \bibinfo
  {author} {\bibfnamefont {S.}~\bibnamefont {Mori}}, \bibinfo {author}
  {\bibfnamefont {Y.}~\bibnamefont {Kousaka}}, \bibinfo {author} {\bibfnamefont
  {J.}~\bibnamefont {Akimitsu}}, \bibinfo {author} {\bibfnamefont
  {S.}~\bibnamefont {Nishihara}}, \bibinfo {author} {\bibfnamefont
  {K.}~\bibnamefont {Inoue}}, \bibinfo {author} {\bibfnamefont {A.~S.}\
  \bibnamefont {Ovchinnikov}},\ and\ \bibinfo {author} {\bibfnamefont
  {J.}~\bibnamefont {Kishine}},\ }\href
  {https://doi.org/10.1103/PhysRevLett.108.107202} {\bibfield  {journal}
  {\bibinfo  {journal} {Physical Review Letters}\ }\textbf {\bibinfo {volume}
  {108}},\ \bibinfo {pages} {107202} (\bibinfo {year} {2012})}\BibitemShut
  {NoStop}%
\bibitem [{\citenamefont {Dai}\ \emph {et~al.}(2019)\citenamefont {Dai},
  \citenamefont {Liu}, \citenamefont {Wang}, \citenamefont {Fan}, \citenamefont
  {Pi}, \citenamefont {Zhang},\ and\ \citenamefont {Zhang}}]{Dai2019}%
  \BibitemOpen
  \bibfield  {author} {\bibinfo {author} {\bibfnamefont {Y.}~\bibnamefont
  {Dai}}, \bibinfo {author} {\bibfnamefont {W.}~\bibnamefont {Liu}}, \bibinfo
  {author} {\bibfnamefont {Y.}~\bibnamefont {Wang}}, \bibinfo {author}
  {\bibfnamefont {J.}~\bibnamefont {Fan}}, \bibinfo {author} {\bibfnamefont
  {L.}~\bibnamefont {Pi}}, \bibinfo {author} {\bibfnamefont {L.}~\bibnamefont
  {Zhang}},\ and\ \bibinfo {author} {\bibfnamefont {Y.}~\bibnamefont {Zhang}},\
  }\href {https://doi.org/10.1088/1361-648x/aafebc} {\bibfield  {journal}
  {\bibinfo  {journal} {Journal of Physics: Condensed Matter}\ }\textbf
  {\bibinfo {volume} {31}},\ \bibinfo {pages} {195803} (\bibinfo {year}
  {2019})}\BibitemShut {NoStop}%
\bibitem [{\citenamefont {Morosan}\ \emph {et~al.}(2007)\citenamefont
  {Morosan}, \citenamefont {Zandbergen}, \citenamefont {Li}, \citenamefont
  {Lee}, \citenamefont {Checkelsky}, \citenamefont {Heinrich}, \citenamefont
  {Siegrist}, \citenamefont {Ong},\ and\ \citenamefont {Cava}}]{Morosan2007}%
  \BibitemOpen
  \bibfield  {author} {\bibinfo {author} {\bibfnamefont {E.}~\bibnamefont
  {Morosan}}, \bibinfo {author} {\bibfnamefont {H.~W.}\ \bibnamefont
  {Zandbergen}}, \bibinfo {author} {\bibfnamefont {L.}~\bibnamefont {Li}},
  \bibinfo {author} {\bibfnamefont {M.}~\bibnamefont {Lee}}, \bibinfo {author}
  {\bibfnamefont {J.~G.}\ \bibnamefont {Checkelsky}}, \bibinfo {author}
  {\bibfnamefont {M.}~\bibnamefont {Heinrich}}, \bibinfo {author}
  {\bibfnamefont {T.}~\bibnamefont {Siegrist}}, \bibinfo {author}
  {\bibfnamefont {N.~P.}\ \bibnamefont {Ong}},\ and\ \bibinfo {author}
  {\bibfnamefont {R.~J.}\ \bibnamefont {Cava}},\ }\href
  {https://doi.org/10.1103/PhysRevB.75.104401} {\bibfield  {journal} {\bibinfo
  {journal} {Physical Review B}\ }\textbf {\bibinfo {volume} {75}},\ \bibinfo
  {pages} {104401} (\bibinfo {year} {2007})}\BibitemShut {NoStop}%
\bibitem [{\citenamefont {Ghimire}\ \emph {et~al.}(2018)\citenamefont
  {Ghimire}, \citenamefont {Botana}, \citenamefont {Jiang}, \citenamefont
  {Zhang}, \citenamefont {Chen},\ and\ \citenamefont {Mitchell}}]{Ghimire2018}%
  \BibitemOpen
  \bibfield  {author} {\bibinfo {author} {\bibfnamefont {N.~J.}\ \bibnamefont
  {Ghimire}}, \bibinfo {author} {\bibfnamefont {A.~S.}\ \bibnamefont {Botana}},
  \bibinfo {author} {\bibfnamefont {J.~S.}\ \bibnamefont {Jiang}}, \bibinfo
  {author} {\bibfnamefont {J.}~\bibnamefont {Zhang}}, \bibinfo {author}
  {\bibfnamefont {Y.-S.}\ \bibnamefont {Chen}},\ and\ \bibinfo {author}
  {\bibfnamefont {J.~F.}\ \bibnamefont {Mitchell}},\ }\href
  {https://doi.org/10.1038/s41467-018-05756-7} {\bibfield  {journal} {\bibinfo
  {journal} {Nature Communications}\ }\textbf {\bibinfo {volume} {9}},\
  \bibinfo {pages} {3280} (\bibinfo {year} {2018})}\BibitemShut {NoStop}%
\bibitem [{\citenamefont {Inoshita}\ \emph {et~al.}(2019)\citenamefont
  {Inoshita}, \citenamefont {Hirayama}, \citenamefont {Hamada}, \citenamefont
  {Hosono},\ and\ \citenamefont {Murakami}}]{Inoshita2019}%
  \BibitemOpen
  \bibfield  {author} {\bibinfo {author} {\bibfnamefont {T.}~\bibnamefont
  {Inoshita}}, \bibinfo {author} {\bibfnamefont {M.}~\bibnamefont {Hirayama}},
  \bibinfo {author} {\bibfnamefont {N.}~\bibnamefont {Hamada}}, \bibinfo
  {author} {\bibfnamefont {H.}~\bibnamefont {Hosono}},\ and\ \bibinfo {author}
  {\bibfnamefont {S.}~\bibnamefont {Murakami}},\ }\href
  {https://doi.org/10.1103/PhysRevB.100.121112} {\bibfield  {journal} {\bibinfo
   {journal} {Physical Review B}\ }\textbf {\bibinfo {volume} {100}},\ \bibinfo
  {pages} {121112(R)} (\bibinfo {year} {2019})}\BibitemShut {NoStop}%
\bibitem [{\citenamefont {Momma}\ and\ \citenamefont
  {Izumi}(2011)}]{Momma2011}%
  \BibitemOpen
  \bibfield  {author} {\bibinfo {author} {\bibfnamefont {K.}~\bibnamefont
  {Momma}}\ and\ \bibinfo {author} {\bibfnamefont {F.}~\bibnamefont {Izumi}},\
  }\href {https://doi.org/10.1107/S0021889811038970} {\bibfield  {journal}
  {\bibinfo  {journal} {Journal of Applied Crystallography}\ }\textbf {\bibinfo
  {volume} {44}},\ \bibinfo {pages} {1272} (\bibinfo {year}
  {2011})}\BibitemShut {NoStop}%
\bibitem [{\citenamefont {Himpsel}(1983)}]{Himpsel1983}%
  \BibitemOpen
  \bibfield  {author} {\bibinfo {author} {\bibfnamefont {F.}~\bibnamefont
  {Himpsel}},\ }\href {https://doi.org/10.1080/00018738300101521} {\bibfield
  {journal} {\bibinfo  {journal} {Advances in Physics}\ }\textbf {\bibinfo
  {volume} {32}},\ \bibinfo {pages} {1} (\bibinfo {year} {1983})}\BibitemShut
  {NoStop}%
\bibitem [{\citenamefont {Yeh}\ and\ \citenamefont {Lindau}(1985)}]{Yeh1985}%
  \BibitemOpen
  \bibfield  {author} {\bibinfo {author} {\bibfnamefont {J.}~\bibnamefont
  {Yeh}}\ and\ \bibinfo {author} {\bibfnamefont {I.}~\bibnamefont {Lindau}},\
  }\href {https://doi.org/https://doi.org/10.1016/0092-640X(85)90016-6}
  {\bibfield  {journal} {\bibinfo  {journal} {Atomic Data and Nuclear Data
  Tables}\ }\textbf {\bibinfo {volume} {32}},\ \bibinfo {pages} {1 } (\bibinfo
  {year} {1985})}\BibitemShut {NoStop}%
\bibitem [{\citenamefont {Clark}(1976)}]{Clark1976}%
  \BibitemOpen
  \bibfield  {author} {\bibinfo {author} {\bibfnamefont {W.~B.}\ \bibnamefont
  {Clark}},\ }\href {https://doi.org/10.1088/0022-3719/9/24/005} {\bibfield
  {journal} {\bibinfo  {journal} {Journal of Physics C: Solid State Physics}\
  }\textbf {\bibinfo {volume} {9}},\ \bibinfo {pages} {L693} (\bibinfo {year}
  {1976})}\BibitemShut {NoStop}%
\bibitem [{\citenamefont {Parkin}\ and\ \citenamefont
  {Beal}(1980)}]{Parkin1980c}%
  \BibitemOpen
  \bibfield  {author} {\bibinfo {author} {\bibfnamefont {S.~S.~P.}\
  \bibnamefont {Parkin}}\ and\ \bibinfo {author} {\bibfnamefont {A.~R.}\
  \bibnamefont {Beal}},\ }\href {https://doi.org/10.1080/01418638008224031}
  {\bibfield  {journal} {\bibinfo  {journal} {Philosophical Magazine B}\
  }\textbf {\bibinfo {volume} {42}},\ \bibinfo {pages} {627} (\bibinfo {year}
  {1980})}\BibitemShut {NoStop}%
\bibitem [{\citenamefont {Battaglia}\ \emph {et~al.}(2007)\citenamefont
  {Battaglia}, \citenamefont {Cercellier}, \citenamefont {Despont},
  \citenamefont {Monney}, \citenamefont {Prester}, \citenamefont {Berger},
  \citenamefont {Forr{\'o}}, \citenamefont {Garnier},\ and\ \citenamefont
  {Aebi}}]{Battaglia2007}%
  \BibitemOpen
  \bibfield  {author} {\bibinfo {author} {\bibfnamefont {C.}~\bibnamefont
  {Battaglia}}, \bibinfo {author} {\bibfnamefont {H.}~\bibnamefont
  {Cercellier}}, \bibinfo {author} {\bibfnamefont {L.}~\bibnamefont {Despont}},
  \bibinfo {author} {\bibfnamefont {C.}~\bibnamefont {Monney}}, \bibinfo
  {author} {\bibfnamefont {M.}~\bibnamefont {Prester}}, \bibinfo {author}
  {\bibfnamefont {H.}~\bibnamefont {Berger}}, \bibinfo {author} {\bibfnamefont
  {L.}~\bibnamefont {Forr{\'o}}}, \bibinfo {author} {\bibfnamefont {M.~G.}\
  \bibnamefont {Garnier}},\ and\ \bibinfo {author} {\bibfnamefont
  {P.}~\bibnamefont {Aebi}},\ }\href
  {https://doi.org/10.1140/epjb/e2007-00188-1} {\bibfield  {journal} {\bibinfo
  {journal} {The European Physical Journal B}\ }\textbf {\bibinfo {volume}
  {57}},\ \bibinfo {pages} {385} (\bibinfo {year} {2007})}\BibitemShut
  {NoStop}%
\bibitem [{\citenamefont {Sirica}\ \emph {et~al.}(2016)\citenamefont {Sirica},
  \citenamefont {Mo}, \citenamefont {Bondino}, \citenamefont {Pis},
  \citenamefont {Nappini}, \citenamefont {Vilmercati}, \citenamefont {Yi},
  \citenamefont {Gai}, \citenamefont {Snijders}, \citenamefont {Das},
  \citenamefont {Vobornik}, \citenamefont {Ghimire}, \citenamefont {Koehler},
  \citenamefont {Li}, \citenamefont {Sapkota}, \citenamefont {Parker},
  \citenamefont {Mandrus},\ and\ \citenamefont {Mannella}}]{Sirica2016}%
  \BibitemOpen
  \bibfield  {author} {\bibinfo {author} {\bibfnamefont {N.}~\bibnamefont
  {Sirica}}, \bibinfo {author} {\bibfnamefont {S.-K.}\ \bibnamefont {Mo}},
  \bibinfo {author} {\bibfnamefont {F.}~\bibnamefont {Bondino}}, \bibinfo
  {author} {\bibfnamefont {I.}~\bibnamefont {Pis}}, \bibinfo {author}
  {\bibfnamefont {S.}~\bibnamefont {Nappini}}, \bibinfo {author} {\bibfnamefont
  {P.}~\bibnamefont {Vilmercati}}, \bibinfo {author} {\bibfnamefont
  {J.}~\bibnamefont {Yi}}, \bibinfo {author} {\bibfnamefont {Z.}~\bibnamefont
  {Gai}}, \bibinfo {author} {\bibfnamefont {P.~C.}\ \bibnamefont {Snijders}},
  \bibinfo {author} {\bibfnamefont {P.~K.}\ \bibnamefont {Das}}, \bibinfo
  {author} {\bibfnamefont {I.}~\bibnamefont {Vobornik}}, \bibinfo {author}
  {\bibfnamefont {N.}~\bibnamefont {Ghimire}}, \bibinfo {author} {\bibfnamefont
  {M.~R.}\ \bibnamefont {Koehler}}, \bibinfo {author} {\bibfnamefont
  {L.}~\bibnamefont {Li}}, \bibinfo {author} {\bibfnamefont {D.}~\bibnamefont
  {Sapkota}}, \bibinfo {author} {\bibfnamefont {D.~S.}\ \bibnamefont {Parker}},
  \bibinfo {author} {\bibfnamefont {D.~G.}\ \bibnamefont {Mandrus}},\ and\
  \bibinfo {author} {\bibfnamefont {N.}~\bibnamefont {Mannella}},\ }\href
  {https://doi.org/10.1103/PhysRevB.94.075141} {\bibfield  {journal} {\bibinfo
  {journal} {Physical Review B}\ }\textbf {\bibinfo {volume} {94}},\ \bibinfo
  {pages} {075141} (\bibinfo {year} {2016})}\BibitemShut {NoStop}%
\bibitem [{Note1()}]{Note1}%
  \BibitemOpen
  \bibinfo {note} {In fact, the mentioned effects are routinely taken into
  account within {\protect \it ab initio} calculations, and were rather
  thoroughly addressed by exploring the results of multiple DFT calculations
  that we made in relation this paper.}\BibitemShut {Stop}%
\bibitem [{\citenamefont {Sirica}\ \emph {et~al.}(2020)\citenamefont {Sirica},
  \citenamefont {Vilmercati}, \citenamefont {Bondino}, \citenamefont {Pis},
  \citenamefont {Nappini}, \citenamefont {Mo}, \citenamefont {Fedorov},
  \citenamefont {Das}, \citenamefont {Vobornik}, \citenamefont {Fujii},
  \citenamefont {Li}, \citenamefont {Sapkota}, \citenamefont {Parker},
  \citenamefont {Mandrus},\ and\ \citenamefont {Mannella}}]{Sirica2020}%
  \BibitemOpen
  \bibfield  {author} {\bibinfo {author} {\bibfnamefont {N.}~\bibnamefont
  {Sirica}}, \bibinfo {author} {\bibfnamefont {P.}~\bibnamefont {Vilmercati}},
  \bibinfo {author} {\bibfnamefont {F.}~\bibnamefont {Bondino}}, \bibinfo
  {author} {\bibfnamefont {I.}~\bibnamefont {Pis}}, \bibinfo {author}
  {\bibfnamefont {S.}~\bibnamefont {Nappini}}, \bibinfo {author} {\bibfnamefont
  {S.~K.}\ \bibnamefont {Mo}}, \bibinfo {author} {\bibfnamefont {A.~V.}\
  \bibnamefont {Fedorov}}, \bibinfo {author} {\bibfnamefont {P.~K.}\
  \bibnamefont {Das}}, \bibinfo {author} {\bibfnamefont {I.}~\bibnamefont
  {Vobornik}}, \bibinfo {author} {\bibfnamefont {J.}~\bibnamefont {Fujii}},
  \bibinfo {author} {\bibfnamefont {L.}~\bibnamefont {Li}}, \bibinfo {author}
  {\bibfnamefont {D.}~\bibnamefont {Sapkota}}, \bibinfo {author} {\bibfnamefont
  {D.~S.}\ \bibnamefont {Parker}}, \bibinfo {author} {\bibfnamefont {D.~G.}\
  \bibnamefont {Mandrus}},\ and\ \bibinfo {author} {\bibfnamefont
  {N.}~\bibnamefont {Mannella}},\ }\href
  {https://doi.org/10.1038/s42005-020-0333-3} {\bibfield  {journal} {\bibinfo
  {journal} {Communications Physics}\ }\textbf {\bibinfo {volume} {3}},\
  \bibinfo {pages} {1} (\bibinfo {year} {2020})}\BibitemShut {NoStop}%
\bibitem [{\citenamefont {Bari\ifmmode \check{s}\else
  \v{s}\fi{}i\ifmmode~\acute{c}\else \'{c}\fi{}}\ \emph
  {et~al.}(2011)\citenamefont {Bari\ifmmode \check{s}\else
  \v{s}\fi{}i\ifmmode~\acute{c}\else \'{c}\fi{}}, \citenamefont
  {Smiljani\ifmmode~\acute{c}\else \'{c}\fi{}}, \citenamefont {Pop\ifmmode
  \check{c}\else \v{c}\fi{}evi\ifmmode~\acute{c}\else \'{c}\fi{}},
  \citenamefont {Bilu\ifmmode \check{s}\else \v{s}\fi{}i\ifmmode~\acute{c}\else
  \'{c}\fi{}}, \citenamefont {Tuti\ifmmode~\check{s}\else \v{s}\fi{}},
  \citenamefont {Smontara}, \citenamefont {Berger}, \citenamefont {Ja\ifmmode
  \acute{c}\else \'{c}\fi{}imovi\ifmmode~\acute{c}\else \'{c}\fi{}},
  \citenamefont {Yuli},\ and\ \citenamefont {Forr\'o}}]{Barisic2011}%
  \BibitemOpen
  \bibfield  {author} {\bibinfo {author} {\bibfnamefont {N.}~\bibnamefont
  {Bari\ifmmode \check{s}\else \v{s}\fi{}i\ifmmode~\acute{c}\else \'{c}\fi{}}},
  \bibinfo {author} {\bibfnamefont {I.}~\bibnamefont
  {Smiljani\ifmmode~\acute{c}\else \'{c}\fi{}}}, \bibinfo {author}
  {\bibfnamefont {P.}~\bibnamefont {Pop\ifmmode \check{c}\else
  \v{c}\fi{}evi\ifmmode~\acute{c}\else \'{c}\fi{}}}, \bibinfo {author}
  {\bibfnamefont {A.}~\bibnamefont {Bilu\ifmmode \check{s}\else
  \v{s}\fi{}i\ifmmode~\acute{c}\else \'{c}\fi{}}}, \bibinfo {author}
  {\bibfnamefont {E.}~\bibnamefont {Tuti\ifmmode~\check{s}\else \v{s}\fi{}}},
  \bibinfo {author} {\bibfnamefont {A.}~\bibnamefont {Smontara}}, \bibinfo
  {author} {\bibfnamefont {H.}~\bibnamefont {Berger}}, \bibinfo {author}
  {\bibfnamefont {J.}~\bibnamefont {Ja\ifmmode \acute{c}\else
  \'{c}\fi{}imovi\ifmmode~\acute{c}\else \'{c}\fi{}}}, \bibinfo {author}
  {\bibfnamefont {O.}~\bibnamefont {Yuli}},\ and\ \bibinfo {author}
  {\bibfnamefont {L.}~\bibnamefont {Forr\'o}},\ }\href
  {https://doi.org/10.1103/PhysRevB.84.075157} {\bibfield  {journal} {\bibinfo
  {journal} {Physical Review B}\ }\textbf {\bibinfo {volume} {84}},\ \bibinfo
  {pages} {075157} (\bibinfo {year} {2011})}\BibitemShut {NoStop}%
\bibitem [{\citenamefont {Popčević}\ \emph {et~al.}(2020)\citenamefont
  {Popčević}, \citenamefont {Batistić}, \citenamefont {Smontara},
  \citenamefont {Velebit}, \citenamefont {Jaćimović}, \citenamefont
  {Martino}, \citenamefont {Živković}, \citenamefont {Tsyrulin},
  \citenamefont {Piatek}, \citenamefont {Berger}, \citenamefont {Sidorenko},
  \citenamefont {Rønnow}, \citenamefont {Barišić}, \citenamefont {Forró},\
  and\ \citenamefont {Tutiš}}]{Popcevic2019}%
  \BibitemOpen
  \bibfield  {author} {\bibinfo {author} {\bibfnamefont {P.}~\bibnamefont
  {Popčević}}, \bibinfo {author} {\bibfnamefont {I.}~\bibnamefont
  {Batistić}}, \bibinfo {author} {\bibfnamefont {A.}~\bibnamefont {Smontara}},
  \bibinfo {author} {\bibfnamefont {K.}~\bibnamefont {Velebit}}, \bibinfo
  {author} {\bibfnamefont {J.}~\bibnamefont {Jaćimović}}, \bibinfo {author}
  {\bibfnamefont {E.}~\bibnamefont {Martino}}, \bibinfo {author} {\bibfnamefont
  {I.}~\bibnamefont {Živković}}, \bibinfo {author} {\bibfnamefont
  {N.}~\bibnamefont {Tsyrulin}}, \bibinfo {author} {\bibfnamefont
  {J.}~\bibnamefont {Piatek}}, \bibinfo {author} {\bibfnamefont
  {H.}~\bibnamefont {Berger}}, \bibinfo {author} {\bibfnamefont {A.~A.}\
  \bibnamefont {Sidorenko}}, \bibinfo {author} {\bibfnamefont {H.~M.}\
  \bibnamefont {Rønnow}}, \bibinfo {author} {\bibfnamefont {N.}~\bibnamefont
  {Barišić}}, \bibinfo {author} {\bibfnamefont {L.}~\bibnamefont {Forró}},\
  and\ \bibinfo {author} {\bibfnamefont {E.}~\bibnamefont {Tutiš}},\
  }\href@noop {} {\bibinfo {title} {Electronic transport and magnetism in the
  alternating stack of metallic and highly frustrated magnetic layers in
  {Co$_{1/3}$NbS$_2$}}} (\bibinfo {year} {2020}),\ \Eprint
  {https://arxiv.org/abs/2003.08127} {arXiv:2003.08127 [cond-mat.str-el]}
  \BibitemShut {NoStop}%
\bibitem [{\citenamefont {Mangelsen}\ \emph {et~al.}(2021)\citenamefont
  {Mangelsen}, \citenamefont {Zimmer}, \citenamefont {N{\"{a}}ther},
  \citenamefont {Mankovsky}, \citenamefont {Polesya}, \citenamefont {Ebert},\
  and\ \citenamefont {Bensch}}]{Mangelsen2021}%
  \BibitemOpen
  \bibfield  {author} {\bibinfo {author} {\bibfnamefont {S.}~\bibnamefont
  {Mangelsen}}, \bibinfo {author} {\bibfnamefont {P.}~\bibnamefont {Zimmer}},
  \bibinfo {author} {\bibfnamefont {C.}~\bibnamefont {N{\"{a}}ther}}, \bibinfo
  {author} {\bibfnamefont {S.}~\bibnamefont {Mankovsky}}, \bibinfo {author}
  {\bibfnamefont {S.}~\bibnamefont {Polesya}}, \bibinfo {author} {\bibfnamefont
  {H.}~\bibnamefont {Ebert}},\ and\ \bibinfo {author} {\bibfnamefont
  {W.}~\bibnamefont {Bensch}},\ }\href
  {https://doi.org/10.1103/PhysRevB.103.184408} {\bibfield  {journal} {\bibinfo
   {journal} {Physical Review B}\ }\textbf {\bibinfo {volume} {103}},\ \bibinfo
  {pages} {184408} (\bibinfo {year} {2021})}\BibitemShut {NoStop}%
\bibitem [{NIS()}]{NIST}%
  \BibitemOpen
  \href@noop {} {}\bibinfo {note} {A. Jablonski, F. Salvat, C. J. Powell, and
  A. Y. Lee, NIST Electron Elastic-Scattering Cross-Section Database Version
  4.0, NIST Standard Reference Database Number 64, National Institute of
  Standards and Technology, Gaithersburg MD, 20899 (2016);
  https://srdata.nist.gov/srd64/, (retrieved February 15th, 2022).}\BibitemShut
  {Stop}%
\bibitem [{\citenamefont {Giannozzi}\ \emph {et~al.}(2009)\citenamefont
  {Giannozzi}, \citenamefont {Baroni}, \citenamefont {Bonini}, \citenamefont
  {Calandra}, \citenamefont {Car}, \citenamefont {Cavazzoni}, \citenamefont
  {Ceresoli}, \citenamefont {Chiarotti}, \citenamefont {Cococcioni},
  \citenamefont {Dabo}, \citenamefont {Corso}, \citenamefont {de~Gironcoli},
  \citenamefont {Fabris}, \citenamefont {Fratesi}, \citenamefont {Gebauer},
  \citenamefont {Gerstmann}, \citenamefont {Gougoussis}, \citenamefont
  {Kokalj}, \citenamefont {Lazzeri}, \citenamefont {Martin-Samos},
  \citenamefont {Marzari}, \citenamefont {Mauri}, \citenamefont {Mazzarello},
  \citenamefont {Paolini}, \citenamefont {Pasquarello}, \citenamefont
  {Paulatto}, \citenamefont {Sbraccia}, \citenamefont {Scandolo}, \citenamefont
  {Sclauzero}, \citenamefont {Seitsonen}, \citenamefont {Smogunov},
  \citenamefont {Umari},\ and\ \citenamefont {Wentzcovitch}}]{Giannozzi2009}%
  \BibitemOpen
  \bibfield  {author} {\bibinfo {author} {\bibfnamefont {P.}~\bibnamefont
  {Giannozzi}}, \bibinfo {author} {\bibfnamefont {S.}~\bibnamefont {Baroni}},
  \bibinfo {author} {\bibfnamefont {N.}~\bibnamefont {Bonini}}, \bibinfo
  {author} {\bibfnamefont {M.}~\bibnamefont {Calandra}}, \bibinfo {author}
  {\bibfnamefont {R.}~\bibnamefont {Car}}, \bibinfo {author} {\bibfnamefont
  {C.}~\bibnamefont {Cavazzoni}}, \bibinfo {author} {\bibfnamefont
  {D.}~\bibnamefont {Ceresoli}}, \bibinfo {author} {\bibfnamefont {G.~L.}\
  \bibnamefont {Chiarotti}}, \bibinfo {author} {\bibfnamefont {M.}~\bibnamefont
  {Cococcioni}}, \bibinfo {author} {\bibfnamefont {I.}~\bibnamefont {Dabo}},
  \bibinfo {author} {\bibfnamefont {A.~D.}\ \bibnamefont {Corso}}, \bibinfo
  {author} {\bibfnamefont {S.}~\bibnamefont {de~Gironcoli}}, \bibinfo {author}
  {\bibfnamefont {S.}~\bibnamefont {Fabris}}, \bibinfo {author} {\bibfnamefont
  {G.}~\bibnamefont {Fratesi}}, \bibinfo {author} {\bibfnamefont
  {R.}~\bibnamefont {Gebauer}}, \bibinfo {author} {\bibfnamefont
  {U.}~\bibnamefont {Gerstmann}}, \bibinfo {author} {\bibfnamefont
  {C.}~\bibnamefont {Gougoussis}}, \bibinfo {author} {\bibfnamefont
  {A.}~\bibnamefont {Kokalj}}, \bibinfo {author} {\bibfnamefont
  {M.}~\bibnamefont {Lazzeri}}, \bibinfo {author} {\bibfnamefont
  {L.}~\bibnamefont {Martin-Samos}}, \bibinfo {author} {\bibfnamefont
  {N.}~\bibnamefont {Marzari}}, \bibinfo {author} {\bibfnamefont
  {F.}~\bibnamefont {Mauri}}, \bibinfo {author} {\bibfnamefont
  {R.}~\bibnamefont {Mazzarello}}, \bibinfo {author} {\bibfnamefont
  {S.}~\bibnamefont {Paolini}}, \bibinfo {author} {\bibfnamefont
  {A.}~\bibnamefont {Pasquarello}}, \bibinfo {author} {\bibfnamefont
  {L.}~\bibnamefont {Paulatto}}, \bibinfo {author} {\bibfnamefont
  {C.}~\bibnamefont {Sbraccia}}, \bibinfo {author} {\bibfnamefont
  {S.}~\bibnamefont {Scandolo}}, \bibinfo {author} {\bibfnamefont
  {G.}~\bibnamefont {Sclauzero}}, \bibinfo {author} {\bibfnamefont {A.~P.}\
  \bibnamefont {Seitsonen}}, \bibinfo {author} {\bibfnamefont {A.}~\bibnamefont
  {Smogunov}}, \bibinfo {author} {\bibfnamefont {P.}~\bibnamefont {Umari}},\
  and\ \bibinfo {author} {\bibfnamefont {R.~M.}\ \bibnamefont {Wentzcovitch}},\
  }\href {https://doi.org/10.1088/0953-8984/21/39/395502} {\bibfield  {journal}
  {\bibinfo  {journal} {Journal of Physics: Condensed Matter}\ }\textbf
  {\bibinfo {volume} {21}},\ \bibinfo {pages} {395502} (\bibinfo {year}
  {2009})}\BibitemShut {NoStop}%
\bibitem [{\citenamefont {Giannozzi}\ \emph {et~al.}(2017)\citenamefont
  {Giannozzi}, \citenamefont {Andreussi}, \citenamefont {Brumme}, \citenamefont
  {Bunau}, \citenamefont {Nardelli}, \citenamefont {Calandra}, \citenamefont
  {Car}, \citenamefont {Cavazzoni}, \citenamefont {Ceresoli}, \citenamefont
  {Cococcioni}, \citenamefont {Colonna}, \citenamefont {Carnimeo},
  \citenamefont {Corso}, \citenamefont {de~Gironcoli}, \citenamefont {Delugas},
  \citenamefont {Jr}, \citenamefont {Ferretti}, \citenamefont {Floris},
  \citenamefont {Fratesi}, \citenamefont {Fugallo}, \citenamefont {Gebauer},
  \citenamefont {Gerstmann}, \citenamefont {Giustino}, \citenamefont {Gorni},
  \citenamefont {Jia}, \citenamefont {Kawamura}, \citenamefont {Ko},
  \citenamefont {Kokalj}, \citenamefont {Küçükbenli}, \citenamefont
  {Lazzeri}, \citenamefont {Marsili}, \citenamefont {Marzari}, \citenamefont
  {Mauri}, \citenamefont {Nguyen}, \citenamefont {Nguyen}, \citenamefont {de-la
  Roza}, \citenamefont {Paulatto}, \citenamefont {Poncé}, \citenamefont
  {Rocca}, \citenamefont {Sabatini}, \citenamefont {Santra}, \citenamefont
  {Schlipf}, \citenamefont {Seitsonen}, \citenamefont {Smogunov}, \citenamefont
  {Timrov}, \citenamefont {Thonhauser}, \citenamefont {Umari}, \citenamefont
  {Vast}, \citenamefont {Wu},\ and\ \citenamefont {Baroni}}]{Giannozzi2017}%
  \BibitemOpen
  \bibfield  {author} {\bibinfo {author} {\bibfnamefont {P.}~\bibnamefont
  {Giannozzi}}, \bibinfo {author} {\bibfnamefont {O.}~\bibnamefont
  {Andreussi}}, \bibinfo {author} {\bibfnamefont {T.}~\bibnamefont {Brumme}},
  \bibinfo {author} {\bibfnamefont {O.}~\bibnamefont {Bunau}}, \bibinfo
  {author} {\bibfnamefont {M.~B.}\ \bibnamefont {Nardelli}}, \bibinfo {author}
  {\bibfnamefont {M.}~\bibnamefont {Calandra}}, \bibinfo {author}
  {\bibfnamefont {R.}~\bibnamefont {Car}}, \bibinfo {author} {\bibfnamefont
  {C.}~\bibnamefont {Cavazzoni}}, \bibinfo {author} {\bibfnamefont
  {D.}~\bibnamefont {Ceresoli}}, \bibinfo {author} {\bibfnamefont
  {M.}~\bibnamefont {Cococcioni}}, \bibinfo {author} {\bibfnamefont
  {N.}~\bibnamefont {Colonna}}, \bibinfo {author} {\bibfnamefont
  {I.}~\bibnamefont {Carnimeo}}, \bibinfo {author} {\bibfnamefont {A.~D.}\
  \bibnamefont {Corso}}, \bibinfo {author} {\bibfnamefont {S.}~\bibnamefont
  {de~Gironcoli}}, \bibinfo {author} {\bibfnamefont {P.}~\bibnamefont
  {Delugas}}, \bibinfo {author} {\bibfnamefont {R.~A.~D.}\ \bibnamefont {Jr}},
  \bibinfo {author} {\bibfnamefont {A.}~\bibnamefont {Ferretti}}, \bibinfo
  {author} {\bibfnamefont {A.}~\bibnamefont {Floris}}, \bibinfo {author}
  {\bibfnamefont {G.}~\bibnamefont {Fratesi}}, \bibinfo {author} {\bibfnamefont
  {G.}~\bibnamefont {Fugallo}}, \bibinfo {author} {\bibfnamefont
  {R.}~\bibnamefont {Gebauer}}, \bibinfo {author} {\bibfnamefont
  {U.}~\bibnamefont {Gerstmann}}, \bibinfo {author} {\bibfnamefont
  {F.}~\bibnamefont {Giustino}}, \bibinfo {author} {\bibfnamefont
  {T.}~\bibnamefont {Gorni}}, \bibinfo {author} {\bibfnamefont
  {J.}~\bibnamefont {Jia}}, \bibinfo {author} {\bibfnamefont {M.}~\bibnamefont
  {Kawamura}}, \bibinfo {author} {\bibfnamefont {H.-Y.}\ \bibnamefont {Ko}},
  \bibinfo {author} {\bibfnamefont {A.}~\bibnamefont {Kokalj}}, \bibinfo
  {author} {\bibfnamefont {E.}~\bibnamefont {Küçükbenli}}, \bibinfo {author}
  {\bibfnamefont {M.}~\bibnamefont {Lazzeri}}, \bibinfo {author} {\bibfnamefont
  {M.}~\bibnamefont {Marsili}}, \bibinfo {author} {\bibfnamefont
  {N.}~\bibnamefont {Marzari}}, \bibinfo {author} {\bibfnamefont
  {F.}~\bibnamefont {Mauri}}, \bibinfo {author} {\bibfnamefont {N.~L.}\
  \bibnamefont {Nguyen}}, \bibinfo {author} {\bibfnamefont {H.-V.}\
  \bibnamefont {Nguyen}}, \bibinfo {author} {\bibfnamefont {A.~O.}\
  \bibnamefont {de-la Roza}}, \bibinfo {author} {\bibfnamefont
  {L.}~\bibnamefont {Paulatto}}, \bibinfo {author} {\bibfnamefont
  {S.}~\bibnamefont {Poncé}}, \bibinfo {author} {\bibfnamefont
  {D.}~\bibnamefont {Rocca}}, \bibinfo {author} {\bibfnamefont
  {R.}~\bibnamefont {Sabatini}}, \bibinfo {author} {\bibfnamefont
  {B.}~\bibnamefont {Santra}}, \bibinfo {author} {\bibfnamefont
  {M.}~\bibnamefont {Schlipf}}, \bibinfo {author} {\bibfnamefont {A.~P.}\
  \bibnamefont {Seitsonen}}, \bibinfo {author} {\bibfnamefont {A.}~\bibnamefont
  {Smogunov}}, \bibinfo {author} {\bibfnamefont {I.}~\bibnamefont {Timrov}},
  \bibinfo {author} {\bibfnamefont {T.}~\bibnamefont {Thonhauser}}, \bibinfo
  {author} {\bibfnamefont {P.}~\bibnamefont {Umari}}, \bibinfo {author}
  {\bibfnamefont {N.}~\bibnamefont {Vast}}, \bibinfo {author} {\bibfnamefont
  {X.}~\bibnamefont {Wu}},\ and\ \bibinfo {author} {\bibfnamefont
  {S.}~\bibnamefont {Baroni}},\ }\href
  {http://stacks.iop.org/0953-8984/29/i=46/a=465901} {\bibfield  {journal}
  {\bibinfo  {journal} {Journal of Physics: Condensed Matter}\ }\textbf
  {\bibinfo {volume} {29}},\ \bibinfo {pages} {465901} (\bibinfo {year}
  {2017})}\BibitemShut {NoStop}%
\bibitem [{\citenamefont {Corso}(2014)}]{DalCorso2014}%
  \BibitemOpen
  \bibfield  {author} {\bibinfo {author} {\bibfnamefont {A.~D.}\ \bibnamefont
  {Corso}},\ }\href {https://doi.org/10.1016/j.commatsci.2014.07.043}
  {\bibfield  {journal} {\bibinfo  {journal} {Computational Materials Science}\
  }\textbf {\bibinfo {volume} {95}},\ \bibinfo {pages} {337} (\bibinfo {year}
  {2014})}\BibitemShut {NoStop}%
\bibitem [{\citenamefont {Perdew}\ \emph {et~al.}(1996)\citenamefont {Perdew},
  \citenamefont {Burke},\ and\ \citenamefont {Ernzerhof}}]{Perdew1996}%
  \BibitemOpen
  \bibfield  {author} {\bibinfo {author} {\bibfnamefont {J.~P.}\ \bibnamefont
  {Perdew}}, \bibinfo {author} {\bibfnamefont {K.}~\bibnamefont {Burke}},\ and\
  \bibinfo {author} {\bibfnamefont {M.}~\bibnamefont {Ernzerhof}},\ }\href
  {https://doi.org/10.1103/PhysRevLett.77.3865} {\bibfield  {journal} {\bibinfo
   {journal} {Physical Review Letters}\ }\textbf {\bibinfo {volume} {77}},\
  \bibinfo {pages} {3865} (\bibinfo {year} {1996})}\BibitemShut {NoStop}%
\bibitem [{\citenamefont {Marzari}\ \emph {et~al.}(1999)\citenamefont
  {Marzari}, \citenamefont {Vanderbilt}, \citenamefont {DeVita},\ and\
  \citenamefont {Payne}}]{Marzari1999}%
  \BibitemOpen
  \bibfield  {author} {\bibinfo {author} {\bibfnamefont {N.}~\bibnamefont
  {Marzari}}, \bibinfo {author} {\bibfnamefont {D.}~\bibnamefont {Vanderbilt}},
  \bibinfo {author} {\bibfnamefont {A.}~\bibnamefont {DeVita}},\ and\ \bibinfo
  {author} {\bibfnamefont {M.~C.}\ \bibnamefont {Payne}},\ }\href
  {https://doi.org/10.1103/PhysRevLett.82.3296} {\bibfield  {journal} {\bibinfo
   {journal} {Physical Review Letters}\ }\textbf {\bibinfo {volume} {82}},\
  \bibinfo {pages} {3296} (\bibinfo {year} {1999})}\BibitemShut {NoStop}%
\bibitem [{SIr()}]{SIref}%
  \BibitemOpen
  \href@noop {} {}\bibinfo {note} {See Supplemental Material for details
  regarding the crystal quality, additional LEED images, the temperature
  dependence of the ARPES spectra, various DFT calculations, and tight binding
  parametrization of the DFT spectra.}\BibitemShut {Stop}%
\bibitem [{\citenamefont {Ehlen}\ \emph {et~al.}(2020)\citenamefont {Ehlen},
  \citenamefont {Hell}, \citenamefont {Marini}, \citenamefont {Hasdeo},
  \citenamefont {Saito}, \citenamefont {Falke}, \citenamefont {Goerbig},
  \citenamefont {Santo}, \citenamefont {Petaccia}, \citenamefont {Profeta},\
  and\ \citenamefont {Grüneis}}]{Ehlen2020}%
  \BibitemOpen
  \bibfield  {author} {\bibinfo {author} {\bibfnamefont {N.}~\bibnamefont
  {Ehlen}}, \bibinfo {author} {\bibfnamefont {M.}~\bibnamefont {Hell}},
  \bibinfo {author} {\bibfnamefont {G.}~\bibnamefont {Marini}}, \bibinfo
  {author} {\bibfnamefont {E.~H.}\ \bibnamefont {Hasdeo}}, \bibinfo {author}
  {\bibfnamefont {R.}~\bibnamefont {Saito}}, \bibinfo {author} {\bibfnamefont
  {Y.}~\bibnamefont {Falke}}, \bibinfo {author} {\bibfnamefont {M.~O.}\
  \bibnamefont {Goerbig}}, \bibinfo {author} {\bibfnamefont {G.~D.}\
  \bibnamefont {Santo}}, \bibinfo {author} {\bibfnamefont {L.}~\bibnamefont
  {Petaccia}}, \bibinfo {author} {\bibfnamefont {G.}~\bibnamefont {Profeta}},\
  and\ \bibinfo {author} {\bibfnamefont {A.}~\bibnamefont {Grüneis}},\ }\href
  {https://doi.org/10.1021/ACSNANO.9B08622/ASSET/IMAGES/ACSNANO.9B08622.SOCIAL.JPEG_V03}
  {\bibfield  {journal} {\bibinfo  {journal} {ACS Nano}\ }\textbf {\bibinfo
  {volume} {14}},\ \bibinfo {pages} {1055} (\bibinfo {year}
  {2020})}\BibitemShut {NoStop}%
\bibitem [{\citenamefont {Hell}\ \emph {et~al.}(2020)\citenamefont {Hell},
  \citenamefont {Ehlen}, \citenamefont {Marini}, \citenamefont {Falke},
  \citenamefont {Senkovskiy}, \citenamefont {Herbig}, \citenamefont {Teichert},
  \citenamefont {Jolie}, \citenamefont {Michely}, \citenamefont {Avila},
  \citenamefont {Santo}, \citenamefont {la~Torre}, \citenamefont {Petaccia},
  \citenamefont {Profeta},\ and\ \citenamefont {Grüneis}}]{Hell2020}%
  \BibitemOpen
  \bibfield  {author} {\bibinfo {author} {\bibfnamefont {M.}~\bibnamefont
  {Hell}}, \bibinfo {author} {\bibfnamefont {N.}~\bibnamefont {Ehlen}},
  \bibinfo {author} {\bibfnamefont {G.}~\bibnamefont {Marini}}, \bibinfo
  {author} {\bibfnamefont {Y.}~\bibnamefont {Falke}}, \bibinfo {author}
  {\bibfnamefont {B.~V.}\ \bibnamefont {Senkovskiy}}, \bibinfo {author}
  {\bibfnamefont {C.}~\bibnamefont {Herbig}}, \bibinfo {author} {\bibfnamefont
  {C.}~\bibnamefont {Teichert}}, \bibinfo {author} {\bibfnamefont
  {W.}~\bibnamefont {Jolie}}, \bibinfo {author} {\bibfnamefont
  {T.}~\bibnamefont {Michely}}, \bibinfo {author} {\bibfnamefont
  {J.}~\bibnamefont {Avila}}, \bibinfo {author} {\bibfnamefont {G.~D.}\
  \bibnamefont {Santo}}, \bibinfo {author} {\bibfnamefont {D.~M.}\ \bibnamefont
  {la~Torre}}, \bibinfo {author} {\bibfnamefont {L.}~\bibnamefont {Petaccia}},
  \bibinfo {author} {\bibfnamefont {G.}~\bibnamefont {Profeta}},\ and\ \bibinfo
  {author} {\bibfnamefont {A.}~\bibnamefont {Grüneis}},\ }\href
  {https://doi.org/10.1038/s41467-020-15130-1} {\bibfield  {journal} {\bibinfo
  {journal} {Nature Communications 2020 11:1}\ }\textbf {\bibinfo {volume}
  {11}},\ \bibinfo {pages} {1} (\bibinfo {year} {2020})}\BibitemShut {NoStop}%
\bibitem [{\citenamefont {Moser}(2017)}]{Moser2017}%
  \BibitemOpen
  \bibfield  {author} {\bibinfo {author} {\bibfnamefont {S.}~\bibnamefont
  {Moser}},\ }\href
  {https://doi.org/https://doi.org/10.1016/j.elspec.2016.11.007} {\bibfield
  {journal} {\bibinfo  {journal} {Journal of Electron Spectroscopy and Related
  Phenomena}\ }\textbf {\bibinfo {volume} {214}},\ \bibinfo {pages} {29 }
  (\bibinfo {year} {2017})}\BibitemShut {NoStop}%
\bibitem [{\citenamefont {Cao}\ \emph {et~al.}(2013)\citenamefont {Cao},
  \citenamefont {Waugh}, \citenamefont {Zhang}, \citenamefont {Luo},
  \citenamefont {Wang}, \citenamefont {Reber}, \citenamefont {Mo},
  \citenamefont {Xu}, \citenamefont {Yang}, \citenamefont {Schneeloch},
  \citenamefont {Gu}, \citenamefont {Brahlek}, \citenamefont {Bansal},
  \citenamefont {Oh}, \citenamefont {Zunger},\ and\ \citenamefont
  {Dessau}}]{Cao2013}%
  \BibitemOpen
  \bibfield  {author} {\bibinfo {author} {\bibfnamefont {Y.}~\bibnamefont
  {Cao}}, \bibinfo {author} {\bibfnamefont {J.~A.}\ \bibnamefont {Waugh}},
  \bibinfo {author} {\bibfnamefont {X.-W.}\ \bibnamefont {Zhang}}, \bibinfo
  {author} {\bibfnamefont {J.-W.}\ \bibnamefont {Luo}}, \bibinfo {author}
  {\bibfnamefont {Q.}~\bibnamefont {Wang}}, \bibinfo {author} {\bibfnamefont
  {T.~J.}\ \bibnamefont {Reber}}, \bibinfo {author} {\bibfnamefont {S.~K.}\
  \bibnamefont {Mo}}, \bibinfo {author} {\bibfnamefont {Z.}~\bibnamefont {Xu}},
  \bibinfo {author} {\bibfnamefont {A.}~\bibnamefont {Yang}}, \bibinfo {author}
  {\bibfnamefont {J.}~\bibnamefont {Schneeloch}}, \bibinfo {author}
  {\bibfnamefont {G.~D.}\ \bibnamefont {Gu}}, \bibinfo {author} {\bibfnamefont
  {M.}~\bibnamefont {Brahlek}}, \bibinfo {author} {\bibfnamefont
  {N.}~\bibnamefont {Bansal}}, \bibinfo {author} {\bibfnamefont
  {S.}~\bibnamefont {Oh}}, \bibinfo {author} {\bibfnamefont {A.}~\bibnamefont
  {Zunger}},\ and\ \bibinfo {author} {\bibfnamefont {D.~S.}\ \bibnamefont
  {Dessau}},\ }\href {https://doi.org/10.1038/nphys2685} {\bibfield  {journal}
  {\bibinfo  {journal} {Nature Physics}\ }\textbf {\bibinfo {volume} {9}},\
  \bibinfo {pages} {499} (\bibinfo {year} {2013})}\BibitemShut {NoStop}%
\bibitem [{\citenamefont {Damascelli}(2004)}]{Damascelli2004}%
  \BibitemOpen
  \bibfield  {author} {\bibinfo {author} {\bibfnamefont {A.}~\bibnamefont
  {Damascelli}},\ }\href {https://doi.org/10.1238/physica.topical.109a00061}
  {\bibfield  {journal} {\bibinfo  {journal} {Physica Scripta}\ }\textbf
  {\bibinfo {volume} {T109}},\ \bibinfo {pages} {61} (\bibinfo {year}
  {2004})}\BibitemShut {NoStop}%
\bibitem [{\citenamefont {Moreschini}\ \emph {et~al.}(2014)\citenamefont
  {Moreschini}, \citenamefont {Moser}, \citenamefont {Ebrahimi}, \citenamefont
  {{Dalla Piazza}}, \citenamefont {Kim}, \citenamefont {Boseggia},
  \citenamefont {McMorrow}, \citenamefont {R{\o}nnow}, \citenamefont {Chang},
  \citenamefont {Prabhakaran}, \citenamefont {Boothroyd}, \citenamefont
  {Rotenberg}, \citenamefont {Bostwick},\ and\ \citenamefont
  {Grioni}}]{Moreschini2014}%
  \BibitemOpen
  \bibfield  {author} {\bibinfo {author} {\bibfnamefont {L.}~\bibnamefont
  {Moreschini}}, \bibinfo {author} {\bibfnamefont {S.}~\bibnamefont {Moser}},
  \bibinfo {author} {\bibfnamefont {A.}~\bibnamefont {Ebrahimi}}, \bibinfo
  {author} {\bibfnamefont {B.}~\bibnamefont {{Dalla Piazza}}}, \bibinfo
  {author} {\bibfnamefont {K.~S.}\ \bibnamefont {Kim}}, \bibinfo {author}
  {\bibfnamefont {S.}~\bibnamefont {Boseggia}}, \bibinfo {author}
  {\bibfnamefont {D.~F.}\ \bibnamefont {McMorrow}}, \bibinfo {author}
  {\bibfnamefont {H.~M.}\ \bibnamefont {R{\o}nnow}}, \bibinfo {author}
  {\bibfnamefont {J.}~\bibnamefont {Chang}}, \bibinfo {author} {\bibfnamefont
  {D.}~\bibnamefont {Prabhakaran}}, \bibinfo {author} {\bibfnamefont {A.~T.}\
  \bibnamefont {Boothroyd}}, \bibinfo {author} {\bibfnamefont {E.}~\bibnamefont
  {Rotenberg}}, \bibinfo {author} {\bibfnamefont {A.}~\bibnamefont
  {Bostwick}},\ and\ \bibinfo {author} {\bibfnamefont {M.}~\bibnamefont
  {Grioni}},\ }\href {https://doi.org/10.1103/PhysRevB.89.201114} {\bibfield
  {journal} {\bibinfo  {journal} {Physical Review B}\ }\textbf {\bibinfo
  {volume} {89}},\ \bibinfo {pages} {201114} (\bibinfo {year}
  {2014})}\BibitemShut {NoStop}%
\bibitem [{\citenamefont {Weber}\ \emph {et~al.}(2018)\citenamefont {Weber},
  \citenamefont {Hott}, \citenamefont {Heid}, \citenamefont {Lev},
  \citenamefont {Caputo}, \citenamefont {Schmitt},\ and\ \citenamefont
  {Strocov}}]{weber2018}%
  \BibitemOpen
  \bibfield  {author} {\bibinfo {author} {\bibfnamefont {F.}~\bibnamefont
  {Weber}}, \bibinfo {author} {\bibfnamefont {R.}~\bibnamefont {Hott}},
  \bibinfo {author} {\bibfnamefont {R.}~\bibnamefont {Heid}}, \bibinfo {author}
  {\bibfnamefont {L.~L.}\ \bibnamefont {Lev}}, \bibinfo {author} {\bibfnamefont
  {M.}~\bibnamefont {Caputo}}, \bibinfo {author} {\bibfnamefont
  {T.}~\bibnamefont {Schmitt}},\ and\ \bibinfo {author} {\bibfnamefont {V.~N.}\
  \bibnamefont {Strocov}},\ }\href {https://doi.org/10.1103/PhysRevB.97.235122}
  {\bibfield  {journal} {\bibinfo  {journal} {Physical Review B}\ }\textbf
  {\bibinfo {volume} {97}},\ \bibinfo {pages} {235122} (\bibinfo {year}
  {2018})}\BibitemShut {NoStop}%
\bibitem [{\citenamefont {Strocov}(2003)}]{Strocov2003}%
  \BibitemOpen
  \bibfield  {author} {\bibinfo {author} {\bibfnamefont {V.~N.}\ \bibnamefont
  {Strocov}},\ }\href {https://doi.org/10.1016 / S0368-2048(03)00054-9}
  {\bibfield  {journal} {\bibinfo  {journal} {Journal of Electron Spectroscopy
  and Related Phenomena}\ }\textbf {\bibinfo {volume} {130}},\ \bibinfo {pages}
  {65} (\bibinfo {year} {2003})}\BibitemShut {NoStop}%
\bibitem [{\citenamefont {Tanuma}\ \emph {et~al.}(1994)\citenamefont {Tanuma},
  \citenamefont {Powell},\ and\ \citenamefont {Penn}}]{Tanuma1994}%
  \BibitemOpen
  \bibfield  {author} {\bibinfo {author} {\bibfnamefont {S.}~\bibnamefont
  {Tanuma}}, \bibinfo {author} {\bibfnamefont {C.~J.}\ \bibnamefont {Powell}},\
  and\ \bibinfo {author} {\bibfnamefont {D.~R.}\ \bibnamefont {Penn}},\ }\href
  {https://doi.org/10.1002/sia.740210302} {\bibfield  {journal} {\bibinfo
  {journal} {Surface and Interface Analysis}\ }\textbf {\bibinfo {volume}
  {21}},\ \bibinfo {pages} {165} (\bibinfo {year} {1994})}\BibitemShut
  {NoStop}%
\bibitem [{\citenamefont {El~Youbi}\ \emph {et~al.}(2021)\citenamefont
  {El~Youbi}, \citenamefont {Jung}, \citenamefont {Richter}, \citenamefont
  {Hricovini}, \citenamefont {Cacho},\ and\ \citenamefont
  {Watson}}]{Youbi2020}%
  \BibitemOpen
  \bibfield  {author} {\bibinfo {author} {\bibfnamefont {Z.}~\bibnamefont
  {El~Youbi}}, \bibinfo {author} {\bibfnamefont {S.~W.}\ \bibnamefont {Jung}},
  \bibinfo {author} {\bibfnamefont {C.}~\bibnamefont {Richter}}, \bibinfo
  {author} {\bibfnamefont {K.}~\bibnamefont {Hricovini}}, \bibinfo {author}
  {\bibfnamefont {C.}~\bibnamefont {Cacho}},\ and\ \bibinfo {author}
  {\bibfnamefont {M.~D.}\ \bibnamefont {Watson}},\ }\href
  {https://doi.org/10.1103/PhysRevB.103.155105} {\bibfield  {journal} {\bibinfo
   {journal} {Physical Review B}\ }\textbf {\bibinfo {volume} {103}},\ \bibinfo
  {pages} {155105} (\bibinfo {year} {2021})}\BibitemShut {NoStop}%
\bibitem [{\citenamefont {Nakayama}\ \emph {et~al.}(2006)\citenamefont
  {Nakayama}, \citenamefont {Miwa}, \citenamefont {Ikuta}, \citenamefont
  {Hinode},\ and\ \citenamefont {Wakihara}}]{Nakayama2006}%
  \BibitemOpen
  \bibfield  {author} {\bibinfo {author} {\bibfnamefont {M.}~\bibnamefont
  {Nakayama}}, \bibinfo {author} {\bibfnamefont {K.}~\bibnamefont {Miwa}},
  \bibinfo {author} {\bibfnamefont {H.}~\bibnamefont {Ikuta}}, \bibinfo
  {author} {\bibfnamefont {H.}~\bibnamefont {Hinode}},\ and\ \bibinfo {author}
  {\bibfnamefont {M.}~\bibnamefont {Wakihara}},\ }\href
  {https://doi.org/10.1021/cm060932n} {\bibfield  {journal} {\bibinfo
  {journal} {Chemistry of Materials}\ }\textbf {\bibinfo {volume} {18}},\
  \bibinfo {pages} {4996} (\bibinfo {year} {2006})}\BibitemShut {NoStop}%
\bibitem [{\citenamefont {Mankovsky}\ \emph {et~al.}(2016)\citenamefont
  {Mankovsky}, \citenamefont {Polesya}, \citenamefont {Ebert},\ and\
  \citenamefont {Bensch}}]{Mankovsky2016}%
  \BibitemOpen
  \bibfield  {author} {\bibinfo {author} {\bibfnamefont {S.}~\bibnamefont
  {Mankovsky}}, \bibinfo {author} {\bibfnamefont {S.}~\bibnamefont {Polesya}},
  \bibinfo {author} {\bibfnamefont {H.}~\bibnamefont {Ebert}},\ and\ \bibinfo
  {author} {\bibfnamefont {W.}~\bibnamefont {Bensch}},\ }\href
  {https://doi.org/10.1103/PhysRevB.94.184430} {\bibfield  {journal} {\bibinfo
  {journal} {Physical Review B}\ }\textbf {\bibinfo {volume} {94}},\ \bibinfo
  {pages} {184430} (\bibinfo {year} {2016})}\BibitemShut {NoStop}%
\bibitem [{\citenamefont {Anisimov}\ \emph {et~al.}(1993)\citenamefont
  {Anisimov}, \citenamefont {Solovyev}, \citenamefont {Korotin}, \citenamefont
  {Czy\ifmmode~\dot{z}\else \.{z}\fi{}yk},\ and\ \citenamefont
  {Sawatzky}}]{Anisimov1993}%
  \BibitemOpen
  \bibfield  {author} {\bibinfo {author} {\bibfnamefont {V.~I.}\ \bibnamefont
  {Anisimov}}, \bibinfo {author} {\bibfnamefont {I.~V.}\ \bibnamefont
  {Solovyev}}, \bibinfo {author} {\bibfnamefont {M.~A.}\ \bibnamefont
  {Korotin}}, \bibinfo {author} {\bibfnamefont {M.~T.}\ \bibnamefont
  {Czy\ifmmode~\dot{z}\else \.{z}\fi{}yk}},\ and\ \bibinfo {author}
  {\bibfnamefont {G.~A.}\ \bibnamefont {Sawatzky}},\ }\href
  {https://doi.org/10.1103/PhysRevB.48.16929} {\bibfield  {journal} {\bibinfo
  {journal} {Phys. Rev. B}\ }\textbf {\bibinfo {volume} {48}},\ \bibinfo
  {pages} {16929} (\bibinfo {year} {1993})}\BibitemShut {NoStop}%
\bibitem [{\citenamefont {Liechtenstein}\ \emph {et~al.}(1995)\citenamefont
  {Liechtenstein}, \citenamefont {Anisimov},\ and\ \citenamefont
  {Zaanen}}]{Liechtenstein1995}%
  \BibitemOpen
  \bibfield  {author} {\bibinfo {author} {\bibfnamefont {A.~I.}\ \bibnamefont
  {Liechtenstein}}, \bibinfo {author} {\bibfnamefont {V.~I.}\ \bibnamefont
  {Anisimov}},\ and\ \bibinfo {author} {\bibfnamefont {J.}~\bibnamefont
  {Zaanen}},\ }\href {https://doi.org/10.1103/PhysRevB.52.R5467} {\bibfield
  {journal} {\bibinfo  {journal} {Phys. Rev. B}\ }\textbf {\bibinfo {volume}
  {52}},\ \bibinfo {pages} {R5467} (\bibinfo {year} {1995})}\BibitemShut
  {NoStop}%
\bibitem [{\citenamefont {Cococcioni}\ and\ \citenamefont
  {de~Gironcoli}(2005)}]{Cococcioni2005}%
  \BibitemOpen
  \bibfield  {author} {\bibinfo {author} {\bibfnamefont {M.}~\bibnamefont
  {Cococcioni}}\ and\ \bibinfo {author} {\bibfnamefont {S.}~\bibnamefont
  {de~Gironcoli}},\ }\href {https://doi.org/10.1103/PhysRevB.71.035105}
  {\bibfield  {journal} {\bibinfo  {journal} {Phys. Rev. B}\ }\textbf {\bibinfo
  {volume} {71}},\ \bibinfo {pages} {035105} (\bibinfo {year}
  {2005})}\BibitemShut {NoStop}%
\bibitem [{\citenamefont {Martin}(2004)}]{Martin2004}%
  \BibitemOpen
  \bibfield  {author} {\bibinfo {author} {\bibfnamefont {R.~M.}\ \bibnamefont
  {Martin}},\ }\href {https://doi.org/10.1017/CBO9780511805769} {\emph
  {\bibinfo {title} {Electronic Structure: Basic Theory and Practical
  Methods}}}\ (\bibinfo  {publisher} {Cambridge University Press},\ \bibinfo
  {year} {2004})\BibitemShut {NoStop}%
\bibitem [{\citenamefont {Tolba}\ \emph {et~al.}(2018)\citenamefont {Tolba},
  \citenamefont {Gameel}, \citenamefont {Ali}, \citenamefont {Almossalami},\
  and\ \citenamefont {Allam}}]{Tolba2018}%
  \BibitemOpen
  \bibfield  {author} {\bibinfo {author} {\bibfnamefont {S.~A.}\ \bibnamefont
  {Tolba}}, \bibinfo {author} {\bibfnamefont {K.~M.}\ \bibnamefont {Gameel}},
  \bibinfo {author} {\bibfnamefont {B.~A.}\ \bibnamefont {Ali}}, \bibinfo
  {author} {\bibfnamefont {H.~A.}\ \bibnamefont {Almossalami}},\ and\ \bibinfo
  {author} {\bibfnamefont {N.~K.}\ \bibnamefont {Allam}},\ }\bibinfo {title}
  {{The DFT+U: Approaches, Accuracy, and Applications}},\ in\ \href
  {https://doi.org/https://doi.org/10.5772/intechopen.72020} {\emph {\bibinfo
  {booktitle} {Density Functional Calculations - Recent Progresses of Theory
  and Application}}}\ (\bibinfo {year} {2018})\BibitemShut {NoStop}%
\bibitem [{\citenamefont {Timrov}\ \emph {et~al.}(2018)\citenamefont {Timrov},
  \citenamefont {Marzari},\ and\ \citenamefont {Cococcioni}}]{Timrov2018}%
  \BibitemOpen
  \bibfield  {author} {\bibinfo {author} {\bibfnamefont {I.}~\bibnamefont
  {Timrov}}, \bibinfo {author} {\bibfnamefont {N.}~\bibnamefont {Marzari}},\
  and\ \bibinfo {author} {\bibfnamefont {M.}~\bibnamefont {Cococcioni}},\
  }\href {https://doi.org/10.1103/PhysRevB.98.085127} {\bibfield  {journal}
  {\bibinfo  {journal} {Phys. Rev. B}\ }\textbf {\bibinfo {volume} {98}},\
  \bibinfo {pages} {085127} (\bibinfo {year} {2018})}\BibitemShut {NoStop}%
\bibitem [{\citenamefont {Juhin}\ \emph {et~al.}(2010)\citenamefont {Juhin},
  \citenamefont {de~Groot}, \citenamefont {Vank\'o}, \citenamefont {Calandra},\
  and\ \citenamefont {Brouder}}]{Juhin2010}%
  \BibitemOpen
  \bibfield  {author} {\bibinfo {author} {\bibfnamefont {A.}~\bibnamefont
  {Juhin}}, \bibinfo {author} {\bibfnamefont {F.}~\bibnamefont {de~Groot}},
  \bibinfo {author} {\bibfnamefont {G.}~\bibnamefont {Vank\'o}}, \bibinfo
  {author} {\bibfnamefont {M.}~\bibnamefont {Calandra}},\ and\ \bibinfo
  {author} {\bibfnamefont {C.}~\bibnamefont {Brouder}},\ }\href
  {https://doi.org/10.1103/PhysRevB.81.115115} {\bibfield  {journal} {\bibinfo
  {journal} {Physical Review B}\ }\textbf {\bibinfo {volume} {81}},\ \bibinfo
  {pages} {115115} (\bibinfo {year} {2010})}\BibitemShut {NoStop}%
\bibitem [{\citenamefont {Chen}\ \emph {et~al.}(2011)\citenamefont {Chen},
  \citenamefont {Wu},\ and\ \citenamefont {Selloni}}]{Chen2011}%
  \BibitemOpen
  \bibfield  {author} {\bibinfo {author} {\bibfnamefont {J.}~\bibnamefont
  {Chen}}, \bibinfo {author} {\bibfnamefont {X.}~\bibnamefont {Wu}},\ and\
  \bibinfo {author} {\bibfnamefont {A.}~\bibnamefont {Selloni}},\ }\href
  {https://doi.org/10.1103/PhysRevB.83.245204} {\bibfield  {journal} {\bibinfo
  {journal} {Physical Review B}\ }\textbf {\bibinfo {volume} {83}},\ \bibinfo
  {pages} {245204} (\bibinfo {year} {2011})}\BibitemShut {NoStop}%
\bibitem [{\citenamefont {Mann}\ \emph {et~al.}(2016)\citenamefont {Mann},
  \citenamefont {Lee}, \citenamefont {Cococcioni}, \citenamefont {Smit},\ and\
  \citenamefont {Neaton}}]{Mann2016}%
  \BibitemOpen
  \bibfield  {author} {\bibinfo {author} {\bibfnamefont {G.~W.}\ \bibnamefont
  {Mann}}, \bibinfo {author} {\bibfnamefont {K.}~\bibnamefont {Lee}}, \bibinfo
  {author} {\bibfnamefont {M.}~\bibnamefont {Cococcioni}}, \bibinfo {author}
  {\bibfnamefont {B.}~\bibnamefont {Smit}},\ and\ \bibinfo {author}
  {\bibfnamefont {J.~B.}\ \bibnamefont {Neaton}},\ }\href
  {https://doi.org/10.1063/1.4947240} {\bibfield  {journal} {\bibinfo
  {journal} {The Journal of Chemical Physics}\ }\textbf {\bibinfo {volume}
  {144}},\ \bibinfo {pages} {174104} (\bibinfo {year} {2016})}\BibitemShut
  {NoStop}%
\bibitem [{\citenamefont {Mostofi}\ \emph {et~al.}(2014)\citenamefont
  {Mostofi}, \citenamefont {Yates}, \citenamefont {Pizzi}, \citenamefont {Lee},
  \citenamefont {Souza}, \citenamefont {Vanderbilt},\ and\ \citenamefont
  {Marzari}}]{Wannier902014}%
  \BibitemOpen
  \bibfield  {author} {\bibinfo {author} {\bibfnamefont {A.~A.}\ \bibnamefont
  {Mostofi}}, \bibinfo {author} {\bibfnamefont {J.~R.}\ \bibnamefont {Yates}},
  \bibinfo {author} {\bibfnamefont {G.}~\bibnamefont {Pizzi}}, \bibinfo
  {author} {\bibfnamefont {Y.-S.}\ \bibnamefont {Lee}}, \bibinfo {author}
  {\bibfnamefont {I.}~\bibnamefont {Souza}}, \bibinfo {author} {\bibfnamefont
  {D.}~\bibnamefont {Vanderbilt}},\ and\ \bibinfo {author} {\bibfnamefont
  {N.}~\bibnamefont {Marzari}},\ }\href
  {https://doi.org/https://doi.org/10.1016/j.cpc.2014.05.003} {\bibfield
  {journal} {\bibinfo  {journal} {Computer Physics Communications}\ }\textbf
  {\bibinfo {volume} {185}},\ \bibinfo {pages} {2309} (\bibinfo {year}
  {2014})}\BibitemShut {NoStop}%
\bibitem [{\citenamefont {Wu}\ \emph {et~al.}(2018)\citenamefont {Wu},
  \citenamefont {Zhang}, \citenamefont {Song}, \citenamefont {Troyer},\ and\
  \citenamefont {Soluyanov}}]{Wu2017}%
  \BibitemOpen
  \bibfield  {author} {\bibinfo {author} {\bibfnamefont {Q.}~\bibnamefont
  {Wu}}, \bibinfo {author} {\bibfnamefont {S.}~\bibnamefont {Zhang}}, \bibinfo
  {author} {\bibfnamefont {H.-F.}\ \bibnamefont {Song}}, \bibinfo {author}
  {\bibfnamefont {M.}~\bibnamefont {Troyer}},\ and\ \bibinfo {author}
  {\bibfnamefont {A.~A.}\ \bibnamefont {Soluyanov}},\ }\href
  {https://doi.org/https://doi.org/10.1016/j.cpc.2017.09.033} {\bibfield
  {journal} {\bibinfo  {journal} {Computer Physics Communications}\ }\textbf
  {\bibinfo {volume} {224}},\ \bibinfo {pages} {405 } (\bibinfo {year}
  {2018})}\BibitemShut {NoStop}%
\bibitem [{\citenamefont {Georges}\ \emph {et~al.}(1996)\citenamefont
  {Georges}, \citenamefont {Kotliar}, \citenamefont {Krauth},\ and\
  \citenamefont {Rozenberg}}]{Georges1996}%
  \BibitemOpen
  \bibfield  {author} {\bibinfo {author} {\bibfnamefont {A.}~\bibnamefont
  {Georges}}, \bibinfo {author} {\bibfnamefont {G.}~\bibnamefont {Kotliar}},
  \bibinfo {author} {\bibfnamefont {W.}~\bibnamefont {Krauth}},\ and\ \bibinfo
  {author} {\bibfnamefont {M.~J.}\ \bibnamefont {Rozenberg}},\ }\href
  {https://doi.org/10.1103/RevModPhys.68.13} {\bibfield  {journal} {\bibinfo
  {journal} {Reviews of Modern Physics}\ }\textbf {\bibinfo {volume} {68}},\
  \bibinfo {pages} {13} (\bibinfo {year} {1996})}\BibitemShut {NoStop}%
\bibitem [{\citenamefont {Kotliar}\ \emph {et~al.}(2006)\citenamefont
  {Kotliar}, \citenamefont {Savrasov}, \citenamefont {Haule}, \citenamefont
  {Oudovenko}, \citenamefont {Parcollet},\ and\ \citenamefont
  {Marianetti}}]{Kotliar2006}%
  \BibitemOpen
  \bibfield  {author} {\bibinfo {author} {\bibfnamefont {G.}~\bibnamefont
  {Kotliar}}, \bibinfo {author} {\bibfnamefont {S.~Y.}\ \bibnamefont
  {Savrasov}}, \bibinfo {author} {\bibfnamefont {K.}~\bibnamefont {Haule}},
  \bibinfo {author} {\bibfnamefont {V.~S.}\ \bibnamefont {Oudovenko}}, \bibinfo
  {author} {\bibfnamefont {O.}~\bibnamefont {Parcollet}},\ and\ \bibinfo
  {author} {\bibfnamefont {C.~A.}\ \bibnamefont {Marianetti}},\ }\href
  {https://doi.org/10.1103/RevModPhys.78.865} {\bibfield  {journal} {\bibinfo
  {journal} {Reviews of Modern Physics}\ }\textbf {\bibinfo {volume} {78}},\
  \bibinfo {pages} {865} (\bibinfo {year} {2006})}\BibitemShut {NoStop}%
\bibitem [{\citenamefont {de' Medici}\ \emph {et~al.}(2011)\citenamefont {de'
  Medici}, \citenamefont {Mravlje},\ and\ \citenamefont
  {Georges}}]{Medici2011}%
  \BibitemOpen
  \bibfield  {author} {\bibinfo {author} {\bibfnamefont {L.}~\bibnamefont {de'
  Medici}}, \bibinfo {author} {\bibfnamefont {J.}~\bibnamefont {Mravlje}},\
  and\ \bibinfo {author} {\bibfnamefont {A.}~\bibnamefont {Georges}},\ }\href
  {https://doi.org/10.1103/PhysRevLett.107.256401} {\bibfield  {journal}
  {\bibinfo  {journal} {Physical Review Letters}\ }\textbf {\bibinfo {volume}
  {107}},\ \bibinfo {pages} {256401} (\bibinfo {year} {2011})}\BibitemShut
  {NoStop}%
\bibitem [{\citenamefont {Kotliar}\ \emph {et~al.}(1988)\citenamefont
  {Kotliar}, \citenamefont {Lee},\ and\ \citenamefont {Read}}]{Kotliar1988}%
  \BibitemOpen
  \bibfield  {author} {\bibinfo {author} {\bibfnamefont {G.}~\bibnamefont
  {Kotliar}}, \bibinfo {author} {\bibfnamefont {P.}~\bibnamefont {Lee}},\ and\
  \bibinfo {author} {\bibfnamefont {N.}~\bibnamefont {Read}},\ }\href
  {https://doi.org/10.1016/0921-4534(88)90714-9} {\bibfield  {journal}
  {\bibinfo  {journal} {Physica C: Superconductivity}\ }\textbf {\bibinfo
  {volume} {153-155}},\ \bibinfo {pages} {538} (\bibinfo {year}
  {1988})}\BibitemShut {NoStop}%
\bibitem [{\citenamefont {Coleman}(1984)}]{Coleman1984}%
  \BibitemOpen
  \bibfield  {author} {\bibinfo {author} {\bibfnamefont {P.}~\bibnamefont
  {Coleman}},\ }\href {https://doi.org/10.1103/PhysRevB.29.3035} {\bibfield
  {journal} {\bibinfo  {journal} {Physical Review B}\ }\textbf {\bibinfo
  {volume} {29}},\ \bibinfo {pages} {3035} (\bibinfo {year}
  {1984})}\BibitemShut {NoStop}%
\bibitem [{\citenamefont {Coleman}(2007)}]{Coleman2007}%
  \BibitemOpen
  \bibfield  {author} {\bibinfo {author} {\bibfnamefont {P.}~\bibnamefont
  {Coleman}},\ }\bibinfo {title} {Heavy fermions: Electrons at the edge of
  magnetism},\ in\ \href
  {https://doi.org/https://doi.org/10.1002/9780470022184.hmm105} {\emph
  {\bibinfo {booktitle} {Handbook of Magnetism and Advanced Magnetic
  Materials}}}\ (\bibinfo  {publisher} {American Cancer Society},\ \bibinfo
  {year} {2007})\BibitemShut {NoStop}%
\bibitem [{\citenamefont {Nikšić}\ \emph {et~al.}(1995)\citenamefont
  {Nikšić}, \citenamefont {Tutiš},\ and\ \citenamefont
  {Barišić}}]{Niksic1995}%
  \BibitemOpen
  \bibfield  {author} {\bibinfo {author} {\bibfnamefont {H.}~\bibnamefont
  {Nikšić}}, \bibinfo {author} {\bibfnamefont {E.}~\bibnamefont {Tutiš}},\
  and\ \bibinfo {author} {\bibfnamefont {S.}~\bibnamefont {Barišić}},\ }\href
  {https://doi.org/https://doi.org/10.1016/0921-4534(94)02380-8} {\bibfield
  {journal} {\bibinfo  {journal} {Physica C: Superconductivity}\ }\textbf
  {\bibinfo {volume} {241}},\ \bibinfo {pages} {247} (\bibinfo {year}
  {1995})}\BibitemShut {NoStop}%
\bibitem [{\citenamefont {Tutiš}\ and\ \citenamefont
  {Barišić}(1994)}]{Tutis1994}%
  \BibitemOpen
  \bibfield  {author} {\bibinfo {author} {\bibfnamefont {E.}~\bibnamefont
  {Tutiš}}\ and\ \bibinfo {author} {\bibfnamefont {S.}~\bibnamefont
  {Barišić}},\ }\href
  {https://doi.org/https://doi.org/10.1016/0921-4534(94)92312-4} {\bibfield
  {journal} {\bibinfo  {journal} {Physica C: Superconductivity}\ }\textbf
  {\bibinfo {volume} {235-240}},\ \bibinfo {pages} {2181} (\bibinfo {year}
  {1994})}\BibitemShut {NoStop}%
\bibitem [{\citenamefont {Tuti{\v{s}}}\ \emph {et~al.}(1996)\citenamefont
  {Tuti{\v{s}}}, \citenamefont {Nik{\v{s}}i{\'{c}}},\ and\ \citenamefont
  {Bari{\v{s}}i{\'{c}}}}]{Tutis1997}%
  \BibitemOpen
  \bibfield  {author} {\bibinfo {author} {\bibfnamefont {E.}~\bibnamefont
  {Tuti{\v{s}}}}, \bibinfo {author} {\bibfnamefont {H.}~\bibnamefont
  {Nik{\v{s}}i{\'{c}}}},\ and\ \bibinfo {author} {\bibfnamefont
  {S.}~\bibnamefont {Bari{\v{s}}i{\'{c}}}},\ }in\ \href
  {https://doi.org/10.1007/BFb0106021} {\emph {\bibinfo {booktitle} {From
  Quantum Mechanics to Technology}}},\ \bibinfo {editor} {edited by\ \bibinfo
  {editor} {\bibfnamefont {Z.}~\bibnamefont {Petru}}, \bibinfo {editor}
  {\bibfnamefont {J.}~\bibnamefont {Przystawa}},\ and\ \bibinfo {editor}
  {\bibfnamefont {K.}~\bibnamefont {Rapcewicz}}}\ (\bibinfo  {publisher}
  {Springer Berlin Heidelberg},\ \bibinfo {address} {Berlin, Heidelberg},\
  \bibinfo {year} {1996})\ pp.\ \bibinfo {pages} {161--175}\BibitemShut
  {NoStop}%
\bibitem [{\citenamefont {Tanaka}\ \emph {et~al.}(2022)\citenamefont {Tanaka},
  \citenamefont {Okazaki}, \citenamefont {Kuroda}, \citenamefont {Noguchi},
  \citenamefont {Arai}, \citenamefont {Minami}, \citenamefont {Ideta},
  \citenamefont {Tanaka}, \citenamefont {Lu}, \citenamefont {Hashimoto},
  \citenamefont {Kandyba}, \citenamefont {Cattelan}, \citenamefont {Barinov},
  \citenamefont {Muro}, \citenamefont {Sasagawa},\ and\ \citenamefont
  {Kondo}}]{Tanaka2022}%
  \BibitemOpen
  \bibfield  {author} {\bibinfo {author} {\bibfnamefont {H.}~\bibnamefont
  {Tanaka}}, \bibinfo {author} {\bibfnamefont {S.}~\bibnamefont {Okazaki}},
  \bibinfo {author} {\bibfnamefont {K.}~\bibnamefont {Kuroda}}, \bibinfo
  {author} {\bibfnamefont {R.}~\bibnamefont {Noguchi}}, \bibinfo {author}
  {\bibfnamefont {Y.}~\bibnamefont {Arai}}, \bibinfo {author} {\bibfnamefont
  {S.}~\bibnamefont {Minami}}, \bibinfo {author} {\bibfnamefont
  {S.}~\bibnamefont {Ideta}}, \bibinfo {author} {\bibfnamefont
  {K.}~\bibnamefont {Tanaka}}, \bibinfo {author} {\bibfnamefont
  {D.}~\bibnamefont {Lu}}, \bibinfo {author} {\bibfnamefont {M.}~\bibnamefont
  {Hashimoto}}, \bibinfo {author} {\bibfnamefont {V.}~\bibnamefont {Kandyba}},
  \bibinfo {author} {\bibfnamefont {M.}~\bibnamefont {Cattelan}}, \bibinfo
  {author} {\bibfnamefont {A.}~\bibnamefont {Barinov}}, \bibinfo {author}
  {\bibfnamefont {T.}~\bibnamefont {Muro}}, \bibinfo {author} {\bibfnamefont
  {T.}~\bibnamefont {Sasagawa}},\ and\ \bibinfo {author} {\bibfnamefont
  {T.}~\bibnamefont {Kondo}},\ }\href
  {https://doi.org/10.1103/PhysRevB.105.L121102} {\bibfield  {journal}
  {\bibinfo  {journal} {Physical Review B}\ }\textbf {\bibinfo {volume}
  {105}},\ \bibinfo {pages} {L121102} (\bibinfo {year} {2022})}\BibitemShut
  {NoStop}%
\bibitem [{\citenamefont {Yang}\ \emph {et~al.}(2022)\citenamefont {Yang},
  \citenamefont {LaBollita}, \citenamefont {Cheng}, \citenamefont {Bhandari},
  \citenamefont {Cochran}, \citenamefont {Yin}, \citenamefont {Hossain},
  \citenamefont {Belopolski}, \citenamefont {Zhang}, \citenamefont {Jiang},
  \citenamefont {Shumiya}, \citenamefont {Multer}, \citenamefont {Liskevich},
  \citenamefont {Usanov}, \citenamefont {Dang}, \citenamefont {Strocov},
  \citenamefont {Davydov}, \citenamefont {Ghimire}, \citenamefont {Botana},\
  and\ \citenamefont {Hasan}}]{Yang2022}%
  \BibitemOpen
  \bibfield  {author} {\bibinfo {author} {\bibfnamefont {X.~P.}\ \bibnamefont
  {Yang}}, \bibinfo {author} {\bibfnamefont {H.}~\bibnamefont {LaBollita}},
  \bibinfo {author} {\bibfnamefont {Z.-J.}\ \bibnamefont {Cheng}}, \bibinfo
  {author} {\bibfnamefont {H.}~\bibnamefont {Bhandari}}, \bibinfo {author}
  {\bibfnamefont {T.~A.}\ \bibnamefont {Cochran}}, \bibinfo {author}
  {\bibfnamefont {J.-X.}\ \bibnamefont {Yin}}, \bibinfo {author} {\bibfnamefont
  {M.~S.}\ \bibnamefont {Hossain}}, \bibinfo {author} {\bibfnamefont
  {I.}~\bibnamefont {Belopolski}}, \bibinfo {author} {\bibfnamefont
  {Q.}~\bibnamefont {Zhang}}, \bibinfo {author} {\bibfnamefont
  {Y.}~\bibnamefont {Jiang}}, \bibinfo {author} {\bibfnamefont
  {N.}~\bibnamefont {Shumiya}}, \bibinfo {author} {\bibfnamefont
  {D.}~\bibnamefont {Multer}}, \bibinfo {author} {\bibfnamefont
  {M.}~\bibnamefont {Liskevich}}, \bibinfo {author} {\bibfnamefont {D.~A.}\
  \bibnamefont {Usanov}}, \bibinfo {author} {\bibfnamefont {Y.}~\bibnamefont
  {Dang}}, \bibinfo {author} {\bibfnamefont {V.~N.}\ \bibnamefont {Strocov}},
  \bibinfo {author} {\bibfnamefont {A.~V.}\ \bibnamefont {Davydov}}, \bibinfo
  {author} {\bibfnamefont {N.~J.}\ \bibnamefont {Ghimire}}, \bibinfo {author}
  {\bibfnamefont {A.~S.}\ \bibnamefont {Botana}},\ and\ \bibinfo {author}
  {\bibfnamefont {M.~Z.}\ \bibnamefont {Hasan}},\ }\href
  {https://doi.org/10.48550/arxiv.2203.00675} {\bibinfo {title} {Visualizing
  the out-of-plane electronic dispersions in an intercalated transition metal
  dichalcogenide}} (\bibinfo {year} {2022}),\ \Eprint
  {https://arxiv.org/abs/arXiv:2203.00675v1} {arXiv:2203.00675v1} \BibitemShut
  {NoStop}%
\end{thebibliography}%

\onecolumngrid

\clearpage
\renewcommand{\thefigure}{S\arabic{figure}}
\section*{ Supplemental Material}
\setcounter{section}{0}

\section{Samples}
    Our ARPES spectra were collected on two samples with different orientations.
   Sample A was oriented for scans along $\Gamma$ - M$_0$ and had slightly lower quality than sample B, oriented for scans along $\Gamma$ - K$_0$ direction.
   The sample quality is reflected in the residual resistivity ratio (RRR - the ratio of electrical resistivity at room temperature and 2 K) as well as the temperature of the magnetic ordering (Fig.~\ref{figS1}).
   Sample A had an RRR of 1.75, while sample B showed a somewhat higher RRR of 2.25. 
   The magnetic ordering temperatures of samples A and sample B are T$_N$ = 26 and 28 K, respectively. 
   The lower quality of sample A is the probable reason why bonding and antibonding bands cannot be unambiguously resolved as separate bands along $\Gamma$ - M$_0$ in Fig.~\ref{fig2} in main text. 
   Also, the sample quality is probably the reason for the $\beta$ band showing less clearly in sample A. 
  

\renewcommand{\thefigure}{S\arabic{figure}}
\setcounter{figure}{0}
  \begin{figure}[h!]
  \includegraphics[width=0.45\textwidth]{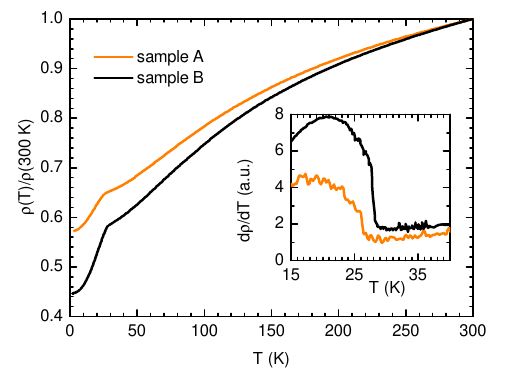}
  \caption{Electrical resistivity curves for samples A and B normalized to their 300 K values. The inset shows the first derivative of resistivities. The start of the sharp upturn of the first derivative marks the transition point}. 
   
  \label{figS1}
  \end{figure}

\section{LEED}


  \begin{figure}[h!]
  \includegraphics[width=0.45\textwidth]{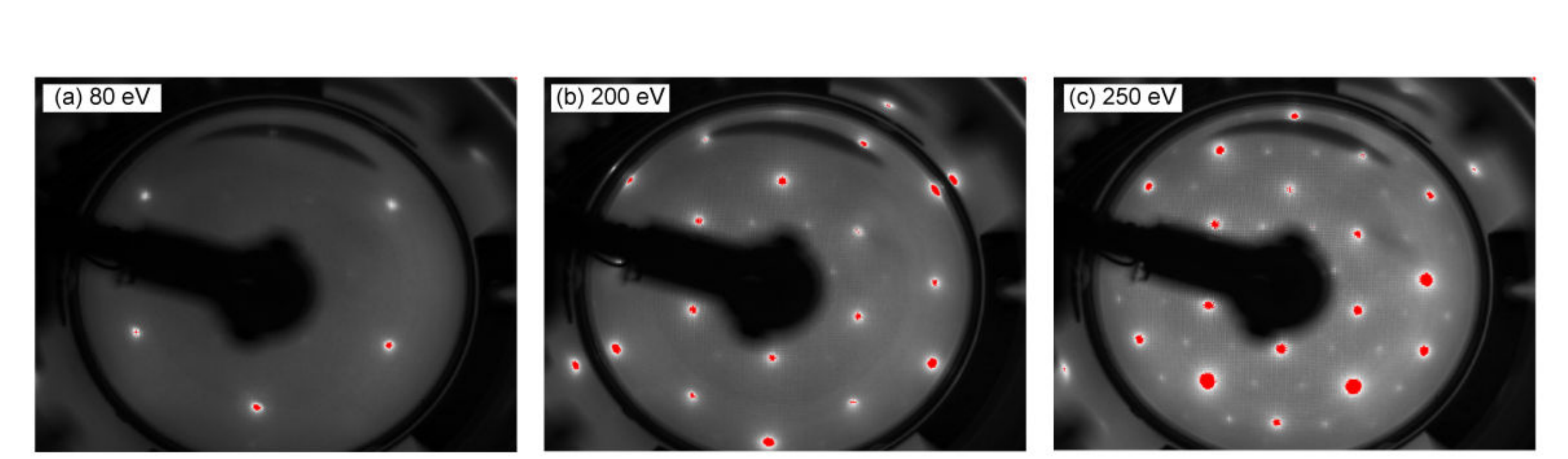}
  \caption{The low-energy electron diffraction (LEED) patterns of Co$_{1/3}$NbS$_2$ obtained by using a) 80 eV, b) 200 eV and c) 250 eV electrons. Black arrows indicate positions of peaks related to ($\sqrt{3}\times\sqrt{3}$)$R30^{\circ}$ superlattice of Co$_{1/3}$NbS$_2$. Red spots indicate the strong signals of 2H-NbS$_2$ lattice. }. 
   
  \label{figLEED}
  \end{figure}

Low-energy electron diffraction (LEED) patterns of freshly cleaved Co$_{1/3}$NbS$_2$ recorded using three different incident electron energies are shown in Fig. ~\ref{figLEED}. The peaks related to parent compound 2H-NbS$_2$ are readily observable at all energies as strong red peaks. The peaks related to Co$_{1/3}$NbS$_2$ super-structure are also discernible at all three panels as indicated with black arrows. However, they are much weaker than those of 2H-NbS$_2$. Primary reason for the difference in intensities between super-structure and parent compound spots is lower concentration of Co ions compared to Nb ions (1:3). In addition, the intensities of all the spots decrease with decreasing the electron energy. It also appears that intensity of super-structure peaks fades more quickly than that of the parent compound. All these phenomena can be associated with the elastic electron scattering cross sections of Co and Nb ions and their electron energy dependence.


\section{Temperature dependence of spectra}


  \begin{figure}[h!]
  \includegraphics[width=\textwidth]{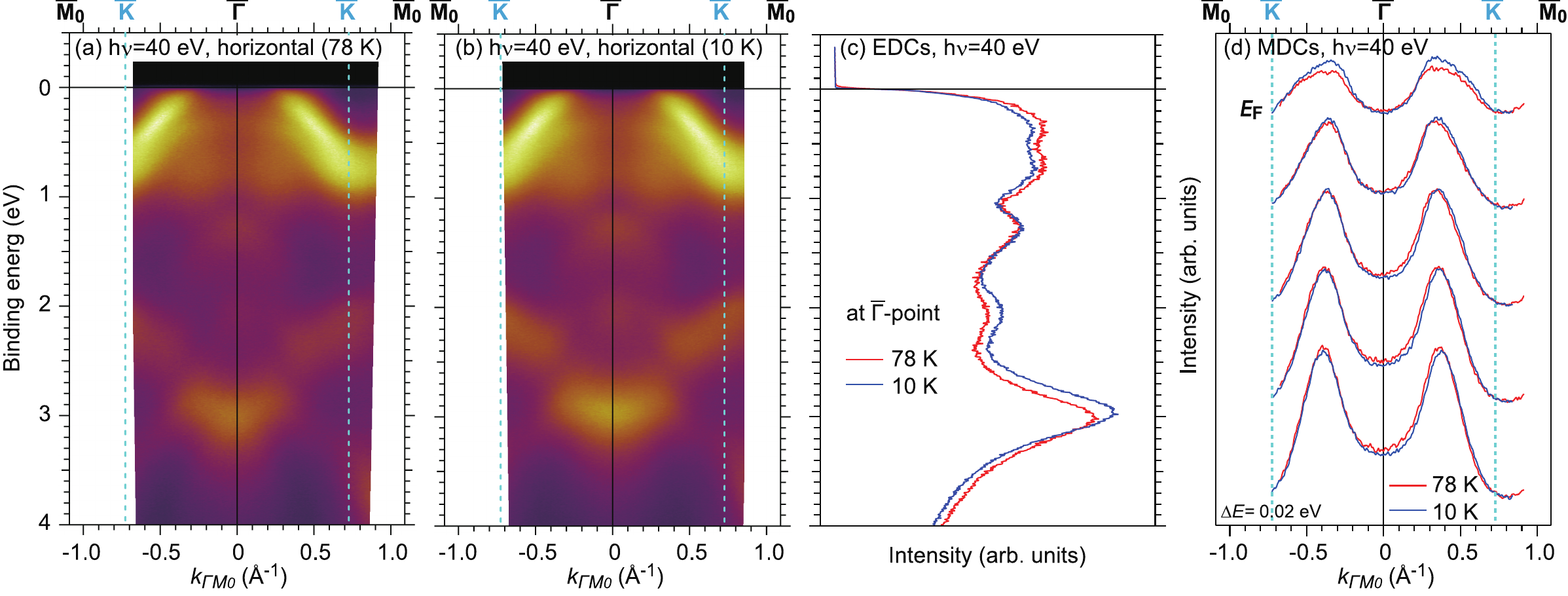}

  \caption{ARPES intensity plots for Co$_{1/3}$NbS$_2$ along the $\Gamma {\rm M_0}$ direction, measured with $h\nu=$ 40 eV using horizontal polarization at (a) 78 K and (b) 10 K. The light-blue dashed lines represent the high symmetry points of the Brillouin zone for the  ($\sqrt{3}\times\sqrt{3}$)$R30^{\circ}$ superlattice. The data are further compared in panels (c) and (d): 
(c) represents the EDCs at the $\Gamma$ point;  
 (d) illustrates MDCs measured at the Fermi level and four other electron energies in steps of 0.02 eV. The curves in red and blue correspond to data at 78 K and 10 K, respectively.} 
  \label{figS2} 
  \end{figure}


  \begin{figure}[h!]
  \includegraphics[width=\textwidth]{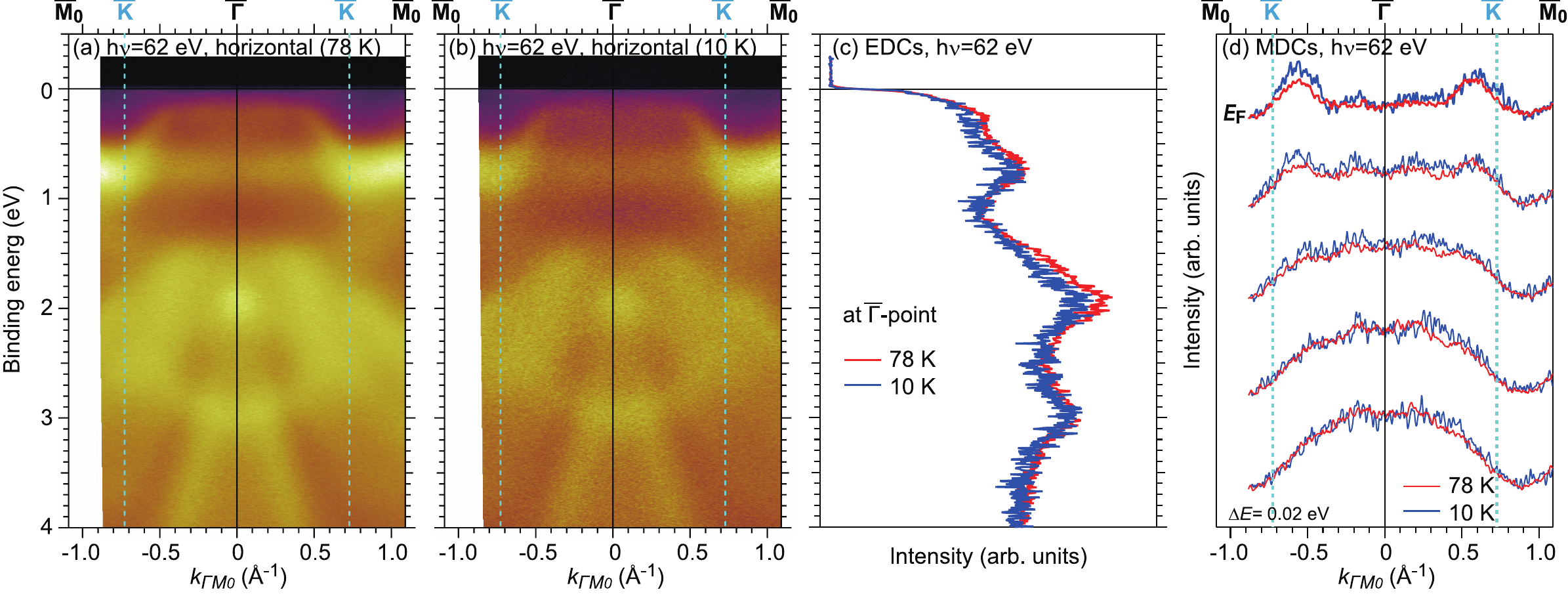}

  \caption{ARPES intensity plots of Co$_{1/3}$NbS$_2$ along the $\Gamma {\rm M_0}$ direction measured with $h\nu=$ 62 eV using horizontal polarization at (a) 78 K and (b) 10 K. Light blue dashed lines represent the high symmetry points of the Brillouin zone  of ($\sqrt{3}\times\sqrt{3}$)$R30^{\circ}$ lattice. (c) and (d) represent temperature dependence of EDCs at $\Gamma$ point and MDCs near $E_{\rm F}$ in steps of 0.02 eV (from top to bottom). Red and blue curves correspond to 78 and 10 K, respectively, taken from the ARPES plots of (a) and (b).}
  \label{figS3} 
  \end{figure}

     Figs.~\ref{figS2} and \ref{figS3} were recorded on the sample A where the $\beta$ band and the antibonding band could not be resolved as separate bands at the Fermi level. 
   However, the data presented in Figs.~\ref{figS2} and \ref{figS3} confirm the absence of temperature variation in the band dispersions upon magnetic transition. 
   While slight transfer of intensity to deeper binding energies at lower temperature appears to occur at $h\nu$ =40 eV the same cannot be argued for  $h\nu$ = 62. 
  Also, the $\beta$ band shows no temperature dependence. Thus, a comparison of 78K data with 10 K data is fully justified. 
  
\clearpage
\section{ Polarization and orientation dependence}


  \begin{figure}[h!]
  \includegraphics[width=\textwidth]{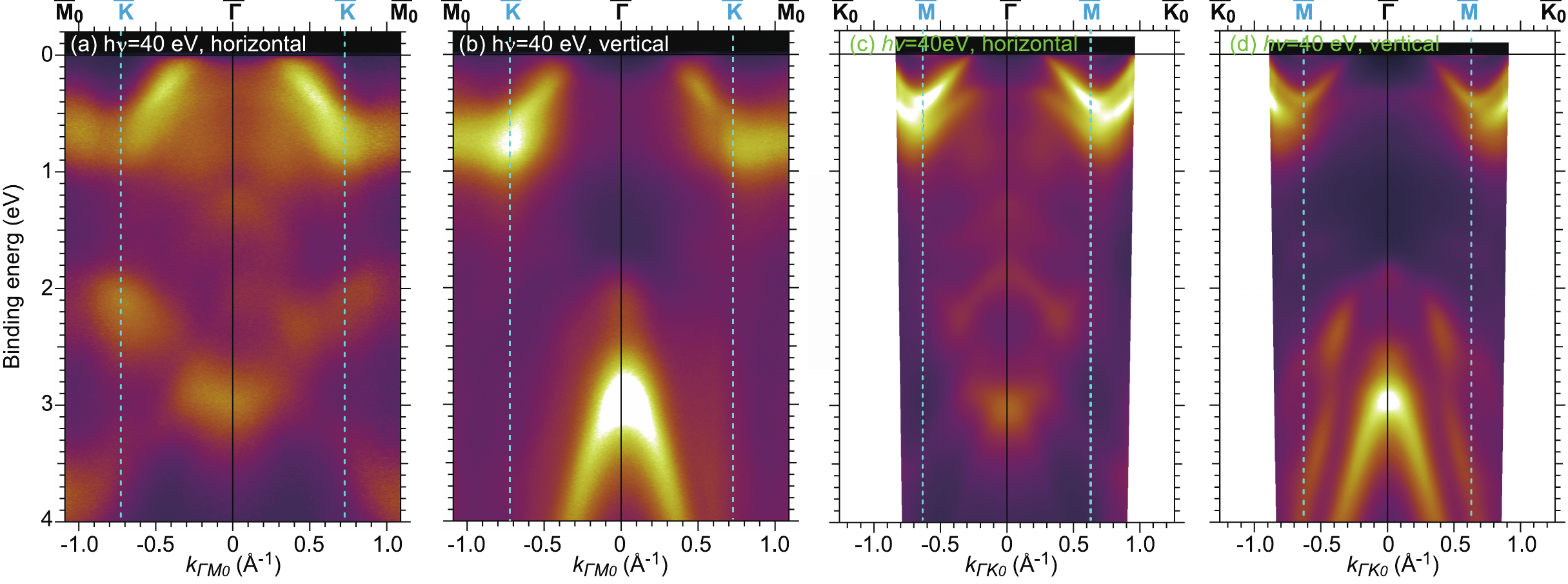}

  \caption{Photon polarization dependent ARPES intensity plots of Co$_{1/3}$NbS$_2$. (a) Horizontal- and (b) vertical-polarization measured with $h\nu$=40 eV at 78 K along the $\Gamma {\rm M_0}$ direction. In order to cover an extended momentum range, three ARPES images collected $\pm15^{\circ}$-frame are stitched. (c) Horizontal- and (d) vertical-polarization measured with $h\nu$=40 eV at 10 K along the $\Gamma {\rm K_0}$ direction. Black solid and light blue dashed lines represent the high symmetry points of the Brillouin zone of the 1$\times$1 and ($\sqrt{3}\times\sqrt{3}$)$R30^{\circ}$ lattices, respectively.  }
  \label{figS4} 
  \end{figure}

   At $h\nu=$ 40 eV photon energy (Fig.~\ref{figS4}), more bands are visible in the horizontal polarization then in the vertical polarization. 
   The same appears in Fig.~\ref{nfig3} for the photon energy of 62 eV.
   The differences in spectra presented in Fig.~\ref{figS4} and Fig.~\ref{nfig3} stems from different photon energy used. 
   While at $h\nu=$ 40 eV Nb orbitals are much more pronounced, at $h\nu=$ 62 eV both, Nb and S, contributions are present resulting in richer spectra.
  Spectra in panels (a) and (b) in Fig.~\ref{figS4} are measured on sample A, whereas the spectra in panels (c) and (d) are measured on sample B. 
   Thus the difference in the quality of the spectra in panels (a) and (b) vs. spectra in panels (c) and (d).


\section{DFT calculations}

\subsection{Wannier90 representation of conduction bands } 

\begin{figure}[h!]
  \includegraphics[width=0.45\textwidth]{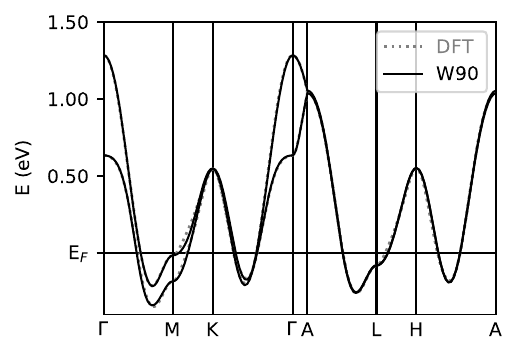}
  \caption{The conducting bands calculated for 2H-NbS$_2$ and their approximation through Wannier90 TB parametrization.} 
  \label{FigNbS2spag}
  \end{figure}

  The DFT calculated conducting bands for 2H-NbS$_2$ upon full lattice relaxation is shown in Fig.~\ref{FigNbS2spag}. The full lines show the tight-binding (TB) approximation  to DFT-calculated bands (dashed lines), obtained through Wannier90 code. The dominant TB parameters include the hybridization integrals between Nb orbitals: in-plane nearest neighbors, $t_1=0.66$ eV; in-plane next-to-nearest neighbors, $t_1=0.104$,  and inter-plane nearest neighbors $t_i= -0.11$ eV
  

\begin{figure}[h!]
  \includegraphics[width=0.45\textwidth]{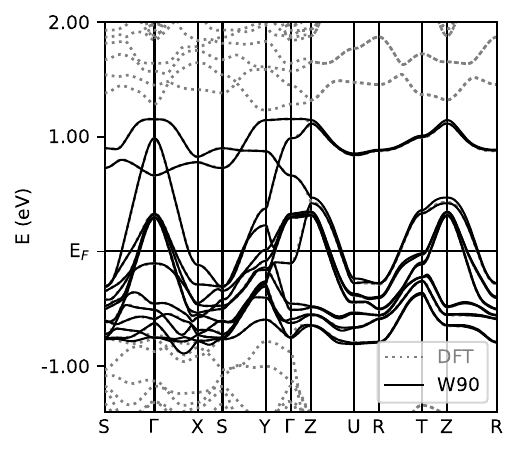}
  \caption{ The electronic bands for Co$_{1/3}$NbS$_2$ calculated in the HOFK antiferromagnetic state (dashed lines), and the some of these bands, as approximated through Wannier90 TB parametrization (full lines).}
  \label{FigAFspag}
  \end{figure}

  The electronic bands for Co$_{1/3}$NbS$_2$ calculated in the HOFK antiferromagnetic state, with fully relaxed crystal structure, are shown in Fig.~\ref{FigAFspag}. 
  The full lines show the tight-binding (TB) parametrization to DFT-calculated bands (shown in dashed lines), obtained through the Wannier90 code. 
  The in-plane hybridization integrals between Nb orbitals are similar in value to those found in 2H-NbS$_2$, with variations within the super-structure, $t_1=( 0.05 \pm 0.01 )$eV, $t_2=( 0.11 \pm 0.008 )$eV. 
  The hybridization integrals between Nb orbitals in adjacent layers vary within the supercell, between -0.029 eV and 0.022 eV. 
  However, the interlayer coupling is dominantly set by a much bigger hybridization integral between nearest Co and Nb orbitals, $t_{Co}\approx 0.23$eV. 
  The variation in Nb orbital energy levels within supercell is 0.04 eV.  The energy level of Co orbital is 0.83 eV above the average energy level of Nb orbitals.


\subsection{Common features of  TB modeling of Co$_{1/3}$NbS$_2$ independent of the type magnetic order}

   Having no DFT+DMFT or any other type of {\it ab-initio} electronic structure calculation that would include spin fluctuations at hand, it is difficult to argue about the degree to which these fluctuations may affect the final spectra. 
  On the other hand, one may gain some insight into how DFT treats the magnetically intercalated TMD's by exploring different magnetic orderings, besides the HOFK AF state. 
  In that respect, we have looked into the DFT electronic structure of several other magnetically ordered states, including variants where their ferromagnetic versions replace the inter-plane or in-plane AF configurations. 
  Some of these configurations emerge with total energies higher than AF HOFK state by only a few meV per Co-atom. 
  We also made Wannier90 TB parametrization for a few of these configurations. 
  Interestingly, the results follow a common pattern. 
  Essentially, the same set of TB parameters apply in whatever magnetic state, with the spatial distribution of TB parameters depending on the magnetic arrangement of Co atoms.   
  The Wannier90 representation of conducting bands results in one effective orbital per Nb for both spin projections and one Co-centered orbital in one spin projection. 
  The Nb orbitals are approximately the same as in pure 2H-NbS$_2$, whereas the spin characters of the Co orbitals that strongly hybridize with Nb orbitals and lie closest to the Fermi level depend on the magnetic configuration.  
  Consistently, these Co orbitals are always approximately 0.8 eV higher in energy than the Nb orbitals making the conduction bands, whereas the hybridization between the Co orbital and the nearest Nb orbitals is relatively large in value, of approximately 0.25 eV.
  This Co-Nb hybridization consistently dominates over the effective hybridization between Nb layers in the intercalated material, whereas the in-layer Nb-Nb hybridizations experience minor changes with respect to those in 2H-NbS$_2$.   
  The NbS$_2$ bands receive approximately 2 electrons from each Co atom;  each Co atom contributes seven Co $d$-bands that lay several eV below the Fermi energy and three Co $d$-bands of the same spin projection above the Fermi level. 
  This consistently implies $S\approx 3/2$ spin state of Co, as showing in magnetic susceptibility measurements.
  The main features of DFT-calculated electronic structures of various magnetic states can be all understood within these terms.


  \begin{figure}[t!] 
  \includegraphics[width=\textwidth]{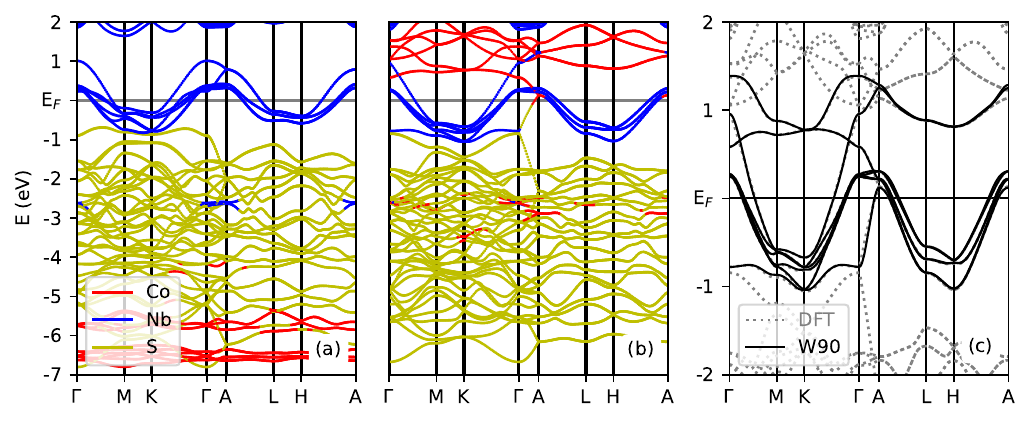}
  
  \caption{(Color online). 
  The electronic bands for Co$_{1/3}$NbS$_2$ calculated in the ferromagnetic (FM) state, with  
   (a) electronic bands for electrons with minority spin projection, and 
  (b) electronic bands for electrons with majority spin projection. 
  Red, blue, and yellow represent bands with dominant contributions coming from Co, Nb, and S orbitals.
  Consistently with $S_z\approx3/2 state$ of Co, all cobalt majority spin-projection bands are positioned deeply below the Fermi level, whereas three cobalt minority spin-projection bands remain above the Fermi level. 
  (c) The Wannier90 TB parametrization (full lines) is shown on the top of the calculated bands for electrons with minority spin projection. 
  The main TB parameters are $t_1\approx 0.06 $eV, $t_2 \approx  0.12 $ eV and $t_{Co}\approx 0.26$eV, close in value to those found for AF HOFK state.  The notation explained in the text.}
  \label{figFM-spag}
  \end{figure}

  The electronic structure of a fully ferromagnetic state is easiest to describe.  This electronic structure is pictured in Fig.~\ref{figFM-spag}.
  In this state, with magnetic moments all pointing in the same direction,  $S_z(Co) \approx+3/2$.
  With two Co atoms per unit cell, all bands Co-bands above the Fermi energy, three per Co atom,  are spin-down. 
  The remaining seven bands per Co atom are found below the Fermi level, with two of them of minority spin and five of majority spin character. 
  The considerable Nb-Co overlap integral provides the dominant interlayer hybridization for spin-down electrons. 
  The corresponding effective interlayer hybridization for spin-up electrons is much smaller, as the spin-up Co-orbitals, positioned several eV below the Fermi energy, do not contribute significantly to the inter-layer hybridization. 
  The generalization to more elaborate magnetic structures is relatively straightforward:  The Co atom in the $S_z \approx +3/2$ spin state provides the orbital and local hybridization path between Nb-layers for $S_z=-1/2$ electrons, but not for  $S_z=+1/2$ electrons. 
  The opposite applies for Co atoms in the $S_z\approx -3/2$ state.  
  Thus, for the electron of a given spin projection, the bridging orbital may be locally present or absent at the Co-site, depending on the $S_z$ component of the local magnetic moment of the Co atom. 
  The spin state of the Co atom determines orbitals and hybridization integrals in its immediate vicinity in a way very much independent of the type of the overall magnetic order.    
  Of course, in the AFM HOFK ordered state, the number of bands increases two times relative to the FM state, as the unit cell doubles, with six Co-bands showing above the Fermi level,  and spin-up and spin-down bands degenerate in energy. 
  Using the recipe described above, one can reproduce to good precision the DFT-calculated energy bands in the vicinity of the Fermi level of whatever linearly-polarized magnetically ordered states, using essentially the same set of tight-binding parameters. 
  One can foresee using the recipe to build the electronic bands in magnetically ordered states that are not linearly polarized, although we have not checked these against the DFT calculations.

  In summary, we may say that DFT calculations treats the spin-ordering on intercalated magnetic ions in a way reminiscent of the Ising field, where only direction matters: 
  Once the direction of the local magnetic moment is set, the self-consistent calculation produces a similar result for minority vs. majority spin projection density, irrespective of the spin sign.
  The electronic band structure then reflects how the particular magnetic ion blocks or assists the hopping in separate spin channels.  
  The interplay of quantum spin fluctuations and conducting electrons, crucial for strongly correlated electron systems, is lost in such approaches.

\end{document}